%% file: main.tex
\ifdefined\fullversion
\else
\def\fullversion{0}    
\fi

\ifdefined\cameraversion
\else
\def\cameraversion{0}    
\fi

\def\allow{1}      

\documentclass[envcountsame,runningheads,notitlepage]{llncs}
\ifnum\fullversion=1
\fi

\usepackage{bm}
\usepackage{chngpage}
\usepackage{algpseudocode}
\usepackage{amsmath}
\usepackage{tikz}
\usepackage{pgfplots}
\pgfplotsset{compat=1.18}
\usetikzlibrary{arrows.meta}

\usepackage{float}
\usepackage{newfloat}
\DeclareFloatingEnvironment[
    fileext=los,
    listname=List of Schemes,
    name=Scheme,
    placement=H,
    within=section,
]{scheme}

\usepackage{fontawesome5}

\newcommand{\emailicon}[1]{%
  \href{mailto:#1}{%
    \textsuperscript{\scriptsize\faIcon[regular]{envelope}}%
  }%
}

\usepackage{adjustbox}
\usepackage{blindtext}
\usepackage{placeins}

\ifnum\cameraversion=0
\fi

\input{tex_files/ZZ_header}
\input{macros}

\title{ECDSA.Fail: Open Autoresearch for Optimizing Elliptic-Curve Point Addition in Shor's Algorithm}
\titlerunning{ECDSA.Fail: Open Autoresearch for Point-Addition Optimization}
\input{authors}
\date{August 2026}
\hypersetup{
  hidelinks,pdftitle={ECDSA.Fail: Open Autoresearch for Optimizing Elliptic-Curve Point Addition in Shor's Algorithm},pdfauthor={ECDSA.Fail Contributors},pdfsubject={Open autoresearch and reversible secp256k1 point-addition circuit optimization},pdfkeywords={ECDSA, secp256k1, Shor's algorithm, reversible circuits, quantum resource estimation, autoresearch}
}

\begin{document}

\maketitle
\vspace{-0.3cm}

\begin{abstract}
\input{sections/00-abstract}

\end{abstract}

\ifnum\cameraversion=0
  \begingroup
  \makeatletter
  \let\l@title\@gobbletwo
  \let\l@author\@gobbletwo
  \fontsize{8.5pt}{10.5pt}\selectfont
  \tableofcontents
  \makeatother
  \endgroup
  \clearpage
\fi

\input{sections/01-introduction}

\input{sections/02-preliminaries}

\input{sections/03-challenge-benchmark}

\input{sections/04-open-autoresearch}

\input{sections/05-results}

\input{sections/06-limitations}

\input{sections/07-conclusion}

\input{sections/08-acknowledgments}

%
%

\bibliographystyle{splncs04}
\bibliography{refs}

%
%

\appendix
\renewcommand{\theHsection}{appendix.\Alph{section}}

\input{sections/09-contributors}




\end{document}

%% file: tex_files/ZZ_header.tex
\usepackage[utf8]{inputenc}
\usepackage[T1]{fontenc}
\usepackage{hyperref}
\usepackage{verbatim}
\usepackage{tikz}
\usetikzlibrary{positioning,calc,decorations.pathreplacing,backgrounds}
\usepackage{pgfplots}
\usepackage{pgfplotstable}
\pgfplotsset{compat=1.18}
\usepackage{subcaption}
\usepackage{forest}
\usepackage{xspace}
\usepackage{amsmath} 
\usepackage{amssymb}
\usepackage{mathtools}
\usepackage{pifont}
\usepackage{etoolbox}
\usepackage[normalem]{ulem}
\usepackage{booktabs}
\usepackage{longtable}
\usepackage{array}
\usepackage[capitalise,noabbrev]{cleveref}
\usepackage{cite}
\usepackage{multibib}
\usepackage{url}
\usepackage{algorithm}
\usepackage{algpseudocode}
\usepackage{paralist}
\usepackage{mathrsfs}
\usepackage{relsize}
\usepackage{stmaryrd}
\usepackage{multirow}
\usepackage[lambda,n,operators]{cryptocode}

\newtoggle{notes}
\toggletrue{notes} 

\newcommand{\F}{\mathbb{F}}

\makeatletter
\DeclareFontFamily{OMX}{MnSymbolE}{}
\DeclareSymbolFont{MnLargeSymbols}{OMX}{MnSymbolE}{m}{n}
\SetSymbolFont{MnLargeSymbols}{bold}{OMX}{MnSymbolE}{b}{n}
\DeclareFontShape{OMX}{MnSymbolE}{m}{n}{
    <-6>  MnSymbolE5
   <6-7>  MnSymbolE6
   <7-8>  MnSymbolE7
   <8-9>  MnSymbolE8
   <9-10> MnSymbolE9
  <10-12> MnSymbolE10
  <12->   MnSymbolE12
}{}
\DeclareFontShape{OMX}{MnSymbolE}{b}{n}{
    <-6>  MnSymbolE-Bold5
   <6-7>  MnSymbolE-Bold6
   <7-8>  MnSymbolE-Bold7
   <8-9>  MnSymbolE-Bold8
   <9-10> MnSymbolE-Bold9
  <10-12> MnSymbolE-Bold10
  <12->   MnSymbolE-Bold12
}{}

\let\llangle\@undefined
\let\rrangle\@undefined
\DeclareMathDelimiter{\llangle}{\mathopen}%
                     {MnLargeSymbols}{'164}{MnLargeSymbols}{'164}
\DeclareMathDelimiter{\rrangle}{\mathclose}%
                     {MnLargeSymbols}{'171}{MnLargeSymbols}{'171}
\makeatother


%% file: macros.tex
\newif\ifdraft
\ifnum\allow=1 \drafttrue \else \draftfalse \fi

\providecolor{Maroon}{HTML}{9E1B32}
\providecolor{Peach}{HTML}{E58F65}
\providecolor{SkyBlue}{HTML}{6FA8DC}    
\providecolor{Bittersweet}{HTML}{FE6F5E}
\providecolor{BenHuangGreen}{HTML}{2E8B57}
\providecolor{Bartosz}{HTML}{AAAC11}
\providecolor{FrancescoBlue}{HTML}{345995}
\providecolor{PierreLucOrange}{HTML}{FCA503}

\ifdraft
  \newcommand{\fixme}[1]{\noindent\colorbox{yellow}{\scriptsize\textbf{FIXME}}\,%
      \textbf{\textcolor{Maroon}{#1}}}
  \newcommand{\todo}[1]{\noindent\colorbox{yellow}{\scriptsize\textbf{TODO}}\,%
      \textbf{\textcolor{Maroon}{#1}}}
  \newcommand{\authornote}[3]{\colorbox{#2}{\scriptsize\textbf{#1}}\,%
      {\textcolor{#2}{#3}}}
  \newcommand{\manuel}[1]{\authornote{MANUEL}{Peach}{#1}}
  \newcommand{\jieyi}[1]{\authornote{JIEYI}{SkyBlue}{#1}}

  \newcommand{\benhuang}[1]{\authornote{BENHUANG}{BenHuangGreen}{#1}}
  
  \newcommand{\francesco}[1]{\authornote{FRANCESCO}{FrancescoBlue}{#1}}
  \newcommand{\pierreluc}[1]{\authornote{PIERRELUC}{PierreLucOrange}{#1}}

\else
  \newcommand{\fixme}[1]{}\newcommand{\todo}[1]{}
  \newcommand{\manuel}[1]{}\newcommand{\jieyi}[1]{}
  \newcommand{\benhuang}[1]{}
  \newcommand{\francesco}[1]{}
  \newcommand{\pierreluc}[1]{}
\fi

\providecommand{\ket}[1]{\ensuremath{\left\lvert #1\right\rangle}}

\providecommand{\challengename}{ECDSA.Fail}

\definecolor{arInk}{HTML}{22262E}   \definecolor{arMuted}{HTML}{6B7280}
\definecolor{arAccent}{HTML}{3D5A80} \definecolor{arAccentBg}{HTML}{E9EEF5}
\definecolor{arWarm}{HTML}{B07D3B}   \definecolor{arWarmBg}{HTML}{F7EEDE}
\definecolor{arBorder}{HTML}{C9CED6} \definecolor{arEdge}{HTML}{7A828E}


%% file: authors.tex
\author{%
\small
Jieyi Long\inst{1}\emailicon{jieyi@thetalabs.org}, Theodore Pender\inst{2},
Zhao Huang\inst{3}, Manuel B. Santos\inst{4},
Samrendra Kumar Singh\inst{5}, Bartosz Naskr\k{e}cki\inst{6,7},
BitWonka\inst{8},
Pierre-Luc Dallaire-Demers\inst{9}, Francesco Giannicola,
Ruben M. L. Paschoarelli\inst{4}, Oli Freuler\inst{5},
Jackie Chia-Hsun Lee\inst{10}, Vasily Gnuchev, Gopi Kannappan, John Boyer,
Xavier Butler, Akash Balasubramani\inst{5}, Jordan Newman,
Bereket Dereje,
Alexander Hertlein\inst{11}, Robert Kodra\inst{2},
Lucas Levy\inst{5}, Shaan Patel, JT Rose, Matt Zweil,
Okechukwu Wisdom\inst{12}, Tarek El-Eter\inst{2}, Edison Lee, Michael Dong\inst{3},
Alan Li\inst{3}, Anto Joseph\inst{13}, Duy Nguyen\inst{13},
Other Leaderboard Contributors%
\thanks{See Appendix~\ref{app:credit-contributors} for the complete contributor list.}, 
Gajesh Naik\inst{13}\emailicon{gajesh@eigenlabs.org},
Gautham Anant\inst{13}\emailicon{gautham@eigenlabs.org},
Soubhik Deb\inst{13}\emailicon{soubhik@eigenlabs.org},
Justin Drake\inst{14}\emailicon{justin@ethereum.org}%
}

\authorrunning{ECDSA.Fail Contributors}

\institute{%
\small
\textsuperscript{1}Theta Labs \quad
\textsuperscript{2}Starknet Foundation \quad
\textsuperscript{3}Brevis \quad
\textsuperscript{4}MultiVM Labs \\
\textsuperscript{5}StarkWare \quad
\textsuperscript{6}Adam Mickiewicz University Pozna\'{n} \quad
\textsuperscript{7}Warsaw University of Technology \quad
\textsuperscript{8}Octav \quad
\textsuperscript{9}Pauli Group \quad
\textsuperscript{10}ScienceVR \\
\textsuperscript{11}Sei Labs \quad
\textsuperscript{12}Stanford Free Systems Lab \quad
\textsuperscript{13}Eigen Labs \quad
\textsuperscript{14}Ethereum Foundation%
}

%% file: sections/00-abstract.tex
We study \emph{Open Autoresearch}, a paradigm in which humans and AI agents address optimization problems by publishing evaluator-verified improvements to a public leaderboard. We instantiate it in {\challengename}\footnote{Project website and the latest results: \url{https://ecdsa.fail}; source repository: \url{https://github.com/Layr-Labs/ecdsafail-challenge}.}, whose participants optimized a reversible \texttt{secp256k1} point-addition circuit, a bottleneck in Shor's algorithm for elliptic-curve cryptography. 
The benchmark minimizes the spacetime-inspired score $S=Q\times T$, where $Q$ is peak logical qubit width and $T$ is average executed Toffoli count. 
Participants reduced $S$ by $86.1\%$, with the best circuit at the data cutoff (26 July 2026) reaching $Q=1{,}151$ and $T=1{,}299{,}453$, giving $Q\times T\approx1.496$B. This is more than $50\%$ below the reported score of Google's point-addition circuit (un)disclosed via a zero-knowledge proof \cite{google26}. Because the benchmark supplies one addend classically, we construct a coherent windowed-addition-compatible variant implementing the single-call interface required by windowed Shor. It uses $Q=1{,}162$ and $T=1{,}684{,}161$; on $100{,}000$ random inputs, its empirical success probability was $\hat p=0.99809$, corresponding under an independently rerunnable per-call sensitivity model to the proxy $Q\times T/\hat p\approx1.961\mathrm{B}$. Its $Q$ and $T$ both lie below the Google and Schrottenloher operating points \cite{google26,s26}, although differing interfaces and accounting conventions preclude formal dominance. After the cutoff, the $Q \times T$ score was further reduced to $1.259\mathrm{B}$, while a separate low-width circuit reached $813$ qubits. The public record shows that AI agents complemented human judgment, providing evidence for open autoresearch on efficiently evaluable, machine-checkable objectives.

%% file: sections/01-introduction.tex
\section{Introduction}
\label{sec:intro}

Quantum computers exploit superposition, entanglement, and interference to obtain asymptotic advantages for certain computational problems~\cite{nielsen-chuang}. That same capability threatens widely deployed public-key cryptography~\cite{shor97}. A \emph{cryptographically relevant quantum computer} (CRQC) could execute Shor's algorithm~\cite{shor97} at cryptographic scale, defeating RSA through integer factorization and elliptic-curve cryptography through the elliptic-curve discrete logarithm problem (ECDLP)~\cite{proos-zalka03}. We refer to the arrival of such a machine as \emph{Q-day}. Although its timing remains uncertain, migration away from vulnerable cryptography is already under way. NIST has standardized post-quantum replacements, and the initial public draft of NIST IR~8547 proposes deprecating classical public-key algorithms at the 112-bit security level after 2030 and disallowing them after 2035~\cite{nistpqc24,nistir8547}. U.S. federal guidance further calls for migrating digital signatures in sensitive federal systems by 2031~\cite{ombm2615}. The threat is particularly consequential for \texttt{secp256k1}, used by Bitcoin and Ethereum for transaction authorization~\cite{nakamoto08,wood-ethereum,sec2,jmv01}: a CRQC capable of solving ECDLP could recover private signing keys from exposed public keys and forge signatures~\cite[pp.~40--41]{digitalassets25}.

When will Q-day come? The answer depends partly on the logical resources required by Shor's ECDLP circuit. Beginning with Proos and Zalka's detailed prime-field construction and Kaye and Zalka's space-efficient reversible techniques~\cite{proos-zalka03,kaye-zalka05}, successive work has reduced these estimates through explicit Toffoli circuits, windowed arithmetic, and architecture-specific implementations~\cite{rnsl17,hjn20,litinski23,grr23}. Google Quantum AI recently reported a further two- to threefold reduction over prior published estimates, attested by a zero-knowledge proof without releasing the circuit; Schrottenloher subsequently obtained comparable figures with an open, fully specified design~\cite{google26,s26}. At the physical layer, Cain et al.\ estimate that a reconfigurable neutral-atom architecture using high-rate qLDPC codes could solve a P-256 discrete logarithm with as few as $10{,}000$ physical qubits, or within days using roughly $26{,}000$, since additional qubits buy parallelism~\cite{cain26atoms}. Under explicit hardware and error-correction assumptions, Dallaire-Demers et al.\ place a full 256-bit ECDLP instance within a 2027--2033 window~\cite{dallaire2026brace}.

Traditionally, similar to other scientific research domains, leading quantum-circuit constructions have been developed by individual specialists or small teams through manual design and heuristic search~\cite{rnsl17,hjn20,litinski23,google26,s26,luo26space}. In recent years, LLM-based agents, including general-purpose coding agents such as Claude Code and Codex, have begun to automate parts of scientific research and broaden access to specialized computational workflows~\cite{aiscientist,alphaevolve,aygun26era,ding25scitoolagent}. Such systems can synthesize research ideas from the literature, modify and execute code, and iteratively improve solutions against quantitative evaluators with limited supervision~\cite{aiscientist,alphaevolve,aygun26era}. Their outputs nevertheless remain hypotheses until checked. With a machine-checkable evaluator, the practical bottleneck can shift from manual implementation toward hypothesis generation and verification.

\begin{figure}[!t]
  \centering
  \includegraphics[width=1.0\linewidth]{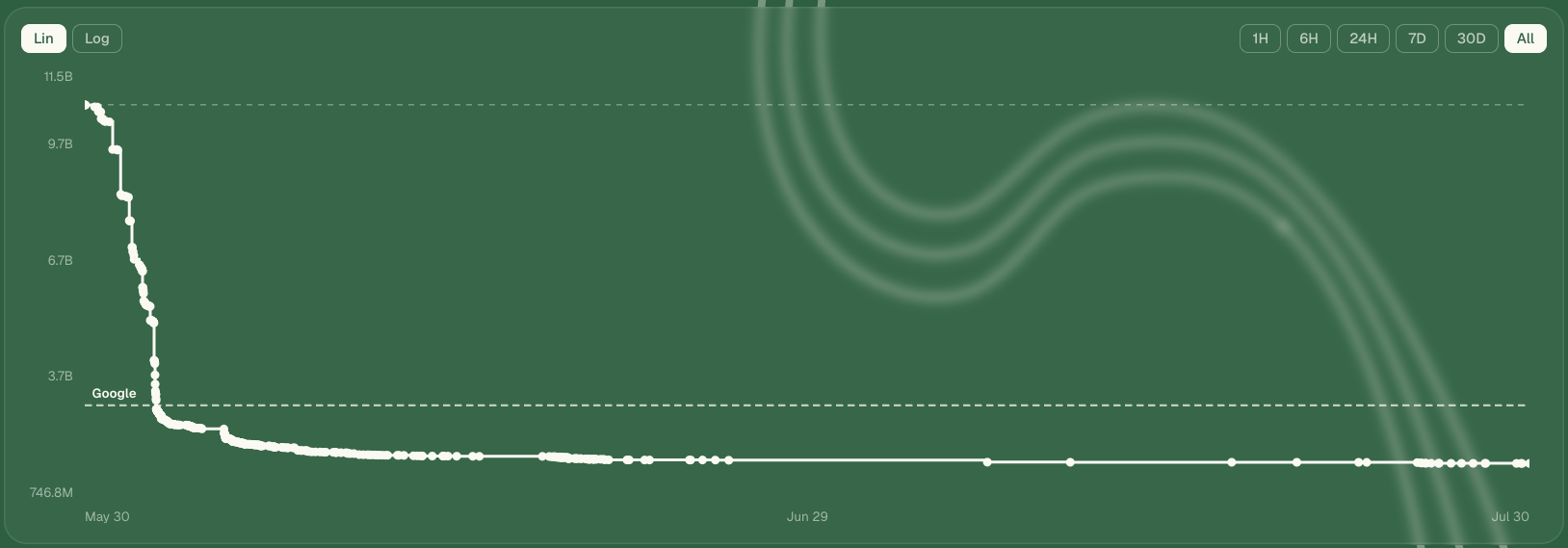}
  \caption{Evolution of the best {\challengename} $Q\times T$ score, from the $10.75$B baseline to $1.496$B at the data cutoff, an $86.1\%$ reduction. The dashed line marks the reported Google/Babbush et al.~\cite{google26}\ operating point of $2.993$B. The challenge score is more than $50\%$ lower, although differing interfaces and accounting conventions make the comparison contextual.}
  \label{fig:qxt-score-trajectory}
\end{figure}


Building on these capabilities, we introduce \emph{Open Autoresearch}: a hard frontier research problem is expressed as a measurable objective with a machine-checkable evaluator, and a heterogeneous, largely self-organizing community of humans and AI agents competes and builds on shared, verified results through public artifacts and a leaderboard. Although developed here for quantum-circuit optimization, the paradigm may apply to other problems with efficiently verifiable objectives. It extends recent AI-driven approaches to evaluator-guided program search, autonomous experimentation, and recent distributed autoresearch systems \cite{funsearch,alphaevolve,aiscientist,liu25autoresearch, karpathy26autoresearch,hyperspace26agi}, while emphasizing two features. First, it is \emph{open}: independent participants share a target and public artifacts, including documented failures, without requiring prior coordination. Second, it is \emph{observable}: progress forms a public sequence of verified, attributable steps, making the research process itself available for analysis. We hypothesize that this combination can help address frontier optimization problems with machine-checkable outcomes.

\begin{figure}[!t]
  \centering
  \includegraphics[width=1.0\linewidth]{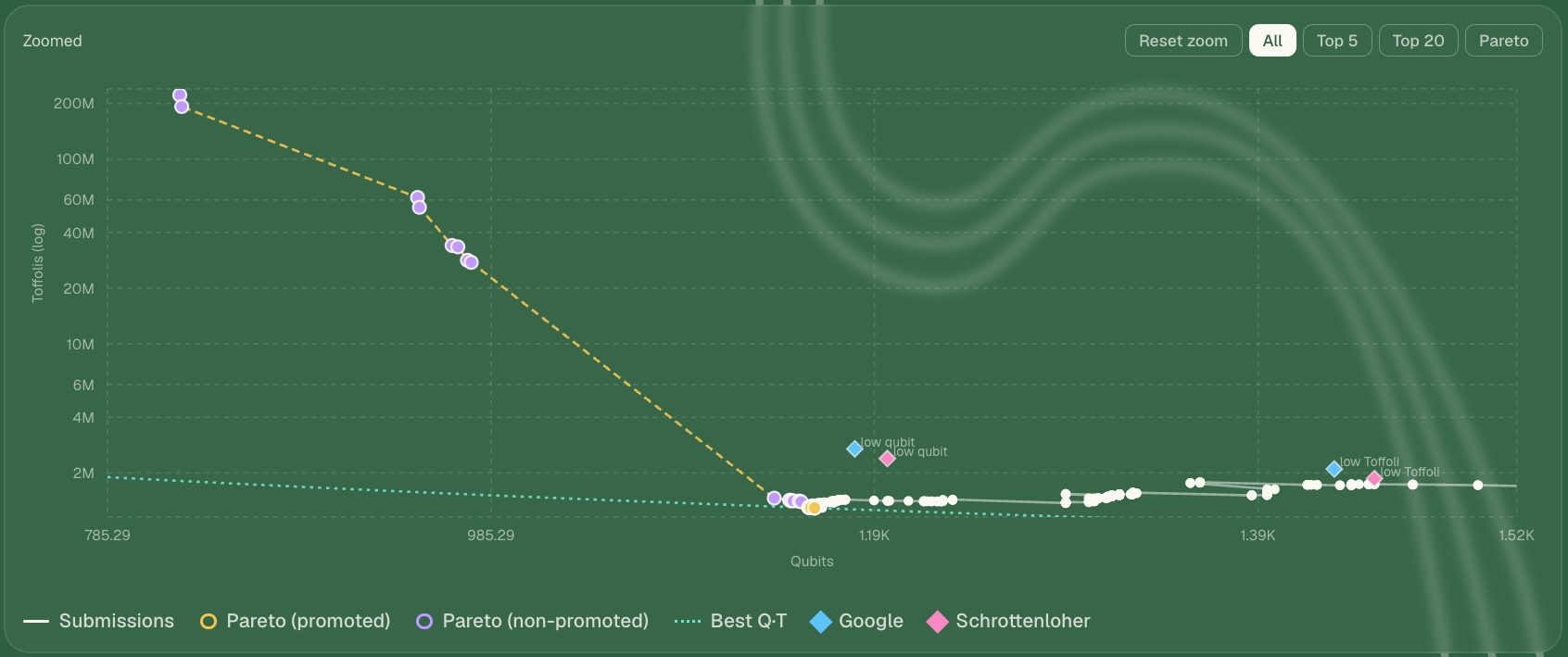}
  \caption{{\challengename} qubit--Toffoli Pareto frontier. White points are challenge submissions; orange and purple circles are promoted and non-promoted admitted Pareto points; the cyan dotted curve marks the best observed $Q\times T$; and diamonds mark the Google/Babbush~\cite{google26} (blue) and Schrottenloher~\cite{s26} (pink) operating points. External points are contextual because interfaces and accounting conventions differ; see \Cref{tab:admitted-pareto,tab:external-pa-comparison}.}
  \label{fig:compare-frontier-with-recent-work}
\end{figure}

Our evidence comes from {\challengename}, an open competition in which more than one hundred participants and their agents optimized a reversible \texttt{secp256k1} mixed point-addition circuit, minimizing $Q\times T$, where $Q$ is peak logical width and $T$ is average executed Toffoli count. As \Cref{fig:qxt-score-trajectory} shows, submissions reduced the score from the $10.75$B baseline ($Q=2{,}715$, $T=3{,}960{,}753$) to $1.496$B ($Q=1{,}151$, $T=1{,}299{,}453$) at the data cutoff (26 July 2026, 09:21:55 UTC), an $86.1\%$ reduction. In \Cref{fig:compare-frontier-with-recent-work}, this circuit lies below the reported Google/Babbush et al.~\cite{google26}\ and Schrottenloher~\cite{s26} operating points on both axes, with a product more than $50\%$ below Google's. To our knowledge, it had the lowest reported numerical \(Q\times T\) among \texttt{secp256k1} point-addition circuits as of the data cutoff. These comparisons are contextual rather than claims of formal dominance because interfaces, correctness assumptions, and accounting conventions differ. In particular, the benchmark supplies one addend classically, whereas windowed Shor selects it coherently. We therefore construct a windowed-addition-compatible single-call variant using $1{,}162$ qubits and approximately $1.68$ million average executed Toffoli gates, also below Google's and Schrottenloher's reported operating points ~\cite{google26,s26}. Like prior constructions~\cite{google26,s26}, our circuits trade a small error rate for lower resources. On a longitudinal sample evaluated over $50{,}000$ pseudorandom inputs, the retry-adjusted score $Q\times T/\hat p$ closely tracks the raw score because the empirical success rate $\hat p$ remains near one (\Cref{sec:QxT-score-evolution}). The score omits Toffoli depth and parallelism, so circuits with identical $Q$ and $T$ score equally even when their schedules differ (\Cref{sec:limitations}).

\subsection{Main Contributions}
\label{sec:contributions}

Our contributions span the research methodology, optimized circuits, and public record of their development.

\begin{itemize}
    \item \textbf{The Open Autoresearch paradigm.} We formulate open autoresearch as a verifier-gated paradigm in which anyone may participate without formal entry requirements, and humans and AI agents optimize a shared objective through a public evaluator, repository, leaderboard, and experimental record. Verified improvements become foundations for further work, while documented failures provide reusable evidence. To the best of our knowledge, {\challengename} is among the first large-scale open-autoresearch projects organized as a public, verifier-gated challenge for independently operated human--agent teams, with more than 100 leaderboard contributors and over 400 promoted submissions at the data cutoff. From the practices observed during {\challengename}, we derive a framework combining parallel search, human steering, verification, shared memory, and evolving agent skills (\Cref{sec:autoresearch-paradigm}). Participant-contributed harnesses and scaffolds are available in a \href{https://github.com/jieyilong/ecdsafail-autoresearch-harness}{public repository}.

    \item \textbf{Frontier-level point-addition circuits.} The {\challengename} challenge optimizes a reversible \texttt{secp256k1} mixed point-addition circuit. At the data cutoff, the best circuit has a $Q\times T$ product more than $50\%$ below the reported Google/Babbush et al.\ point, with reported $Q$ and $T$ below both the Google/Babbush et al.\ and Schrottenloher constructions~\cite{google26,s26} (\Cref{sec:results-summary}). The improvement remains substantial under the retry-adjusted proxy $Q\times T/\hat p$ (\Cref{sec:QxT-score-evolution}). Because accounting conventions differ, these comparisons are contextual rather than claims of formal dominance. Additional operating points characterize the Pareto frontier and low-width regime:
    \begin{itemize}
        \item The best $Q\times T$ submission at the data cutoff, \href{https://github.com/Layr-Labs/ecdsafail-challenge/commit/ae7676ddb082a83b83b48a4a773987b32c5e6009}{\texttt{8e9c9a2}}, uses $Q=1{,}151$ and $T=1{,}299{,}453$ (\Cref{sec:best-QxT-scoring-circuit}) and achieved a score of $1.496$B; its arithmetic and principal optimizations are detailed in \href{https://github.com/jieyilong/ecdsafail-supplemental-materials/blob/ef7ecd2b27cf27881874468e003e53e16ba48075/circuit_details/QxT_track_circuit_8e9c9a2/QxT_track_circuit_8e9c9a2_details.pdf}{Supplemental Note D}. Based on this circuit, we additionally construct a coherent windowed-addition-compatible variant with $Q=1{,}162$ and $T=1{,}684{,}161$ (\Cref{sec:windowed-addition-compatible-circuit}), documented in \href{https://github.com/jieyilong/ecdsafail-supplemental-materials/blob/ef7ecd2b27cf27881874468e003e53e16ba48075/circuit_details/windowed_QxT_track_circuit_8e9c9a2/windowed_QxT_track_circuit_8e9c9a2_details.pdf}{Supplemental Note F}. After the data cutoff, submission \href{https://github.com/Layr-Labs/ecdsafail-challenge/commit/897dda2b0cf267151ecd973252d2a5078cbf1b63}{\texttt{3616dbf}} introduced a novel comparison-free ``ping-pong'' GCD technique (\Cref{sec:ping-pong-dialog-gcd}), which further reduced the benchmark score to approximately $1.259$B, using $Q=1{,}321$ and $T=952{,}707$. This was the first result reported with fewer than one million average executed Toffoli gates and represents a $15.9\%$ reduction in $Q\times T$ relative to the cutoff incumbent.
        \item At the cutoff, the best low-$Q$ submission, \href{https://github.com/Layr-Labs/ecdsafail-challenge/commit/e6d9fa95ab81a3414517030c974e75349912e7ca}{\texttt{b6f2b0a}}, used $825$ qubits (\Cref{sec:best-low-Q-circuit}; \href{https://github.com/jieyilong/ecdsafail-supplemental-materials/blob/ef7ecd2b27cf27881874468e003e53e16ba48075/circuit_details/low_Q_track_circuit_b6f2b0a/low_Q_track_circuit_b6f2b0a_details.pdf}{Supplemental Note E}). A later submission, \href{https://github.com/Layr-Labs/ecdsafail-challenge/commit/364b042990a3e49d66357aa3d14c1f01722ee6c5}{\texttt{b4a51d2}}, reported $813$ qubits, $22$ below Luo et al.'s recent $835$-qubit result~\cite{luo26space}.
    \end{itemize}

    \item \textbf{Source-auditable architectures and optimizations.} We provide reproducible implementations and technical descriptions of the resulting circuits. They build on the record-and-replay Euclidean architecture (a.k.a dialog-GCD) originally introduced by Khattar et al.~\cite{ksgz25}, together with the TrailMix implementation substrate~\cite{trailmix}. Principal optimizations include jump-2 Euclidean steps, compressed transcript codecs, ping-pong dialog-GCD, Karatsuba squaring with pseudo-Mersenne modulo, register-shared extended Euclidean inversion, constant propagation, and dead-code elimination (\Cref{sec:jump-two-gcd,sec:base-five-codec,sec:ping-pong-dialog-gcd,sec:specialized-squaring,sec:register-shared-eea,sec:consprop,sec:dead-code-elimination}). A complete catalog appears in \href{https://github.com/jieyilong/ecdsafail-supplemental-materials/blob/ef7ecd2b27cf27881874468e003e53e16ba48075/appendices/anatomy_of_agent_identified_quantum_circuit_optimizations.pdf}{Supplemental Note B}.

    \item \textbf{Longitudinal and reproducible evidence.} We reconstruct the improvement trajectory, Pareto frontier, low-width branches, and three successive resource regimes, and classify $400$ scored source commits by their primary optimization mechanism (\Cref{sec:circuit-trajectory}). We release circuit sources, submission histories, frozen leaderboard data, operating-point manifests, and reconstruction procedures (\Cref{sec:reproducibility}); the per-commit ledger appears in \href{https://github.com/jieyilong/ecdsafail-supplemental-materials/blob/ef7ecd2b27cf27881874468e003e53e16ba48075/appendices/optimization_census_of_accepted_competition_source_commits.pdf}{Supplemental Note C}.
\end{itemize}

\subsection{Related Work}
\label{sec:related-work}

\paragraph{Resource estimation and circuit optimization for the ECDLP.} Quantum ECDLP resource estimates progressed from early circuit constructions~\cite{proos-zalka03,kaye-zalka05} to explicit Toffoli estimates and windowed arithmetic~\cite{rnsl17,hjn20}, followed by architecture-specific analyses and improved point-addition designs~\cite{litinski23,grr23,google26,s26,luo26,cain26atoms,garn-kan25}. Google attests its unpublished circuit through a zero-knowledge proof~\cite{google26}, whereas several recent designs are open~\cite{s26,luo26,garn-kan25}. These constructions were each developed by one researcher or team. In contrast, our public competition reaches comparable operating points through contributions from independent human--agent teams and releases the resulting source history. TrailMix likewise provides inspectable circuits but follows one maintained implementation rather than a community search~\cite{trailmix}. Our designs recombine reversible primitives and arithmetic architectures from this literature~\cite{cuccaro04,parent2017karatsuba,bernstein-yang19,ksgz25,khattargidney2025,kornerup2025spooky}.

\paragraph{Machine learning for circuit and algorithm optimization.} AlphaTensor, AlphaDev, AlphaTensor-Quantum, and reinforcement-learning methods apply trained optimizers to algorithm or circuit search~\cite{alphatensor,alphadev,alphatensor-quantum,fosel21}, extending classical superoptimization~\cite{massalin1987,schkufza2013stoke}. We instead use off-the-shelf agents, human steering, and a public leaderboard, without task-specific training. Because learned results may be rediscovered by conventional search or improved manually~\cite{kauers2023flip,neri2023alphadev}, we explicitly audit provenance and distinguish invention from rediscovery.


\paragraph{Agents and autoresearch under verifiable objectives.} FunSearch, AlphaEvolve, and The AI Scientist use generate--evaluate--iterate loops for program search and automated research \cite{funsearch,alphaevolve,aiscientist}. Liu et al.\ explicitly formulated Agent-Based Auto Research as a multi-agent framework spanning the scientific lifecycle \cite{liu25autoresearch}. Karpathy's \texttt{autoresearch} subsequently popularized a compact loop in which an agent repeatedly modifies a constrained implementation, evaluates it under a fixed objective and budget, and retains improvements \cite{karpathy26autoresearch}. Shortly thereafter, Hyperspace AI distributed this loop over an open peer-to-peer network whose agents shared experiments and accumulated progress through gossip, replicated leaderboards, and public source artifacts \cite{hyperspace26agi}. {\challengename} builds on this lineage but instantiates a complementary organizational model: independently operated human--agent teams use heterogeneous models and workflows while interoperating through a common domain-specific evaluator, public repository, and shared frontier.

\paragraph{Open, collaborative, and crowdsourced problem solving.} Distributed problem solving appears in open-innovation contests, crowd science, collaborative mathematics, and games such as Foldit~\cite{terwiesch-xu08,franzoni-sauermann14,gowers-nielsen09,cooper10foldit}. Open autoresearch extends this model by placing AI agents alongside human participants, targeting frontier computational-science problems, and preserving the public search record rather than only the winning result.

\subsection{Organization of the Paper}
\label{sec:paper-organization}

The remainder of the paper is organized as follows. \Cref{sec:prelims} reviews the necessary background. \Cref{sec:challenge-and-benchmark} defines the challenge, evaluator, resource model, and correctness scope. \Cref{sec:autoresearch-paradigm} develops the open autoresearch paradigm, \Cref{sec:results} presents the optimized circuits, principal techniques, resource trajectories, and reproducibility procedures, and \Cref{sec:limitations} discusses limitations and responsible interpretation. \href{https://github.com/jieyilong/ecdsafail-supplemental-materials/blob/ef7ecd2b27cf27881874468e003e53e16ba48075/appendices/detailed_results_and_circuit_improvement_trajectory.pdf}{Supplemental Note A}, \href{https://github.com/jieyilong/ecdsafail-supplemental-materials/blob/ef7ecd2b27cf27881874468e003e53e16ba48075/appendices/anatomy_of_agent_identified_quantum_circuit_optimizations.pdf}{Supplemental Note B}, and \href{https://github.com/jieyilong/ecdsafail-supplemental-materials/blob/ef7ecd2b27cf27881874468e003e53e16ba48075/appendices/optimization_census_of_accepted_competition_source_commits.pdf}{Supplemental Note C} provide detailed trajectory analysis, an optimization catalog, and a source-commit census, respectively. Circuit-level analyses of the $Q\times T$-track submission \texttt{8e9c9a2}, the $825$-qubit submission \texttt{b6f2b0a}, and the coherent windowed-addition variant of \texttt{8e9c9a2} appear in \href{https://github.com/jieyilong/ecdsafail-supplemental-materials/blob/ef7ecd2b27cf27881874468e003e53e16ba48075/circuit_details/QxT_track_circuit_8e9c9a2/QxT_track_circuit_8e9c9a2_details.pdf}{Supplemental Note D}, \href{https://github.com/jieyilong/ecdsafail-supplemental-materials/blob/ef7ecd2b27cf27881874468e003e53e16ba48075/circuit_details/low_Q_track_circuit_b6f2b0a/low_Q_track_circuit_b6f2b0a_details.pdf}{Supplemental Note E}, and \href{https://github.com/jieyilong/ecdsafail-supplemental-materials/blob/ef7ecd2b27cf27881874468e003e53e16ba48075/circuit_details/windowed_QxT_track_circuit_8e9c9a2/windowed_QxT_track_circuit_8e9c9a2_details.pdf}{Supplemental Note F}.

%% file: sections/02-preliminaries.tex
\section{Preliminaries}
\label{sec:prelims}

\subsection{Quantum Computing Background}
\label{sec:qc-background}

\paragraph{Qubits and quantum registers.}
A \emph{qubit} is the quantum analogue of a bit: a normalized vector in $\mathbb{C}^2$
spanned by the computational basis states $\lvert 0\rangle$ and $\lvert 1\rangle$. A
\emph{register} of $m$ qubits lives in the $2^m$-dimensional space spanned by the basis
states $\lvert x\rangle$ for $x \in \{0,1\}^m$, and its general (pure) state is a
\emph{superposition} $\sum_x \alpha_x \lvert x\rangle$ with
$\sum_x \lvert\alpha_x\rvert^2 = 1$. Measuring the register in the computational basis
returns $x$ with probability $\lvert\alpha_x\rvert^2$ and collapses the state onto
$\lvert x\rangle$. Throughout this paper registers encode $256$-bit integers. These registers are most
often the $\mathbb{F}_p$ coordinates of an elliptic-curve point. So ``the accumulator
holds a superposition of curve points'' means that two $256$-qubit coordinate registers
are jointly in a superposition of basis states, each encoding one affine point.
A quantum circuit can coherently transform a superposition of exponentially many computational-basis states, but their amplitudes cannot be read out individually: measurement yields only a sampled classical outcome. Quantum algorithms must therefore use interference to concentrate probability on outcomes that encode the desired global structure.

\paragraph{Reversible computation.}
During coherent portions of a quantum computation, the evolution of the quantum registers is governed by unitary operators~\cite{nielsen-chuang}. Practical circuits may also include measurement and classically controlled feed-forward, which are described by quantum instruments rather than by a single unitary acting on the algorithmic registers. Because unitary transformations are invertible, coherent arithmetic must be expressed using reversible gates that preserve sufficient information to recover their inputs. The resulting constraints are not unique to quantum computation: analogous requirements arise in classical reversible computing and in constant-time cryptographic implementations designed to resist timing attacks. They nevertheless become especially consequential when data may be in superposition. First, control flow depending on quantum data must be represented by coherently controlled operations within a fixed circuit schedule. An inherently data-dependent procedure, such as the variable-length GCD iteration used for modular inversion, must therefore be unrolled or padded to a fixed bound or iteration budget, with each branch replaced by controlled operations, yielding a circuit whose shape covers all supported inputs~\cite{rnsl17}. Second, intermediate quantum results cannot be irreversibly discarded while they may remain entangled with the live state; they must instead be reversibly uncomputed or removed through measurement-based uncomputation with appropriate outcome-dependent corrections.

\paragraph{Uncomputation and cleanliness.}
Scratch (\emph{ancilla}) qubits hold intermediate values during a computation. If an
ancilla is discarded or reused while still correlated with the data, it stays \emph{entangled} with the output registers and destroys the interference
on which the surrounding algorithm depends. The standard remedy, due to
Bennett~\cite{bennett73}, is to \emph{uncompute}: after the result is copied out, the
computation of the scratch is run in reverse so every ancilla returns to
$\lvert 0\rangle$, roughly doubling the gate cost. \emph{Measurement-based
uncomputation}~\cite{gidney18} replaces part of this inverse computation by measuring
ancillas in the X basis and applying classically conditioned Clifford corrections; it
saves Toffoli gates, but an incomplete correction leaves a residual \emph{relative
phase} on surviving branches. Beyond computing the right classical function, a circuit
intended to run inside a larger coherent algorithm must therefore also be clean in two
quantum senses: every ancilla returns to $\lvert 0\rangle$, and no residual phase
remains. The benchmark evaluator of \Cref{sec:challenge-and-benchmark} checks these as
three separate channels; the two quantum channels need dedicated checks precisely
because they are invisible to purely classical testing of input--output behavior.

\paragraph{Quantum table lookup.}
A recurring primitive is reading a \emph{classical} table at a \emph{quantum} address:
given precomputed constants $\mathcal{T}[0], \dots, \mathcal{T}[2^w - 1]$, a lookup (often called QROM, for
quantum read-only memory) implements
$\lvert j\rangle \lvert 0\rangle \mapsto \lvert j\rangle \lvert \mathcal{T}[j]\rangle$. A
$2^w$-entry lookup costs $O(2^w)$ Toffoli gates, while the subsequent \emph{unlookup} can
be made substantially cheaper by measurement-based
techniques~\cite{gidney19windowed,litinski23}. The canonical usage pattern is
\emph{lookup~$\to$~use~$\to$~unlookup}: the table value is loaded into a temporary
register, consumed by the arithmetic that needs it, and immediately uncomputed, so the
loaded data is live only briefly and can be scheduled away from the circuit's qubit
high-water mark. This pattern is the interface through which windowed variants of Shor's
algorithm (\Cref{sec:ecc-quantum}) consume elliptic-curve point tables, and it reappears
in this paper wherever a circuit must handle data selected by a quantum index.

\paragraph{Fault-tolerant cost accounting.}
Large-scale cryptanalytic computations are expected to require quantum error correction, whose cost model is sharply non-uniform. In many surface-code architectures, Clifford operations are comparatively inexpensive, whereas non-Clifford gates consume costly distilled resource states~\cite{bk05}. For example, a conventional exact Clifford+$T$ decomposition of a Toffoli gate uses seven single-qubit $T$ gates, although measurement-assisted constructions can reduce this requirement~\cite{jones13}. Non-Clifford operations therefore capture an essential component of a circuit's nonlinear computational work. Although magic-state distillation has historically been expected to dominate physical cost, practical performance may also be limited by routing, storage access, decoding, classical feed-forward, and available parallelism. The executed Toffoli count is consequently a useful but architecture-dependent work proxy rather than a literal time axis. Similarly, quantum error correction gives each live logical qubit a substantial physical footprint, so the peak logical width $Q$ lower-bounds the required logical storage without by itself determining the physical machine size. Writing $T$ for the average executed Toffoli count, the challenge defines
\begin{equation}
  S = Q \times T.
  \label{eq:qt-proxy}
\end{equation}
This product provides a compact, spacetime-inspired benchmark that rewards reductions in both logical width and nonlinear work. It is not a physical-resource estimate: it omits circuit depth, parallelism, routing, memory access, error-correction overhead, and other architecture-specific costs. Finer-grained alternatives include Toffoli depth, active-volume models, and architecture-specific gate schedules. Nevertheless, $S$ captures a central width--work trade-off while remaining inexpensive and straightforward to evaluate under fixed benchmark rules, which motivates its use here. \Cref{sec:challenge-and-benchmark} specifies the accounting convention, including the distinction between emitted and \emph{executed} Toffoli gates.

\paragraph{Kickmix circuits.}
Generic quantum circuits can generate dense entanglement, making them hard to simulate on a classical computer.
However, the circuits in this paper live in a restricted class, called \emph{kickmix}~\cite{google26}, that can be checked efficiently classically. 
By definition, a kickmix circuit is built from only three kinds of
operation~\cite{google26}: reversible permutation gates (X, CNOT, and Toffoli), which map one
computational-basis state to another; diagonal phase gates (such as Z, CZ, and CCZ),
which leave a basis state fixed and only attach a phase to it; and mid-circuit
measurement in the X basis followed by classically conditioned phase corrections, which
is the measurement-based uncomputation described above.
What these operations have in
common is that none of them spreads a basis state across the full $2^m$-dimensional
space. A computational-basis input therefore stays on a sparse support
of only a few basis states throughout the circuit, so the circuit never builds the dense
entanglement that makes general quantum computation hard to simulate. This is what lets the evaluator
of \Cref{sec:challenge-and-benchmark} verify value correctness, ancilla cleanliness, and
phase cleanliness on every test input rather than sampling or trusting the submitter.

\subsection{Elliptic Curve Cryptography and Quantum Attacks}
\label{sec:ecc-quantum}

\paragraph{Elliptic curves and the group law.}
An elliptic curve \cite{silverman2009arithmetic} over a prime field $\F_p$ (in short Weierstrass form, $p>3$) is the set of points $(x,y)$ satisfying $E:y^2=x^3+ax+b$, $4a^3+27b^2\neq 0$, together with a point at infinity $\mathcal{O}$. The points form an abelian group under the chord-and-tangent law with identity $\mathcal{O}$: for a point $R=(x_R,y_R)$ and another point $A=(x_A,y_A)$ with $A\neq\pm R$, their sum $R'=(x_{R'},y_{R'})=R+A$ is
\begin{equation}
  \lambda = \frac{y_R-y_A}{x_R-x_A}, \qquad
  x_{R'} = \lambda^2-x_R-x_A, \qquad
  y_{R'} = \lambda(x_A-x_{R'})-y_A,
  \label{eq:group-law}
\end{equation}
with $\lambda=(3x_R^2+a)/(2y_R)$ for doubling ($A=R$). All arithmetic is in $\F_p$. The division in $\lambda$ is a \emph{modular inverse} $u\mapsto u^{-1}\bmod p$. Classically, it is a minor cost, but in the reversible setting of \Cref{sec:qc-background} it becomes the dominant expense of a point addition, because inversion has no branch-free single-pass circuit and must be realized as an unrolled, input-oblivious GCD iteration that is then uncomputed (\Cref{sec:challenge-and-benchmark}). This paper concerns \texttt{secp256k1}, the curve $y^2=x^3+7$ over $p=2^{256}-2^{32}-977$~\cite{sec2}, whose group of $\mathbb{F}_{p}$-rational points is cyclic of $256$-bit prime order $r$, generated by a standardized base point $G$. Throughout, $n=256$ denotes the field bit-width; for \texttt{secp256k1} the group order satisfies $\lceil\log_2 r\rceil=n$ as well, so the exponent registers are likewise $n$ qubits wide.

\paragraph{ECDLP and ECDSA.}
The \emph{elliptic-curve discrete logarithm problem} (ECDLP) asks, given the generator $G$ and a public point $P=[k]G$, to recover the scalar $k$. The forward direction costs $O(n)$ point additions, but the best known classical inverters (Pollard's rho, baby-step giant-step) require $\Theta(\sqrt{r})\approx2^{128}$ group operations for \texttt{secp256k1}, which is far beyond feasibility. This asymmetry underpins ECDSA~\cite{jmv01}, whose security rests on the hardness of ECDLP: against a properly implemented scheme, the best known classical key-recovery attack is to solve ECDLP for the signer's private key. ECDSA over \texttt{secp256k1} is the signature scheme of Bitcoin, Ethereum, and much of the surrounding blockchain ecosystem, which is why concrete quantum resource estimates for this specific curve carry direct economic significance.

\paragraph{Shor's algorithm for ECDLP.}
Shor's algorithm extends to discrete logarithms in finite abelian groups, including elliptic-curve groups, and therefore solves ECDLP in polynomial time~\cite{shor97}. Proos and Zalka subsequently provided the first detailed quantum circuits and concrete resource analysis for ECDLP over prime-field elliptic curves~\cite{proos-zalka03}. In the idealized hidden-subgroup formulation, the exponent registers range over $\mathbb{Z}_r$, where $r$ is the order of $G$, and the probe function is
\begin{equation}
    f(u,v)=[u]G+[v]P=[u+kv]G.
    \label{eq:probe}
\end{equation}
Because $f(u-k,v+1)=f(u,v)$, the private key is encoded by the period direction $(-k,1)$. Preparing a uniform superposition over $\mathbb{Z}_r^2$, evaluating
\begin{equation}
    U_f:\ket{u,v}\ket{\mathcal O}
    \longmapsto
    \ket{u,v}\ket{[u]G+[v]P},
\end{equation}
and applying inverse Fourier transforms over $\mathbb{Z}_r$ produces characters in the annihilator of this period. Thus, in the idealized model, the measured pair $(c,d)$ satisfies
\begin{equation}
    d\equiv kc\pmod r,
\end{equation}
from which $k=d\,c^{-1}\bmod r$ can be recovered when $c$ is invertible.

A practical implementation stores the exponent values in $n$-qubit registers whose computational basis ranges over $\mathbb{Z}_{2^n}$. Applying $H^{\otimes n}$ prepares each register in a uniform superposition. Because $2^n$ generally differs from the subgroup order $r$, the measured Fourier outcomes do not satisfy the ideal relation $d\equiv kc\pmod r$ exactly; instead, they cluster near values consistent with that relation. Classical postprocessing extracts candidate values of $k$, which are verified by checking whether $P=[k]G$. A semiclassical Fourier transform can reduce the required control-register storage, but it does not literally collapse two existing registers into one qubit. Rather, it evaluates the Fourier transform iteratively by preparing a reusable control qubit, using it to control the elliptic-curve arithmetic, applying the appropriate phase rotation, measuring it, and then repeating the process. In either formulation, the controlled elliptic-curve arithmetic dominates the resource cost: evaluating $U_f$ requires scalar multiplication by both $G$ and $P$ and therefore accounts for most of the logical width and Toffoli count. \Cref{fig:windowed-shor-ecdlp-overview} shows the coherent-register formulation, while the windowed point additions used within its scalar multiplications are detailed in \Cref{fig:two-methods}.

\paragraph{Two methods for scalar multiplication.}
The two scalar multiplications inside $U_f$, namely $[u]G$ and $[v]P$, decompose into sequences of point additions that follow one of two standard methods: \emph{double-and-add} or \emph{windowing} (\Cref{fig:two-methods}). These methods expose different point-addition interfaces, a distinction that matters throughout this paper.

\input{figs/shor-windowed-ecdlp-overview}

\input{figs/point-addition-methods}

In the original double-and-add method~\cite{proos-zalka03,rnsl17}, let $(s,B)$ denote either $(u,G)$ or $(v,P)$ and write $s=\sum_i s_i2^i$. At step $i$, the fixed, classically known point $A_i=[2^i]B$ is added to the accumulator $R$, controlled by the scalar bit $s_i$. The required primitive is therefore a controlled mixed elliptic-curve point addition
\begin{equation}
  U_{A_i}:\ket{s_i}\ket{R}
  \longmapsto
  \ket{s_i}\ket{R+[s_i]A_i},
  \label{eq:mixed-add}
\end{equation}
whose addend $A_i$ is a classical constant fixed at circuit-construction time. Across the two $n$-qubit exponent registers $U$ and $V$ of the probe function~\eqref{eq:probe}, this primitive is executed $2n$ times~\cite{rnsl17}. Writing $\mathrm{PA}^{\mathrm{Toff}}_n$ for the Toffoli count of one point addition, the full double-and-add Shor ECDLP estimate is
\begin{equation}
  \mathrm{ECDLP}^{\mathrm{Toff}}_n
  \approx 2n\cdot\mathrm{PA}^{\mathrm{Toff}}_n.
  \label{eq:daa-shor}
\end{equation}

The windowed approach~\cite{hjn20} trades point additions for table lookups. Rather than consuming one exponent bit at a time, it groups the bits into $w$-bit values called \emph{windows}. Following previous work~\cite{google26,litinski23}, we use $w=16$. For either $(s,B)=(u,G)$ or $(s,B)=(v,P)$, with $w\mid n$, scalar multiplication can be written as
\begin{equation}
  [s]B
  =
  \sum_{\substack{0\leq i<n\\i\equiv0\!\!\!\pmod w}}
  [s_{i:i+w}2^i]B,
  \label{eq:windowed-scalar}
\end{equation}
where $s_{i:i+w}$ is the $w$-bit window of $s$ beginning at bit $i$. For each window position $i$, let $B_i=[2^i]B$ and classically precompute the table
\begin{equation}
  \mathcal{T}^{(i)}_B
  =
  \bigl[\,[0]B_i,\,[1]B_i,\,\ldots,\,[2^w-1]B_i\,\bigr].
\end{equation}
For the quantum window value $j=s_{i:i+w}$, the lookup selects the addend $A_i=\mathcal{T}^{(i)}_B[j]$, and the windowed point-addition primitive realizes
\begin{equation}
  U_{\mathcal{T}^{(i)}_B}:\ket{j}\ket{R}
  \longmapsto
  \ket{j}\ket{R+A_i},
  \qquad
  A_i=\mathcal{T}^{(i)}_B[j].
  \label{eq:windowed-add}
\end{equation}
Here, the window index $j$ is quantum data stored in exponent register $U$ or $V$. Implementing this operation as $2^w$ separately controlled mixed adders, one for each table entry, would inflate the Toffoli cost by a factor of $2^w$ and is therefore prohibitive. Instead, the selected addend $A_i$ is materialized using a QROM lookup at address $\ket{j}$, added coherently to the accumulator, and then uncomputed through the lookup/use/unlookup pattern of \Cref{sec:qc-background}.

Grouping the exponent bits into windows reduces the conceptual number of point additions for the two scalar multiplications from $2n$ to $2n/w$. The resource estimate applies two additional schedule-boundary optimizations. First, the point selected by the initial lookup is written directly into the accumulator, avoiding one point addition. Second, classical postprocessing permits three terminal point additions to be omitted. The resulting windowed schedule therefore contains $2n/w-4$ point-addition calls. These four savings are not consequences of windowing itself: analogous optimizations also shorten the non-windowed schedule. For $n=256$, classical postprocessing saves roughly $48$ bitwise additions, while direct lookup of the starting point saves roughly another $16$. Including an additional QROM cost of $3\cdot2^w$ Toffoli gates per retained windowed addition gives the full Shor ECDLP estimate~\cite{google26,litinski23}
\begin{equation}
  \mathrm{ECDLP}^{\mathrm{Toff}}_n
  \approx
  \bigl(\mathrm{PA}^{\mathrm{Toff}}_n+3\cdot2^w\bigr)
  \left(\frac{2n}{w}-4\right).
  \label{eq:full-attack}
\end{equation}
For $n=256$ and $w=16$, windowing first reduces the conceptual schedule from $512$ bitwise additions to $32$ windowed additions; direct initialization and classical postprocessing then reduce the implemented estimate to $28$ point-addition calls, each using tables with $2^{16}$ entries. \Cref{fig:two-methods} contrasts the two point-addition interfaces at the circuit level.

\paragraph{Elliptic-curve point-addition blueprint.} Elliptic-curve point addition is defined by the group law in \eqref{eq:group-law}. Previous constructions~\cite{proos-zalka03,rnsl17,s26} realize this operation through the sequence of in-place field-arithmetic transformations given in \Cref{alg:point-addition}.

As observed by Proos and Zalka~\cite[Sec.~4.3]{proos-zalka03}, the key to an in-place reversible implementation is that the slope $\lambda$ can be recovered from either the input points $R$ and $A$ or the output point $R'=R+A$:
\begin{equation}
  \lambda
  = \frac{y_R-y_A}{x_R-x_A}
  = -\frac{y_{R'}+y_A}{x_{R'}-x_A}
  \pmod p.
  \label{eq:output-slope}
\end{equation}
The coordinate updates in \Cref{alg:point-addition} follow directly from the group law and \eqref{eq:output-slope}. Because $x_{R'}=\lambda^2-x_R-x_A$ and $X$ initially holds $x_R-x_A$, Steps~3--4 give $X=(x_R-x_A)+3x_A-\lambda^2=x_A-x_{R'}$. The final update in Step~7, $X\gets x_A-X$, therefore produces $x_{R'}$. Similarly, \eqref{eq:output-slope} implies $\lambda(x_A-x_{R'})=y_{R'}+y_A$, so Steps~5--6 transform $Y$ from $\lambda$ into $y_{R'}$. Further details on the quantum operations used to implement these transformations are provided in \href{https://github.com/jieyilong/ecdsafail-supplemental-materials/blob/ef7ecd2b27cf27881874468e003e53e16ba48075/appendices/anatomy_of_agent_identified_quantum_circuit_optimizations.pdf}{Supplemental Note B: List of Agent-Identified Optimizations}.

\begin{algorithm}[t]
  \caption{In-place generic addition of a classically specified addend $A$ to a quantum accumulator $R$, realizing $R'=R+A$. Registers $X$ and $Y$ hold field elements and are recycled in place.}
  \label{alg:point-addition}
  \begin{algorithmic}[1]
    \Statex \textbf{Input:} $R=(x_R,y_R)\in E(\F_p)$ and $A=(x_A,y_A)\in E(\F_p)$.
    \Statex \textbf{Assumptions:} $R,A\neq\mathcal O$ and $R\notin\{A,-A,-2A\}$; equivalently, $x_R\neq x_A$ and $x_A\neq x_{R'}$, where $R'=R+A$.
    \Statex \textbf{Output:} The coordinate registers contain $R'=(x_{R'},y_{R'})=R+A$.
    \State $X \gets x_R-x_A$;\quad $Y \gets y_R-y_A$
      \Comment{$X=d_x,\;Y=d_y$}
    \State $Y \gets YX^{-1}$
      \Comment{$Y=\lambda$}
    \State $X \gets X+3x_A$
      \Comment{$X=x_R+2x_A$}
    \State $X \gets X-Y^2$
      \Comment{$X=x_A-x_{R'}$}
    \State $Y \gets YX$
      \Comment{$Y=\lambda(x_A-x_{R'})$}
    \State $Y \gets Y-y_A$
      \Comment{$Y=y_{R'}$}
    \State $X \gets x_A-X$
      \Comment{$X=x_{R'}$}
  \end{algorithmic}
\end{algorithm}

\paragraph{Dialog-GCD record and replay.} An efficient way to implement the in-place multiplication and division in \Cref{alg:point-addition} is Schrottenloher's record-and-replay Euclidean construction~\cite{s26}, based on the dialog representation of Khattar et al.~\cite{ksgz25}. Rather than maintaining B\'ezout coefficients throughout the Euclidean descent, its record phase evolves only two shrinking integers and stores the branch decisions needed for later reconstruction. In this paragraph and the related optimization subsections, $u$ and $v$ locally denote Euclidean operands rather than the exponent variables of Shor's algorithm.

\begin{algorithm}[t]
  \caption{Dialog-GCD record phase, based on Schrottenloher~\cite[Alg.~2]{s26}.}
  \label{alg:dialog-gcd-record}
  \begin{algorithmic}[1]
    \Statex \textbf{Input:} $(u,v)=(p,x)$ with $x\in\F_p^{*}$ and odd $u$; fixed ordinary-divstep budget $L_{\mathrm{div}}$
    \Statex \textbf{Output:} coherent transcript $\tau=(\sigma_0,\ldots,\sigma_{L_{\mathrm{div}}-1})$
    \State $\tau\gets()$
    \For{$i=0,\ldots,L_{\mathrm{div}}-1$}
      \State $b_i\gets[v\text{ is odd}]$
        \Comment{subtraction flag}
      \State $s_i\gets b_i\wedge[u>v]$
        \Comment{swap flag}
      \State $\sigma_i\gets(b_i,s_i)$;\quad
             $\tau\gets\tau\mathbin{\Vert}\sigma_i$
      \If{$s_i$}
        \State $(u,v)\gets(v,u)$
      \EndIf
      \If{$b_i$}
        \State $v\gets v-u$
      \EndIf
      \State $v\gets v/2$
    \EndFor
    \Statex \textbf{Successful endpoint:} $(u,v)=(1,0)$
  \end{algorithmic}
\end{algorithm}

\Cref{alg:dialog-gcd-record} provides the pseudo code. When $b_i=0$, $v$ is already even; when $b_i=1$, the optional swap leaves two odd operands and their difference is even. Thus, the final division by two is exact in either case. Because $x$ may be quantum data, the symbols $\sigma_i$ form a coherent quantum transcript $\tau$ and cannot be measured or discarded.

For the transcript generated by a nonzero $x$, let $D_x$ denote the resulting linear Euclidean transformation. Successful convergence gives $D_x(p,x)=(1,0)$. Reverse traversal of $\tau$ applies the corresponding modular updates to the B\'ezout or payload registers, yielding $D_x(0,y)=(yx^{-1},0)$ and $D_x^{-1}(y,0)=(0,yx)$. Stage~2 uses $x=d_x$ and $y=d_y$ to obtain $\lambda=d_y d_x^{-1}$, while Stage~5 uses the inverse transformation for in-place multiplication. The optimized iteration schedule and transcript representation used by the final circuit are described in \Cref{sec:jump-two-gcd,sec:base-five-codec}.

%% file: figs/shor-windowed-ecdlp-overview.tex
\begin{figure}[!t]
  \centering
  \makebox[\linewidth][c]{%
    \resizebox{1.15\linewidth}{!}{%
      \begin{tikzpicture}[
        x=1cm,
        y=1cm,
        shorwire/.style={black, line width=0.8pt},
        shorgate/.style={draw=black, line width=0.8pt, fill=blue!6,
          minimum height=7mm, rounded corners=1pt, inner sep=3pt,
          align=center, font=\normalsize},
        shorlookup/.style={draw=black, line width=0.8pt, fill=orange!12,
          minimum height=6mm, rounded corners=1pt, inner sep=3pt,
          align=center, font=\normalsize},
        shoradd/.style={draw=black, line width=0.8pt, fill=violet!8,
          minimum height=7mm, rounded corners=1pt, inner sep=3pt,
          align=center, font=\normalsize},
        shorlabel/.style={font=\large},
      ]
        \path[use as bounding box] (0.2,0.25) rectangle (28.6,11.0);

        \node[shorlabel, anchor=east] at (1.55,9.55)
          {$U:\ \ket{0}^{\otimes256}$};
        \node[shorlabel, anchor=east] at (1.55,7.85)
          {$V:\ \ket{0}^{\otimes256}$};

        \node[shorgate, minimum width=19mm] (fu) at (2.65,9.55)
          {$H^{\otimes256}$};
        \node[shorgate, minimum width=19mm] (fv) at (2.65,7.85)
          {$H^{\otimes256}$};

        \node[shorgate, fill=violet!8, minimum width=17mm] (splitu)
          at (4.65,9.55) {split $U$\\$16\times16$};
        \node[shorgate, fill=violet!8, minimum width=17mm] (splitv)
          at (4.65,7.85) {split $V$\\$16\times16$};

        \draw[shorwire] (1.60,9.55) -- (fu.west);
        \draw[shorwire] (1.60,7.85) -- (fv.west);
        \draw[shorwire] (fu.east) -- (splitu.west);
        \draw[shorwire] (fv.east) -- (splitv.west);

        \draw[black!60, line width=0.8pt, dashed, rounded corners=2pt]
          (5.70,0.40) rectangle (21.35,10.55);
        \node[font=\Large] at (13.53,10.15)
          {$U_f$: coherent windowed double-scalar multiplication};

        \draw[shorwire]
          (splitu.east) -- (5.90,9.55) -- (5.90,7.65);
        \foreach \yy in {9.05,8.35,7.65}{
          \draw[shorwire] (5.90,\yy) -- (6.85,\yy);
          \draw[shorwire] (21.05,\yy) -- (22.00,9.55);
        }

        \draw[shorwire]
          (splitv.east) -- (5.55,7.85) -- (5.55,5.15);
        \foreach \yy in {6.55,5.85,5.15}{
          \draw[shorwire] (5.55,\yy) -- (6.85,\yy);
          \draw[shorwire] (21.05,\yy) -- (22.00,7.85);
        }

        \foreach \yy in {9.05,8.35,7.65}{
          \draw[shorwire] (6.85,\yy) -- (21.05,\yy);
        }

        \node[font=\normalsize, anchor=south east, inner sep=0pt]
          at (6.82,9.10) {$\ket{u_0}$};
        \node[font=\normalsize, anchor=south east, inner sep=0pt]
          at (6.82,8.40) {$\ket{u_1}$};

        \foreach \xx in {6.42,6.54,6.66}{
          \fill[black] (\xx,8.17) circle[radius=0.027];
        }

        \node[font=\normalsize, anchor=south east, inner sep=0pt]
          at (6.82,7.70) {$\ket{u_{15}}$};

        \foreach \yy in {9.05,8.35,7.65}{
          \draw[shorwire] (6.94,\yy-0.12) -- (7.18,\yy+0.12);
          \node[font=\small] at (7.06,\yy+0.31) {$16$};
        }

        \node[shorlookup] (LuZero) at (8.10,9.05) {Lookup};
        \node[shorlookup] (LuOne)  at (10.40,8.35) {Lookup};
        \node[shorlookup] (LuLast) at (12.70,7.65) {Lookup};

        \foreach \yy in {6.55,5.85,5.15}{
          \draw[shorwire] (6.85,\yy) -- (21.05,\yy);
        }

        \node[font=\normalsize, anchor=south east, inner sep=0pt]
          at (6.82,6.60) {$\ket{v_0}$};
        \node[font=\normalsize, anchor=south east, inner sep=0pt]
          at (6.82,5.90) {$\ket{v_1}$};

        \foreach \xx in {6.42,6.54,6.66}{
          \fill[black] (\xx,5.68) circle[radius=0.027];
        }

        \node[font=\normalsize, anchor=south east, inner sep=0pt]
          at (6.82,5.20) {$\ket{v_{15}}$};

        \foreach \yy in {6.55,5.85,5.15}{
          \draw[shorwire] (6.94,\yy-0.12) -- (7.18,\yy+0.12);
          \node[font=\small] at (7.06,\yy+0.31) {$16$};
        }

        \node[shorlookup] (LvZero) at (14.90,6.55) {Lookup};
        \node[shorlookup] (LvOne)  at (17.20,5.85) {Lookup};
        \node[shorlookup] (LvLast) at (19.50,5.15) {Lookup};

        \node[shorlabel, anchor=east] at (1.65,1.85)
          {$R:\ \ket{\mathcal O}$};
        \draw[shorwire] (1.70,1.85) -- (24.20,1.85);
        \draw[shorwire] (2.05,1.73) -- (2.30,1.97);
        \node[font=\small] at (2.18,2.22) {$2\times256$};

        \node[shoradd, minimum width=17mm] (AuZero) at (8.10,1.85)
          {$+[u_0]G$};
        \node[shoradd, minimum width=19mm] (AuOne) at (10.40,1.85)
          {$+[u_1 2^{16}]G$};
        \node[shoradd, minimum width=21mm] (AuLast) at (12.70,1.85)
          {$+[u_{15}2^{240}]G$};

        \node[shoradd, minimum width=17mm] (AvZero) at (14.90,1.85)
          {$+[v_0]P$};
        \node[shoradd, minimum width=19mm] (AvOne) at (17.20,1.85)
          {$+[v_1 2^{16}]P$};
        \node[shoradd, minimum width=21mm] (AvLast) at (19.50,1.85)
          {$+[v_{15}2^{240}]P$};

        \draw[shorwire] (LuZero.south) -- (AuZero.north);
        \draw[shorwire] (LuOne.south)  -- (AuOne.north);
        \draw[shorwire] (LuLast.south) -- (AuLast.north);
        \draw[shorwire] (LvZero.south) -- (AvZero.north);
        \draw[shorwire] (LvOne.south)  -- (AvOne.north);
        \draw[shorwire] (LvLast.south) -- (AvLast.north);

        \draw[
          decorate,
          decoration={brace, amplitude=4pt},
          line width=0.7pt
        ] (13.75,1.25) -- (7.20,1.25)
          node[midway, below=5pt, font=\normalsize]
          {$16$ $G$-window additions};

        \draw[
          decorate,
          decoration={brace, amplitude=4pt},
          line width=0.7pt
        ] (20.55,1.25) -- (14.00,1.25)
          node[midway, below=5pt, font=\normalsize]
          {$16$ $P$-window additions};

        \node[shorgate, fill=violet!8, minimum width=16mm] (mergeu)
          at (22.75,9.55) {merge $U$};
        \node[shorgate, fill=violet!8, minimum width=16mm] (mergev)
          at (22.75,7.85) {merge $V$};

        \node[shorgate, minimum width=18mm, font=\small] (ifu)
          at (24.65,9.55) {$\operatorname{QFT}^{-1}_{2^{256}}$};
        \node[shorgate, minimum width=18mm, font=\small] (ifv)
          at (24.65,7.85) {$\operatorname{QFT}^{-1}_{2^{256}}$};

        \node[shorgate, minimum width=17mm] (mu)
          at (26.75,9.55) {measure $s$};
        \node[shorgate, minimum width=17mm] (mv)
          at (26.75,7.85) {measure $t$};

        \draw[shorwire] (mergeu.east) -- (ifu.west);
        \draw[shorwire] (mergev.east) -- (ifv.west);
        \draw[shorwire] (ifu.east) -- (mu.west);
        \draw[shorwire] (ifv.east) -- (mv.west);

        \node[shorgate, minimum width=22mm] at (23.30,1.85)
          {measure or\\discard $R$};

        \node[
          shorlookup,
          text width=31mm,
          minimum height=27mm
        ] (post) at (25.35,5.10)
          {After all windows:\\
           $R=[u]G+[v]P$\\
           $=[u+kv]G$.\\
           Fourier samples satisfy\\
           $t\approx ks\pmod r$,\\
           from which $k$ is recovered\\
           and verified by $P=[k]G$.};

        \draw[dashed, ->, line width=0.7pt]
          (mu.east) -- (post.north east);
        \draw[dashed, ->, line width=0.7pt]
          (mv.east) -- (post.south east);
      \end{tikzpicture}%
    }%
  }
  \caption{Placement of the point-addition kernel within a power-of-two implementation of windowed Shor ECDLP. The coherent exponent registers are prepared over $\mathbb{Z}_{2^{256}}$ and split into $16$-bit windows. Each window coherently selects a precomputed multiple of $G$ or $P$ and adds it to the accumulator, yielding $R=[u]G+[v]P$ before inverse power-of-two Fourier transforms and measurement. Unlike the idealized formulation over $\mathbb{Z}_r^2$, the measured Fourier samples need not satisfy the annihilator relation exactly; they concentrate near it and require classical postprocessing followed by verification of the recovered candidate $k$. The figure shows the coherent-register formulation; an iterative semiclassical implementation can instead serialize the Fourier controls using a recycled control qubit. The braces show the $32$ conceptual window additions, while boundary optimizations reduce the estimate used in this paper to $28$ point-addition calls. Submission \texttt{8e9c9a2} implements the mixed point-addition kernel, and \Cref{sec:windowed-addition-compatible-circuit} implements and evaluates a windowed-addition-compatible variant that coherently supplies the addend through QROM. Instantiating the complete $28$-call window schedule, all shifted tables, the Fourier transforms, and the classical postprocessing remains future work.}
  \label{fig:windowed-shor-ecdlp-overview}
\end{figure}
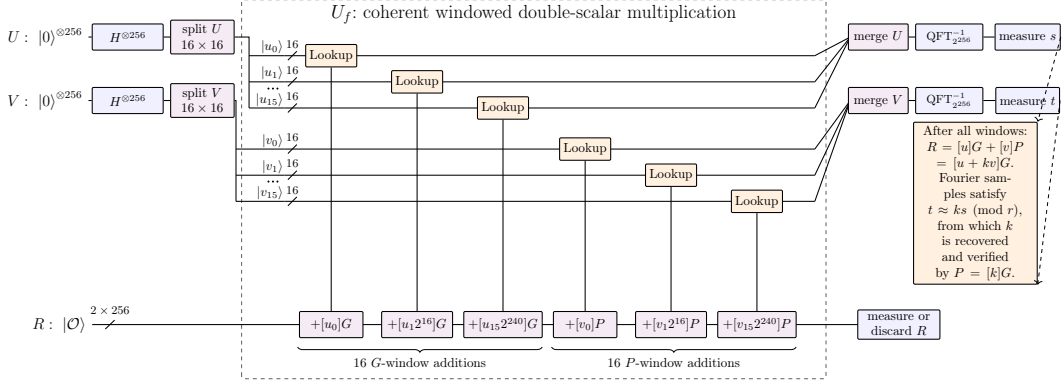

%% file: figs/point-addition-methods.tex
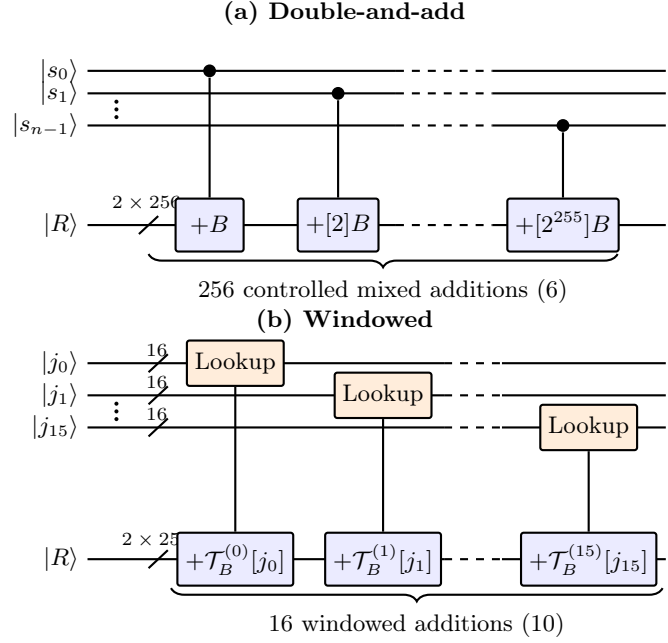
\begin{figure}[t!]
  \centering
  \begin{tikzpicture}[
    scale=0.85,
    wire/.style={thick},
    gate/.style={draw, thick, fill=blue!8, minimum height=7mm,
                 minimum width=9mm, rounded corners=1pt, inner sep=3pt},
    lookup/.style={draw, thick, fill=orange!14, minimum height=6mm,
                   rounded corners=1pt, inner sep=3pt},
    ctrl/.style={circle, fill=black, inner sep=1.7pt},
    lab/.style={font=\small},
  ]
    \node[font=\bfseries] at (5.6,3.9) {(a) Double-and-add};

    \foreach \yy in {3.0,2.65,2.15,0.6}{
      \draw[wire] (1.6,\yy) -- (6.4,\yy);
      \draw[wire, dashed] (6.4,\yy) -- (7.9,\yy);
      \draw[wire] (7.9,\yy) -- (10.6,\yy);
    }
    \node[lab, left] at (1.6,3.0) {$\ket{s_0}$};
    \node[lab, left] at (1.6,2.65) {$\ket{s_1}$};
    \node[lab, left] at (1.6,2.15) {$\ket{s_{n-1}}$};
    \node[lab, left] at (1.6,0.6) {$\ket{R}$};
    \fill (2.05,2.52) circle (0.9pt);
    \fill (2.05,2.40) circle (0.9pt);
    \fill (2.05,2.28) circle (0.9pt);

    \draw[thick] (2.4,0.46) -- (2.7,0.74);
    \node[lab] at (2.52,0.94) {\scriptsize $2 \times 256$};

    \node[gate] (g0) at (3.5,0.6) {$+B$};
    \node[gate] (g1) at (5.5,0.6) {$+[2]B$};
    \node[gate] (g2) at (9.0,0.6) {$+[2^{255}]B$};

    \node[ctrl] (c0) at (3.5,3.0) {};
    \draw[wire] (c0) -- (g0.north);
    \node[ctrl] (c1) at (5.5,2.65) {};
    \draw[wire] (c1) -- (g1.north);
    \node[ctrl] (c2) at (9.0,2.15) {};
    \draw[wire] (c2) -- (g2.north);

    \draw[decorate, decoration={brace, amplitude=5pt}, thick]
      (9.85,0.15) -- (2.55,0.15)
      node[lab, midway, below=6pt]
      {$256$ controlled mixed additions~\eqref{eq:mixed-add}};

    \node[font=\bfseries] at (5.6,-0.9) {(b) Windowed};

    \foreach \yy in {-1.55,-2.05,-2.55,-4.6}{
      \draw[wire] (1.6,\yy) -- (7.15,\yy);
      \draw[wire, dashed] (7.15,\yy) -- (8.05,\yy);
      \draw[wire] (8.05,\yy) -- (10.6,\yy);
    }
    \node[lab, left] at (1.6,-1.55) {$\ket{j_0}$};
    \node[lab, left] at (1.6,-2.05) {$\ket{j_1}$};
    \node[lab, left] at (1.6,-2.55) {$\ket{j_{15}}$};
    \node[lab, left] at (1.6,-4.6) {$\ket{R}$};
    \fill (2.05,-2.18) circle (0.9pt);
    \fill (2.05,-2.30) circle (0.9pt);
    \fill (2.05,-2.42) circle (0.9pt);

    \draw[thick] (2.55,-1.69) -- (2.85,-1.41);
    \node[lab] at (2.68,-1.34) {\scriptsize $16$};
    \draw[thick] (2.55,-2.19) -- (2.85,-1.91);
    \node[lab] at (2.68,-1.84) {\scriptsize $16$};
    \draw[thick] (2.55,-2.69) -- (2.85,-2.41);
    \node[lab] at (2.68,-2.34) {\scriptsize $16$};
    \draw[thick] (2.55,-4.74) -- (2.85,-4.46);
    \node[lab] at (2.67,-4.3) {\scriptsize $2 \times 256$};

    \node[gate] (t0) at (3.9,-4.6)
      {$+\mathcal{T}_{B}^{(0)}[j_0]$};
    \node[gate] (t1) at (6.2,-4.6)
      {$+\mathcal{T}_{B}^{(1)}[j_1]$};
    \node[gate] (tw) at (9.4,-4.6)
      {$+\mathcal{T}_{B}^{(15)}[j_{15}]$};

    \node[lookup] (l0) at (3.9,-1.55) {\footnotesize Lookup};
    \node[lookup] (l1) at (6.2,-2.05) {\footnotesize Lookup};
    \node[lookup] (lw) at (9.4,-2.55) {\footnotesize Lookup};
    \draw[wire] (l0.south) -- (t0.north);
    \draw[wire] (l1.south) -- (t1.north);
    \draw[wire] (lw.south) -- (tw.north);

    \draw[decorate, decoration={brace, amplitude=5pt}, thick]
      (10.55,-5.05) -- (2.9,-5.05)
      node[lab, midway, below=6pt]
      {$16$ windowed additions~\eqref{eq:windowed-add}};
  \end{tikzpicture}
  \caption{The two point-addition interfaces used to compute one scalar multiplication $[s]B$, where $(s,B)$ denotes either $(u,G)$ or $(v,P)$. \textbf{(a)} In the double-and-add approach, each scalar bit $s_i$ controls a mixed addition of the fixed, classically known point $[2^i]B$ to the accumulator $\ket{R}$; this addend is fixed at circuit-construction time and does not occupy a quantum register. \textbf{(b)} In the windowed approach, the scalar is divided into $16$-bit words $j_i=s_{16i:16i+16}$. Each $\ket{j_i}$ indexes a QROM lookup into the precomputed table $\mathcal{T}_{B}^{(i)}$, and the selected point $\mathcal{T}_{B}^{(i)}[j_i]$ is added coherently to $\ket{R}$. The selected addend is therefore quantum data entangled with the window register.}
  \label{fig:two-methods}
\end{figure}

%% file: sections/03-challenge-benchmark.tex
\section{The {\challengename} Challenge and the Benchmark}
\label{sec:challenge-and-benchmark}

{\challengename}~\cite{ecdsafail-web,ecdsafail-github} is an open competition to minimize the resources of a reversible \texttt{secp256k1} point-addition circuit; the curve underlies Bitcoin, Ethereum, and related blockchain systems~\cite{nakamoto08,wood-ethereum}. Submissions are scored by the fault-tolerant spacetime proxy in \eqref{eq:qt-proxy} and accepted only if they are value-correct, ancilla-clean, and phase-clean on an evaluator-generated test set. This section defines the task and its relationship to Shor's algorithm, the circuit model, cost metric, and evaluator. Throughout, $n=256$ is the field bit-width and $p=2^{256}-2^{32}-977$ is the \texttt{secp256k1} prime.

\subsection{The Point-Addition Task}
\label{sec:task}

As described in \Cref{sec:ecc-quantum}, the cost of applying Shor's algorithm to \texttt{secp256k1} is dominated by the two scalar multiplications, $[u]G$ and $[v]P$, in the probe function~\eqref{eq:probe}. These decompose into repeated elliptic-curve point additions, making point addition a principal contributor to the logical width and Toffoli count and motivating {\challengename} to isolate and optimize this primitive.

For \texttt{secp256k1}, $E:y^2=x^3+7$ over $\F_p$, the benchmark supplies a quantum accumulator $R=(x_R,y_R)$ in two $256$-qubit registers and a classically specified addend $A=(x_A,y_A)$. It asks the circuit to implement the reversible in-place map
\begin{equation}
  U_A:\ket{x_R}\ket{y_R}
  \longmapsto
  \ket{x_{R'}}\ket{y_{R'}},
  \qquad
  R'=R+A,
  \label{eq:task}
\end{equation}
using the generic affine group law in \eqref{eq:group-law}. Computing its slope requires $(x_R-x_A)^{-1}\bmod p$, the dominant reversible operation. The circuit's generic-affine domain is $R,A\neq\mathcal O$ and $R\notin\{A,-A,-2A\}$. The cases $R=A$ and $R=-A$ make the input-side denominator vanish, requiring point doubling or producing $\mathcal O$, respectively. The additional case $R=-2A$ gives $R'=-A$ and hence $x_A-x_{R'}=0$, which lies outside the domain of the reversible in-place multiplication used later in the circuit. The evaluator explicitly filters identity inputs and $x_R=x_A$, but does not separately filter $R=-2A$; the latter occurs with probability approximately $1/r$ under random-scalar sampling and therefore has negligible probability in the finite validation set. Circuit correctness is nevertheless not claimed for this case. The quantum inputs \texttt{target\_x} and \texttt{target\_y} initially hold $R$, while the classical inputs \texttt{offset\_x} and \texttt{offset\_y} specify $A$; the circuit overwrites the target registers with $R'$. The scored primitive is the uncontrolled map in \eqref{eq:task}.

\paragraph{Compatibility with the full Shor circuit.} Because the benchmark addend $A$ is classically specified, \eqref{eq:task} closely matches the mixed addition used in double-and-add (\Cref{sec:ecc-quantum}), where each call adds a fixed point $A_i=[2^i]B$ for $B\in\{G,P\}$. Double-and-add additionally controls each call on a scalar bit, so integration requires restoring the global control omitted by the benchmark. Standard techniques for doing so are given in~\cite{rnsl17,hjn20}.

The windowed method presents a different interface. As discussed in \Cref{sec:ecc-quantum}, it instead selects each addend coherently through QROM, for example $A_i=\mathcal{T}^{(i)}_G[u_i]$ or $\mathcal{T}^{(i)}_P[v_i]$, making its coordinates live quantum data. The benchmark circuit is therefore not directly drop-in compatible and must adopt the QROM lookup/use/unlookup interface of \Cref{sec:qc-background}. We implement this adaptation in \Cref{sec:windowed-addition-compatible-circuit}: the resulting variant supplies the addend coherently from a quantum-addressed classical table and clears the lookup workspace by exact reversible replay, adding $11$ logical qubits and approximately $29.6\%$ average executed Toffoli per addition.

\begin{table}[t]
  \centering
  \small
  \setlength{\tabcolsep}{5pt}
  \renewcommand{\arraystretch}{1.2}
  \begin{tabular}{@{}>{\raggedright\arraybackslash}p{1.7cm}%
                    >{\raggedright\arraybackslash}p{4.8cm}%
                    >{\raggedright\arraybackslash}p{4.8cm}@{}}
    \toprule
    \textbf{Type} & \textbf{What is checked} & \textbf{Why it is required} \\
    \midrule
    \texttt{cls} &
    Outputs equal the affine sum $R'=R+A$ on every test. &
    Verifies the intended classical map. \\[2pt]

    \texttt{anc} &
    All scratch and non-output registers return to $\lvert 0\rangle$. &
    Prevents garbage entanglement with the outer computation. \\[2pt]

    \texttt{pha} &
    Surviving branches retain no relative phase after uncomputation or measurement-based cleanup. &
    Preserves the interference required by Shor's algorithm. \\
    \bottomrule
  \end{tabular}
  \vspace{5pt}
  \caption{Correctness checks; a clean submission reports $0/0/0$ on every test.}
  \label{tab:channels}
\end{table}

\subsection{Circuit Model and Cost Metric}
\label{sec:cost}

A submission represents a reversible point-addition map in the kickmix model of \Cref{sec:qc-background}. Its serialized instruction set includes $X$, $Z$, CNOT, CZ, SWAP, CCX, CCZ, qubit reset, and HMR, a demolition measurement in the $X$ basis. Standalone Hadamard gates and persistent $\ket{+}$-state preparation are excluded. This restriction is essential to the tractability of verification: CCX together with the Hadamard gate is universal for quantum computation~\cite{aharonov03}, so allowing unrestricted Hadamards would make the language expressive enough to encode general BQP computations, beyond the scope of the benchmark's efficient classical verifier. Indeed, even reusable $\ket{+}$-state preparation, combined with the existing kickmix operations and HMR, can synthesize a standalone Hadamard. Hadamard is therefore permitted only within HMR, where it is immediately followed by destructive measurement and does not leave a reusable coherent superposition. Each executed CCX or CCZ counts as one Toffoli-equivalent operation. Under the fault-tolerant cost model of \Cref{sec:qc-background}, Toffoli count serves as an architecture-dependent proxy for non-Clifford work, while peak qubit width measures logical space, so the benchmark reports two resources:
\begin{itemize}
  \item \textbf{Peak qubits} $Q$: the maximum number of logical qubits simultaneously live at any point in the circuit.

  \item \textbf{Average executed Toffoli} $T$: the expected number of executed Toffoli-equivalent gates, averaging data-dependent firing probabilities over the evaluator's input distribution~\cite{google26}. It counts Toffoli operations, not magic-state $T$ gates, and differs from the static gate count and circuit depth.
\end{itemize}
The benchmark scores the product of these two quantities,
\begin{equation}
  S = Q \times T,
  \label{eq:score}
\end{equation}
Lower scores are better. The product approximates fault-tolerant spacetime volume, with $Q$ representing space and $T$ the non-Clifford time or distillation load. Because point addition recurs throughout Shor's algorithm, improvements to this primitive reduce the whole-algorithm resource estimate. In addition to this headline metric, the challenge also maintains a low-qubit track that minimizes $Q$ independently of $T$ for qubit-constrained settings. Together, the two objectives chart the $(Q,T)$ Pareto frontier, whose circuits cannot improve one resource without increasing the other; much of this work seeks to advance that frontier.

\subsection{Correctness, Verification, and Scoring}
\label{sec:evaluator}

Acceptance requires passing the evaluator's three correctness checks (\Cref{tab:channels}): \texttt{cls} verifies the classical function on sampled inputs, while \texttt{anc} and \texttt{pha} detect unclean ancillas and residual phases that classical tests cannot observe but that would corrupt Shor's interference.

\paragraph{Test points.}
Following the verifiable fuzz-testing methodology of~\cite{google26}, the evaluator avoids a fixed public test set by seeding SHAKE256 with the serialized submitted circuit and expanding its output into $9{,}024$ curve points. The resulting test set is deterministic but circuit-dependent and becomes known only after the circuit is fixed. Its support-selection and nonce-hunting limitations are discussed in \Cref{sec:limitations}.


\paragraph{Approximation and nonce hunting.}
Exactness is not required: the reference circuits of Babbush et al.~\cite{google26} and Schrottenloher~\cite{s26} are also approximate, with a bounded failure fraction of at most approximately $2^{-13.3}$~\cite{s26}. Submissions can therefore trade accuracy for lower $Q\times T$ by truncating comparisons, narrowing carries, or removing rarely executed gates.

Because the evaluator derives its $9{,}024$ test inputs from a SHAKE256 hash of the serialized operation stream, functionally equivalent emitted circuits can induce different validation sets. The implementation encodes a $48$-bit nonce in a fixed $96$-operation identity tail: for each bit $b_i$, it applies Pauli $X$ twice to either $\texttt{target\_x}[0]$ or $\texttt{target\_x}[1]$, according to the value of $b_i$. Since $X^2=I$, this preserves the circuit function, Toffoli count, and peak width, while the changed target operands alter the hash. Submitters can search these nonces for an error-free test set, or clean validation \emph{island}. For a circuit with error rate $\epsilon$, the clean-nonce density, or \emph{landability}, is approximately $(1-\epsilon)^{9024}$ under the random-oracle model and therefore decreases exponentially with $\epsilon$. In particular, $(1-0.01)^{9024}\approx2^{-130.8}$, so hiding a $1\%$ error rate would require more than $2^{128}$ candidates on average and is infeasible within the $48$-bit nonce space~\cite{google26}. To assess improvements independently of nonce-selected support, we reevaluate a longitudinal sample on $50{,}000$ common pseudorandom inputs and report $Q\times T/\hat p$, where $\hat p$ is the empirical success rate; because $\hat p$ remains near one, the retry-adjusted trajectory closely follows the raw score (\Cref{sec:QxT-score-evolution}).

\paragraph{Anchors.}
The baseline has $Q=2{,}715$ and $T=3{,}960{,}753$, giving $Q\times T\approx1.07\times10^{10}$. External space-optimized reference points include Schrottenloher's $Q=1{,}192$, $T=2^{21.19}\approx2.4M$ circuit~\cite{s26} and the Babbush et al.\ estimate of $Q=1{,}175$, $T=2^{21.36}\approx2.69M$~\cite{google26}. \Cref{sec:results} analyzes the improvement trajectory, Pareto frontier, and cross-work comparisons.

\subsection{Formulating the Circuit Optimization Problem}

Under the circuit model and evaluator above, {\challengename} solves

\begin{equation*}
\min_{C\in\mathcal K} Q(C)T(C)
\qquad\text{subject to}\qquad
\operatorname{cls}(C)=\operatorname{anc}(C)=\operatorname{pha}(C)=0,
\end{equation*}

\noindent where $\mathcal K$ contains the circuits admitted by the kickmix language and locked harness. The task combines valid reversible-circuit synthesis with minimization of nonlinear gate cost and live quantum memory. Related function, unitary, and state minimization variants admit QCMA formulations and search-to-decision reductions~\cite{chia2021quantum}; exact reversible Toffoli minimization is NP-hard~\cite{vandewetering2024optimising}, while subspace-equivalent minimization of several gate counts and depths is co-NQP-hard~\cite{kjelstrom2026exact}. Width optimization adds a hard scheduling problem: under spooky pebbling, minimizing quantum space on general dependency DAGs is PSPACE-hard to approximate, and the corresponding decision problem is PSPACE-complete~\cite{kornerup2025spooky,quist2025tradeoffs}. These results make efficient exact global optimization implausible and motivate \Cref{sec:autoresearch-paradigm}, where agent-driven optimizations accumulate evaluator-verified improvements without exhaustive synthesis.

%% file: sections/04-open-autoresearch.tex
\definecolor{arInk}{HTML}{22262E}
\definecolor{arMuted}{HTML}{6B7280}
\definecolor{arAccent}{HTML}{3D5A80}
\definecolor{arAccentBg}{HTML}{E9EEF5}
\definecolor{arWarm}{HTML}{B07D3B}
\definecolor{arWarmBg}{HTML}{F7EEDE}
\definecolor{arBorder}{HTML}{C9CED6}
\definecolor{arEdge}{HTML}{7A828E}

\section{The Open Autoresearch Paradigm}
\label{sec:autoresearch-paradigm}

This section defines open autoresearch and its discovery loop (\Cref{sec:autoresearch-overview}), describes its {\challengename} implementation (\Cref{sec:challenge-autoresearch-harness}), summarizes observed participant practices (\Cref{sec:autoresearch-techniques}), and consolidates them into a reusable framework (\Cref{sec:consolidated-autoresearch-framework}).

\subsection{Overview of Open Autoresearch}
\label{sec:autoresearch-overview}

\input{figs/open-autoresearch-discovery-loop}

We define \emph{Open Autoresearch} as a process in which a heterogeneous community of humans and AI agents addresses a hard problem formulated as a measurable objective~\cite{liu25autoresearch,karpathy26autoresearch,hyperspace26agi}. An \emph{AI agent} is an LLM coupled to tools and persistent state, enabling it to act, observe machine feedback, and revise its next step~\cite{yao23react,schick23toolformer,shinn23reflexion,long2023treeofthought}. \emph{Autoresearch} delegates repeated hypothesis generation, implementation, and evaluation to these agents under human supervision. Its enabling condition is a \emph{verifiable objective}: a computable measure that efficiently scores candidates and checks their validity. {\challengename} instantiates this paradigm for quantum-circuit optimization, but the same structure applies to other domains~\cite{alphatensor,alphadev}.

Participants may range from domain experts to enthusiasts with unrelated backgrounds, combining human judgment with agents that survey literature, test candidates, and record successes and failures. The model is open~\cite{terwiesch-xu08}: participants need not know one another, share an institution, or coordinate beforehand.

\Cref{fig:autoresearch-loop} depicts the generic discovery loop. Participants select a base, propose a change, and submit it for scoring and validation. The base may be the current best, another Pareto point, or an off-frontier design chosen to escape a local optimum. Valid improvements become public bases for subsequent work; failed candidates are revised and their diagnoses retained as negative evidence. Shared objectives, evaluators, and artifacts thereby turn independent attempts into a cumulative search.

Optional communication through email, forums, messaging platforms, and GitHub accelerates the spread of new techniques. Independent branches using different models, tools, backgrounds, and objectives also make the search resilient to plateaus: an existing or newly joined contributor may unlock progress after the dominant route stalls, after which the accepted improvement becomes available for community-wide adoption and refinement.

Public, verified optimization leaderboards predate open autoresearch. For example, communities such as Opus Magnum already maintain cumulative, multi-objective records of submitted solutions~\cite{opusmagnumleaderboard}. The distinction here is the integration of tool-using AI agents with a machine-checkable scientific evaluator, versioned source and provenance, and shared experiment memory that includes contextualized failures. Existing leaderboard communities could adopt agent-readable evaluator interfaces, automated change summaries, and structured negative-result records to support more autonomous cumulative search.

\subsection{The {\challengename} Open Autoresearch Harness}
\label{sec:challenge-autoresearch-harness}

Launched by Eigen Labs in late May 2026, {\challengename} applies open autoresearch to a reversible \texttt{secp256k1} mixed point-addition circuit. Its primary score is $S=Q\times T$ (\Cref{sec:cost}), supplemented by low-qubit and low-Toffoli tracks that expose other Pareto operating points. The benchmark and results appear in \Cref{sec:challenge-and-benchmark,sec:results}.

The harness implements the discovery loop through a shared repository, locked evaluator, command-line interface, and public leaderboard. Under fixed scoring rules, the command-line tool reports $Q$, $T$, and $Q\times T$ and checks classical output, ancilla cleanliness, and phase correctness. Promoted improvements add their score, source, participant attribution, and declared AI model to the public record, allowing others to synchronize or selectively adopt their techniques.

Participation requires only repository access; Slack and Telegram provide optional discussion and debugging. Contributors may edit circuits directly or delegate implementation and testing to coding agents. Over approximately eight weeks, more than one hundred participants using diverse models produced over four hundred promoted submissions while pursuing frontier, independent, and alternative-resource branches.

The harness thus provides a stable verification boundary and a cumulative record of accepted circuits and provenance. The trajectory combined incremental refinements with architectural changes to inversion, encoding, and register allocation. \Cref{sec:results} analyzes these circuits, while \Cref{sec:autoresearch-techniques} examines the practices that produced them.

\subsection{{\challengename} Autoresearch Techniques}
\label{sec:autoresearch-techniques}

This subsection examines how {\challengename} participants organized AI agents to propose, implement, and verify circuit optimizations. It draws on contributed harnesses, prompts, scripts, experiment logs, and public discussions; representative raw harnesses and scaffolds are available in this \href{https://github.com/jieyilong/ecdsafail-autoresearch-harness}{supplemental repository}. 

\subsubsection{Overview of the Autoresearch Scaffold and Methodology}
\label{sec:techniques-scaffold}

Building on the challenge harness described above, participants added scripts for parameter sweeps, small-domain exhaustive tests, counterexample-guided search, regression testing, and experiment tracking. Git worktrees or isolated clones kept concurrent experiments separate from the accepted baseline and made candidates reproducible.

Two agent-organization patterns recurred. In the \emph{research-and-engineering} pattern, research agents generated hypotheses and ran inexpensive experiments, while an engineering agent converted promising results into clean patches. In the \emph{orchestrator-and-workers} pattern, a coordinator assigned independent mechanisms to narrowly scoped agents and synthesized their findings. Both limited individual-agent context while preserving shared state through hypothesis queues, iteration logs, scoreboards, and regression records. Human instructions ranged from high-level objectives to detailed specifications of roles, branches, validation stages, and synthesis procedures.

A common workflow also emerged. Participants synchronized with the public frontier, inspected source changes and submission notes, and selected a baseline for the target resource regime. Agents proposed circuit-level changes, estimated or measured their local costs, tested them cheaply, and then evaluated the complete point-addition circuit. Validation progressed from smoke tests through regression checks to the official evaluator, with failures classified by classical output, phase, and ancilla cleanliness. Promising candidates were rebuilt cleanly and compared with the live frontier before submission.

Experiment records included rejected variants, their assumptions, and observed failure modes, allowing later agents to avoid repetition or revisit ideas whose prerequisites had changed. Humans generally remained outside the implementation loop, choosing resource regimes, allocating compute, supplying domain knowledge, redirecting stalled searches, and judging whether score changes reflected genuine improvements. The principal practices for parallel search, failure diagnosis, and human--agent collaboration are discussed next.

\subsubsection{Observed Autoresearch Practices}
\label{sec:techniques-tricks}

Within the common scaffold described above, several practices recurred across successful research efforts. These methods span the organization of agent labor, the selection and diagnosis of experiments, and the accumulation of knowledge across sessions and participants. These practices are descriptive rather than prescriptive: no participant used all of them, and successful efforts combined them in different ways. The value of the open framework is precisely that it permits heterogeneous search strategies.

\begin{itemize}
  \item \textbf{Reusable and evolving agent skills.} Participants packaged recurring procedures as versioned skills that agents could invoke across experiments. These included general methods for autonomous iteration, literature review, and self-critique, as well as challenge-specific procedures for reducing peak width, lowering Toffoli count, diagnosing reversible-circuit failures, searching for admissible evaluator supports, and reconstructing frontier changes. When a new technique or failure mode was identified, its lessons could be incorporated into the relevant skill, allowing methodology to improve together with the circuit frontier.

  \item \textbf{Persistent loops and goals.} Participants often assigned long-running objectives through persistence mechanisms such as Claude Code's \texttt{/loop} command and Codex's \texttt{/goal} command, allowing agents to conduct repeated propose--implement--evaluate cycles with limited supervision. Persistence was paired with checkpoints that assessed novelty, progress, and diminishing returns. A stalled campaign could then be redirected, supplied with additional context, moved to another operating point, or terminated rather than continuing indefinitely along an unproductive route.

  \item \textbf{Parallel, role-separated, and clean-context agents.} Participants commonly separated structural research, implementation, scoring, and independent verification across agents. Orchestrators dispatched clean-context workers, each scoped to one hypothesis, literature question, or failure-mode analysis and asked to return a compact conclusion rather than an extended narrative. Some teams also conducted route bake-offs in which several candidate architectures received comparable initial budgets and only the most promising were explored further, approximating a beam search over circuit variants. Agents from different model families (e.g. GPT, Claude, Grok, GLM, and Kimi) sometimes reviewed one another's conclusions, and every claimed improvement was treated as a hypothesis until independently reproduced in an isolated worktree. This organization broadened exploration while limiting context contamination and shared blind spots.

  \item \textbf{Summarizing and selectively adopting other participants' work.} Dedicated watcher agents compared adjacent public submissions, reconstructed the responsible source changes, and distinguished transferable circuit techniques from configuration-specific effects. Useful findings were summarized in research notes and incorporated into reusable skills. Active agents did not necessarily rebase whenever the frontier changed: participants often preserved independent branches and adopted a public improvement only when it strengthened the circuit family under investigation. This balance enabled rapid cross-pollination without collapsing the search onto a single trajectory.

  \item \textbf{Human steering and non-monotonic exploration.} Human intervention was most valuable in selecting research directions, supplying domain constraints, recognizing stagnation, and deciding whether a score change represented a meaningful circuit improvement. Participants sometimes directed agents to restructure an entire component while tolerating temporary increases in $Q$, $T$, or validation failures, and only afterward repair and optimize the new design. A related strategy asked what prerequisites would have to hold for a proposed optimization to work and then searched for transformations that established those prerequisites. Such non-monotonic exploration exposed circuit families that strict hill climbing would reject prematurely.

  \item \textbf{Shared experiment memory and structured reflection.} Participants maintained append-only records containing each hypothesis, implementation, resource measurement, failure mode, and proposed next step. Failed directions were documented together with their assumptions rather than declared universally invalid, allowing later agents to revisit them after changes in the circuit architecture or resource regime. Periodic reflection passes summarized what had worked, identified recurring obstacles, and transferred insights between circuit components. These records reduced duplicated effort and preserved knowledge across context resets, model changes, and participant boundaries.

  \item \textbf{Literature-guided hypothesis generation.} Agents searched, indexed, and summarized relevant academic and technical literature, extracting candidate mechanisms and known limitations. Historically, Khattar et al.~\cite{ksgz25} introduced the underlying record-and-replay dialog-GCD architecture. However, the challenge record contains no independent agent-derived transfer of this architecture to ECDLP point addition; submissions implementing dialog-GCD appeared only after Schrottenloher published its detailed ECDLP adaptation~\cite{s26}, which enabled substantial reductions in $Q\times T$. Schrottenloher's paper appeared in early June 2026, only days after the challenge began on May 30, 2026, when participation remained limited and sophisticated literature-search and cross-domain retrieval scaffolds had not yet been developed. This chronology does not isolate whether the missed transfer resulted from model capability, limited search effort, differences in terminology and problem formulation, or immature retrieval tooling. It nevertheless identifies cross-domain retrieval and abstraction as important capabilities to strengthen. Once the connection became available, agents rapidly implemented, tested, and refined the architecture against the shared evaluator, illustrating the complementary roles of human research and agent-scale experimentation.
      
  \item \textbf{Staged validation and mechanism-level diagnosis.} Candidate changes were tested first on inexpensive instances and targeted invariants before being subjected to the complete evaluator. When a circuit failed, participants separated classical-output, phase, and ancilla-cleanliness errors and used first-failure traces, error histograms, comparison outcomes, and carry-width measurements to identify the responsible mechanism. The smallest faulty component was then repaired before further parameter or evaluator-support searches were attempted. This approach prevented computational effort from being spent compensating for an unidentified structural error.

  \item \textbf{Exact transformations and circuit fusion.} Participants frequently prompted agents to seek algebraically exact transformations that reduced cost without introducing additional approximation error. Fusion was particularly productive: adjacent arithmetic blocks were combined so that intermediate values, controls, or cleanup operations could be eliminated, for example by replacing a sequence of modular additions and negations with one equivalent constant update. Because such transformations preserve correctness for all inputs under their stated preconditions, they can be composed and evaluated without relying on support-specific correctness behavior.

  \item \textbf{Alternative search-and-distill strategies.} At least one participant used the evaluator to generate low-cost correct subcircuits and trained a smaller model on the resulting examples. Increasing model size, adding reinforcement learning, and eliciting additional reasoning did not overcome the observed performance plateau. This negative result suggested that tool access and evaluator-guided iteration may be more important for this task than scaling a model trained to reproduce a fixed procedure.
\end{itemize}

\subsubsection{Why It Works and What We Have Learned}
\label{sec:techniques-why}

The {\challengename} experience highlights four conditions that supported sustained human--agent progress and delimit where open autoresearch is most applicable.

\begin{itemize}
  \item \textbf{A hard-to-solve but efficiently verifiable objective.} Discovering efficient reversible circuits is difficult, but their resource usage and sampled correctness can be evaluated automatically and reproducibly. Agents can therefore explore speculative changes while a common evaluator rejects contract violations and measures successful candidates. This paradigm is less suitable when evaluation is unreliable, uninformative, or too expensive for repeated experimentation.

  \item \textbf{Cumulative search through openness and diversity.} Public circuits, diffs, notes, and reusable skills preserved both accepted improvements and failed experiments. Participants could adopt promoted circuits while maintaining independent branches and operating points. This combination proved important when progress stalled: apparent plateaus were repeatedly broken by existing or newly joined contributors using different models, tools, expertise, or circuit families. The resulting trajectory was therefore a sequence of public handoffs among partially independent searches rather than the uninterrupted optimization path of one participant or agent.

  \item \textbf{Complementary human and agent capabilities.} Agents excelled at implementation, repeated experimentation, local diagnosis, and incremental refinement. Humans selected objectives, interpreted resource behavior, challenged assumptions, and supplied architectural insights. In particular, the dialog-GCD route used the record-and-replay Euclidean architecture introduced by Khattar et al.~\cite{ksgz25}. Agents accelerated its implementation and refinement and occasionally identified structural transformations that produced substantial resource reductions. The evidence therefore supports a division of labor combining human conceptual direction with agent-scale execution.

  \item \textbf{The objective shapes the research trajectory.} A scalar score rewards immediate progress but may suppress architectures that require temporary regressions. Separate low-qubit and low-Toffoli tracks mitigated this tendency by exposing different regions of the $(Q,T)$ Pareto frontier. Future systems could also reward frontier expansion, reproducible negative results, or promising structural changes. Evaluators and incentives therefore influence which discoveries are pursued and preserved.
\end{itemize}

Open autoresearch is thus best suited to problems with reliable, inexpensive evaluation, public intermediate artifacts, and enough freedom for participants to pursue independent routes.

\begin{algorithm}[!t]
\footnotesize
\caption{Consolidated {\challengename} open autoresearch framework}
\label{alg:consolidated}

\begin{algorithmic}[0]
\Require $\mathsf{track}\in\{S,Q,T,\text{Pareto}\}$; agent token budget
\Require skills $\mathcal{S}$; memory $\mathcal{M}$
\Require shared repository and leaderboard $\mathcal{L}$
\end{algorithmic}

\noindent\textbf{Process A: Asynchronous monitoring and steering}
{%
\algrenewcommand\alglinenumber[1]{\scriptsize A#1:}
\begin{algorithmic}[1]
\While{the campaign remains active}
  \State monitor $\mathcal{L}$ and relevant research literature
  \State incorporate useful findings into $\mathcal{S},\mathcal{M}$
  \State \textbf{human-in-the-loop:} steer the agents as needed (reprioritize / course-correct)
  \State update $\mathcal{M}$ with human steering
\EndWhile
\end{algorithmic}
}

\vspace{3pt}

\noindent\textbf{Process B: Candidate-development cycle}
\par
{%
\algrenewcommand\alglinenumber[1]{\scriptsize B#1:}
\begin{algorithmic}[1]
\While{agent token budget remains}
  \State synchronize with $\mathcal{L}$; refresh $\mathcal{S},\mathcal{M}$
  \State choose a frontier or off-frontier base solution $B$
  \State profile $B$; generate hypotheses $\mathcal{H}$ from $\mathcal{S},\mathcal{M}$
  \ForAll{$h\in\mathcal{H}$ in parallel}
    \State launch a fresh-context subagent with a worktree
    \State assign $h$ and the current $\mathcal{S}$ to the subagent
    \State implement and refine $h$ until stalled
    \State test $\mathrm{cls}/\mathrm{pha}/\mathrm{anc}$; repair or discard
    \State for approximate candidates, triage landability
    \State search nonces only when viable
    \State run the full evaluator; record the outcome
  \EndFor
  \State select the best validated candidate $C^\star$
  \State rebuild and remeasure $C^\star$
  \State submit $C^\star$ if it improves the frontier
  \State on promotion, publish $C^\star$ as a shared base
  \State distill all outcomes into $\mathcal{S}$ and $\mathcal{M}$
\EndWhile
\end{algorithmic}
}

\smallskip
\noindent\textbf{Output:} promoted circuits, evolved skills, and contextualized experiment records.
\end{algorithm}

\subsection{Consolidated {\challengename} Autoresearch Framework}
\label{sec:consolidated-autoresearch-framework}

The practices in \Cref{sec:autoresearch-techniques} emerged independently and were combined differently across participants; no participant used the complete set. \Cref{alg:consolidated} synthesizes them into two concurrent processes. Process~A monitors the public frontier and research literature, incorporates useful findings into shared skills and memory, and permits human reprioritization or course correction. Process~B runs a token-bounded candidate-development cycle that synchronizes with the public record, selects a frontier or off-frontier base, profiles its bottlenecks, and explores hypotheses in parallel using fresh-context agents and isolated worktrees.

The framework separates speculative generation from evidential acceptance. Workers may pursue aggressive rewrites, approximate constructions, or temporarily inferior intermediate designs, but each candidate undergoes classical-output, phase, and ancilla-cleanliness checks. Invalid branches are repaired or discarded; approximate candidates additionally undergo landability triage and nonce search only when viable. Surviving candidates are evaluated under the full benchmark, after which the best is rebuilt, remeasured, and submitted only if it advances the selected frontier.

Progress compounds both within and across participants. Parallel workers broaden each local search, while promoted circuits become shared bases for the community. Skills and experiment memory retain useful mechanisms, measured outcomes, and contextualized failures, allowing later searches to avoid repetition or revisit ideas under changed prerequisites. The framework is therefore not a mandatory strategy but a synthesis of observed practices adaptable to other problems with machine-checkable objectives, inexpensive experiments, and shareable intermediate artifacts.

%% file: figs/open-autoresearch-discovery-loop.tex
\begin{figure}[!t]
\centering
{\hyphenpenalty=10000\exhyphenpenalty=10000%
\begin{tikzpicture}[
  scale=1.00,
  every node/.style={text=arInk},
  font=\small,
  stg/.style={rectangle, rounded corners=4pt, line width=0.6pt, align=center,
              inner sep=5pt, text width=32mm, minimum height=13mm},
  act/.style={stg, draw=arBorder, fill=white},
  shared/.style={stg, draw=arAccent, fill=arAccentBg},
  gate/.style={rectangle, rounded corners=11pt, draw=arWarm, fill=arWarmBg,
               line width=0.6pt, align=center, inner sep=5pt,
               text width=30mm, minimum height=13mm},
  note/.style={rectangle, rounded corners=7pt, draw=arWarm, fill=arWarmBg,
               line width=0.5pt, align=center, inner sep=4pt, text width=27mm},
  flow/.style={->, line width=0.7pt, draw=arEdge,
               shorten >=1.5pt, shorten <=1.5pt},
  ret/.style={flow, draw=arWarm!85!black},
  lbl/.style={font=\footnotesize\bfseries, text=arInk, inner sep=2.5pt,
              fill=white, rounded corners=1pt}
]
\node[shared] (pool)    at ( 90:3.7)
  {\textbf{Shared repository}\\ \textbf{\& leaderboard}\\[1pt]
   {\scriptsize\color{arMuted} pool of promoted base solutions}};
\node[act]    (base)    at ( 30:3.7)
  {\textbf{Pick a base solution}\\[1pt]
   {\scriptsize\color{arMuted} SOTA, a Pareto point,\\ or off-frontier}};
\node[act]    (propose) at (-30:3.7)
  {\textbf{Propose a change}\\[1pt]
   {\scriptsize\color{arMuted} human $+$ agent:\\ read, generate, test}};
\node[shared] (eval)    at (-90:3.7)
  {\textbf{Machine checkable evaluator}\\[1pt]
   {\scriptsize\color{arMuted} score the objective,\\ then check validity}};
\node[gate]   (gate)    at (-150:3.7)
  {\textbf{Valid}\\ \textbf{improvement?}};
\node[act]    (promote) at ( 150:3.7)
  {\textbf{Submit \& promote}\\[1pt]
   {\scriptsize\color{arMuted} a new base for everyone}};
\node[note]   (learn)   at (0.15,0.55)
  {{\scriptsize\color{arMuted} log the negative\\ result, then revise}};
\draw[flow] (pool)    to[bend left=13] (base);
\draw[flow] (base)    to[bend left=13] (propose);
\draw[flow] (propose) to[bend left=13] (eval);
\draw[flow] (eval)    to[bend left=13] (gate);
\draw[flow] (gate)    to[bend left=13] node[lbl,pos=0.5]{yes} (promote);
\draw[flow] (promote) to[bend left=13] (pool);
\draw[ret]  (gate)  to[bend left=22] node[lbl,pos=0.35]{no} (learn);
\draw[ret]  (learn) to[bend left=22] (propose);
\end{tikzpicture}}
\caption{The generic \textit{open autoresearch discovery loop}. Independent human--agent teams select a current-best, Pareto, or deliberately off-frontier base from the shared repository, propose a change, and submit it to a machine-checkable evaluator. Verified improvements become shared bases; rejected candidates are logged with their diagnoses and revised. This makes the search cumulative without requiring prior coordination beyond the shared objective, evaluator, and public artifacts.}
\label{fig:autoresearch-loop}
\end{figure}
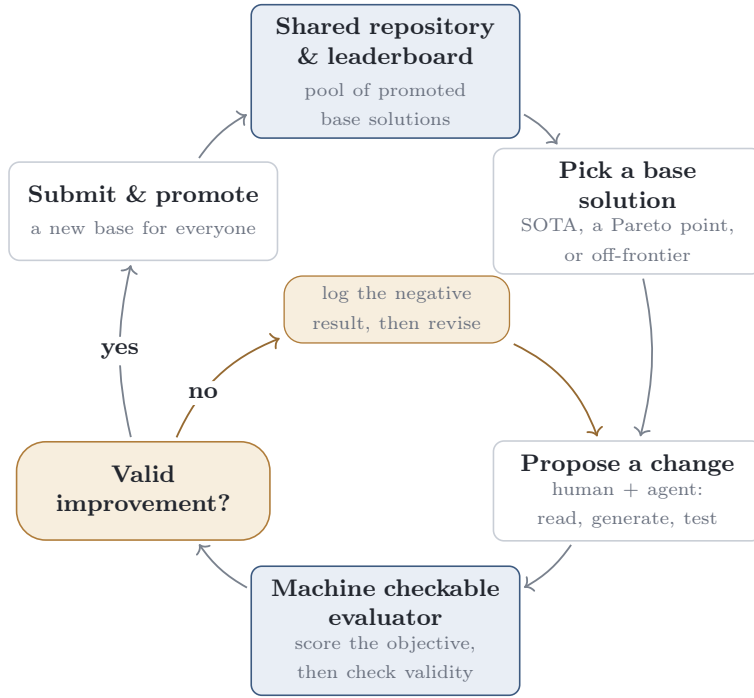

%% file: sections/05-results.tex
\section{Results and Findings}
\label{sec:results}

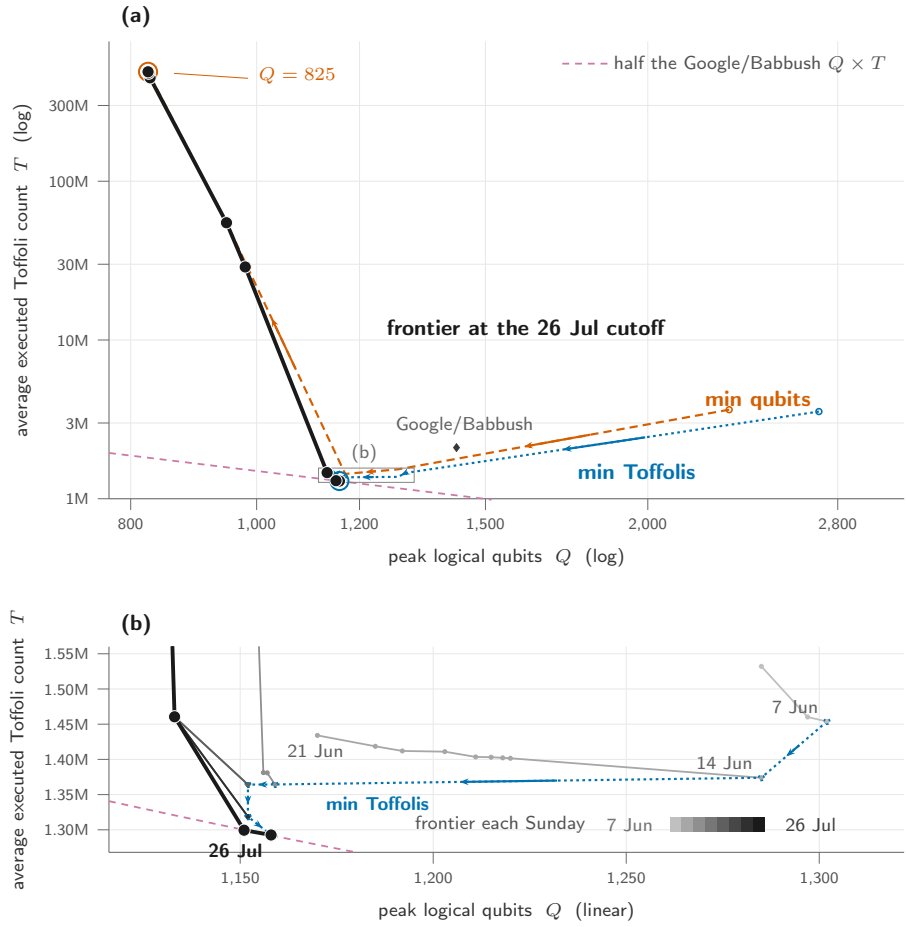
\begin{figure}[!tp]
  \centering
  \input{figs/pareto_frontier}
  \caption{Evolution of the resource frontier. \textbf{(a)} The Pareto frontier at the 26 July data cutoff in the $(Q,T)$ resource plane, with peak logical qubits $Q$ on the horizontal axis, average executed Toffoli count $T$ on the vertical axis, and both axes shown on logarithmic scales. The thick black line and filled circles mark the admitted nondominated operating points at the cutoff, where \emph{admitted} means that the circuit passes all $9{,}024$ Fiat--Shamir-derived validation inputs. The orange dashed and blue dotted trajectories trace the minimum-width (``min qubits'') and minimum-Toffoli (``min Toffolis'') records over time; arrows indicate temporal progression rather than circuit ancestry. The diamond marks the reported Google/Babbush low-gate operating point~\cite{google26}, and the magenta dashed line is the constant-$Q\times T$ locus equal to one half of its reported product, crossed by the frontier at submission \texttt{8e9c9a2} with $1{,}151$ qubits and $1{,}299{,}453$ Toffoli gates. \textbf{(b)} A magnified view, on linear axes, of the low-Toffoli region enclosed by the gray box labeled (b) in panel (a). Gray polylines show the reconstructed Pareto frontier at successive Sunday checkpoints before the cutoff, with darker shades denoting later dates; the thick black line is the frontier at the cutoff. The blue dotted trajectory and arrows show the evolution of the minimum-Toffoli record. Several later weekly frontiers nearly coincide in this region, while the minimum-width trajectory leaves the displayed range after 21 June. Lines connecting operating points are guides to the eye.}
  \label{fig:frontier-reconstruction}
\end{figure}

This section presents the challenge's principal technical and empirical results. We summarize the final operating points, Pareto frontier, external comparisons, and supporting evidence (\Cref{sec:results-summary}); describe the resulting circuit architectures (\Cref{sec:final-circuit-designs}) and key optimizations (\Cref{sec:key-circuit-optimization-techniques}); trace their evolution across submissions (\Cref{sec:circuit-trajectory}); and provide the artifacts and procedures for reproduction (\Cref{sec:reproducibility}).

\subsection{Overall Summary of Results}
\label{sec:results-summary}

As noted in \Cref{sec:cost}, {\challengename} ranks circuits by $S=Q\times T$, where $Q$ is peak logical width and $T$ is the average executed Toffoli count over the submission-dependent validation support. Because participants also optimized these resources separately, the submissions define a $(Q,T)$ Pareto frontier rather than a single optimum. We present the baseline and final operating points (\Cref{sec:headline-operating-points}) and compare them with prior point-addition constructions (\Cref{sec:external-pa-comparison}).

\begin{table}[t]
  \centering
  \small
  \setlength{\tabcolsep}{4pt}
  \begin{tabular}{@{}lrr>{\raggedright\arraybackslash}p{1.45cm}>{\raggedright\arraybackslash}p{3.5cm}@{}}
    \toprule
    \textbf{Submission} & \textbf{$Q$} & \textbf{$T$} & \textbf{Status} & \textbf{Interpretation} \\
    \midrule
    \texttt{b6f2b0a} & $825$ & $489{,}161{,}900$ & Admitted & Lowest observed width. \\
    \texttt{f1d8707} & $826$ & $479{,}337{,}428$ & Admitted & Difference from q825: $-9.82$M Toffoli. \\
    \texttt{39a9b5f} & $828$ & $448{,}102{,}916$ & Admitted & Difference from q826: $-31.23$M Toffoli. \\
    \texttt{a3b0148} & $948$ & $54{,}781{,}961$ & Admitted & Observed exit from the extreme-width cluster. \\
    \texttt{4352cfb} & $980$ & $28{,}847{,}321$ & Admitted & Lowest observed $T$ below $1{,}000$ qubits. \\
    \texttt{1c0e0e9} & $1{,}133$ & $1{,}460{,}511$ & Admitted & Bridge to the product-efficient regime. \\
    \texttt{8e9c9a2} & $1{,}151$ & $1{,}299{,}453$ & Promoted & Product-score incumbent and first $50\%$ crossing. \\
    \texttt{a536a48} & $1{,}158$ & $1{,}292{,}651$ & Promoted & Lowest-$T$ promoted point at the cutoff. \\
    \bottomrule
  \end{tabular}
  \vspace{5pt}
  \caption{Archive-conditioned Pareto set at the data cutoff. Adjacent
  differences are descriptive secants between distinct circuit families, not
  causal marginal effects. ``Admitted'' denotes an archived public clean-run
  report included under the manuscript's curated selection, not official
  promotion.}
  \label{tab:admitted-pareto}
\end{table}

\subsubsection{The Baseline and the Pareto Frontier at the Data Cutoff}
\label{sec:headline-operating-points}

The baseline circuit, \texttt{30c8ded}, used $2{,}715$ logical qubits and $3{,}960{,}753$ average executed Toffoli gates. The best-scoring circuit, \texttt{8e9c9a2}, reduced these values to $1{,}151$ and $1{,}299{,}453$, respectively, lowering $Q$ by $57.6\%$, $T$ by $67.2\%$, and $Q\times T$ by $86.1\%$, from $10.75$B to $1.496$B, a $7.19$-fold improvement. A separate low-width branch reached $Q=825$ at approximately $489$M Toffoli gates. \Cref{fig:frontier-reconstruction} visualizes the resulting frontier and its evolution over time, while \Cref{tab:admitted-pareto} lists the eight admitted Pareto points at the cutoff.

The frontier contains two promoted points and six admitted low-width circuits preserved in public branches and submission notes. Here \emph{admitted} means that a circuit passes all $9{,}024$ Fiat--Shamir-derived validation inputs. The reported $T$ is the challenge's average executed-Toffoli metric, not a static gate count, $T$-depth, or fault-tolerant hardware cost.

The cost of reducing width varies sharply across the frontier: moving from $Q=828$ to $Q=826$ adds $31.23$M Toffoli gates, whereas moving from $Q=1{,}158$ to $Q=1{,}151$ adds only $6{,}802$. These differences are descriptive secants between distinct circuit routes, not causal marginal costs. In particular, the product-efficient branch uses dialog-GCD record and replay, whereas the sub-$1{,}000$-qubit branch uses register-shared shrunken-PZ inversion. The sparsely populated region between them may therefore reflect this architectural discontinuity rather than a fundamental width boundary; a hybrid or otherwise undiscovered construction could smooth the apparent transition. Additional operating points and fixed-budget comparisons appear in \href{https://github.com/jieyilong/ecdsafail-supplemental-materials/blob/ef7ecd2b27cf27881874468e003e53e16ba48075/appendices/detailed_results_and_circuit_improvement_trajectory.pdf}{Supplemental Note A}.

\begin{table}[tbp]
  \centering
  \scriptsize
  \setlength{\tabcolsep}{2.2pt}
  \renewcommand{\arraystretch}{1.15}
  \makebox[\linewidth][c]{%
  \begin{tabular}{@{}
    >{\raggedright\arraybackslash}p{2.65cm}
    >{\raggedright\arraybackslash}p{1.65cm}
    >{\raggedleft\arraybackslash}p{1.05cm}
    >{\raggedleft\arraybackslash}p{1.25cm}
    >{\raggedright\arraybackslash}p{2.75cm}
    >{\raggedleft\arraybackslash}p{2.05cm}
    >{\raggedright\arraybackslash}p{1.35cm}@{}}
    \toprule
    \textbf{Artifact} &
    \textbf{Interface} &
    \textbf{$Q$} &
    \shortstack[c]{\textbf{Reported}\\\textbf{$T$}} &
    \textbf{Error rate} &
    \shortstack[c]{\textbf{Retry-adjusted}\\$Q\times T/\hat p$} &
    \shortstack[c]{\textbf{Window}\\\textbf{support}} \\
    \midrule

    Google/Babbush low-$Q$ \cite{google26}
      & Window-indexed
      & \mbox{$\leq 1{,}175$}
      & \mbox{$\leq 2.7\,\mathrm{M}$}
      & $0/9{,}024$ FS
      & \mbox{$\approx 3.17\,\mathrm{B}^{*}$}
      & Yes \\

    Google/Babbush low-gate \cite{google26}
      & Window-indexed
      & \mbox{$\leq 1{,}425$}
      & \mbox{$\leq 2.1\,\mathrm{M}$}
      & $0/9{,}024$ FS
      & \mbox{$\approx 2.99\,\mathrm{B}^{*}$}
      & Yes \\

    Schrottenloher space-optimized \cite{s26}
      & Window-indexed
      & \mbox{$1{,}208^{\dagger}$}
      & \mbox{$2.59\,\mathrm{M}^{\dagger}$}
      & Reported $\epsilon\leq 2^{-13.3}$; $10{,}000$ tests
      & \mbox{$\approx 3.13\,\mathrm{B}^{*}$}
      & Yes \\

    Schrottenloher gate-optimized \cite{s26}
      & Window-indexed
      & \mbox{$1{,}462^{\dagger}$}
      & \mbox{$2.06\,\mathrm{M}^{\dagger}$}
      & Reported $\epsilon\leq 2^{-13.3}$; $10{,}000$ tests
      & \mbox{$\approx 3.01\,\mathrm{B}^{*}$}
      & Yes \\

    TrailMix \texttt{jump-lowqubit} \cite{trailmix}
      & Mixed; classical addend
      & \mbox{$1{,}169$}
      & \mbox{$\approx 2.09\,\mathrm{M}$}
      & $0/9{,}000$ FS
      & \mbox{$\approx 2.44\,\mathrm{B}^{*}$}
      & Not demonstrated \\

    TrailMix \texttt{shrunken-PZ} \cite{trailmix}
      & Mixed; classical addend
      & \mbox{$1{,}050$}
      & \mbox{$\approx 32.3\,\mathrm{M}$}
      & $0/9{,}000$ FS
      & \mbox{$\approx 33.9\,\mathrm{B}^{*}$}
      & Not demonstrated \\

    {\challengename} \texttt{8e9c9a2}
      & Mixed; classical addend
      & \mbox{$1{,}151$}
      & \mbox{$1.30\,\mathrm{M}$}
      & $0/9{,}024$ FS; $0.192\%$ on $100{,}000$ independent cases
      & \mbox{$\approx 1.499\,\mathrm{B}$}
      & No \\

    {\challengename} windowed \texttt{8e9c9a2}
      & Window-indexed
      & \mbox{$1{,}162$}
      & \mbox{$1.68\,\mathrm{M}$}
      & $0.191\%$ on $100{,}000$ independent cases
      & \mbox{$\approx 1.961\,\mathrm{B}$}
      & Yes \\

    {\challengename} \texttt{b6f2b0a}
      & Mixed; classical addend
      & \mbox{$825$}
      & \mbox{$\approx 489\,\mathrm{M}$}
      & $0/9{,}024$ FS
      & \mbox{$\approx 403\,\mathrm{B}^{*}$}
      & Not demonstrated \\

    {\challengename} \texttt{3616dbf} (post-cutoff)
      & Mixed; classical addend
      & \mbox{$1{,}321$}
      & \mbox{$\approx0.953\,\mathrm{M}$}
      & $0/9{,}024$ FS
      & \mbox{$\approx1.259\,\mathrm{B}^{*}$}
      & Not demonstrated \\
      
    \bottomrule
  \end{tabular}%
  }
  \vspace{5pt}
  \caption{Reported resources, correctness evidence, and retry-adjusted spacetime-resource proxies for representative \texttt{secp256k1} point-addition circuits. Here $\mathrm{B}=10^9$. Under an independently rerunnable classical-success model, a circuit with empirical success probability $\hat p$ requires $1/\hat p$ executions on average to obtain one successful result. Accordingly, $Q\times T/\hat p$ provides a sensitivity-adjusted proxy when success probabilities and failure events are defined and measured comparably. Among the window-supporting circuits listed in the table, the {\challengename} windowed \texttt{8e9c9a2} construction has the lowest tabulated $Q\times T/\hat p$, at approximately $1.961\,\mathrm{B}$, while supporting the required single-call coherent window-indexed point-addition interface. To avoid penalizing external circuits for which no common-corpus estimate is available, rows marked ${}^{*}$ conservatively assume zero error, or $\hat p=1$; these are therefore optimistic estimates, and any nonzero error would increase the required work. The Google/Babbush entries remain upper bounds because their published $Q$ and $T$ values are themselves upper thresholds. FS denotes finite Fiat--Shamir support, whereas ``independent'' denotes a corpus selected independently of the circuit hash. Failure definitions and accounting conventions differ across sources, so the comparison remains contextual. Window support denotes a coherent window-indexed point-addition interface, not a complete implementation of Shor's algorithm. ${}^{\dagger}$The Schrottenloher values include the separately reported $w=16$ overhead of $16$ qubits and $3\cdot2^{16}$ Toffoli gates; the corresponding arithmetic-core values are $(1{,}192,2.39\,\mathrm{M})$ and $(1{,}446,1.86\,\mathrm{M})$.}
  \label{tab:external-pa-comparison}
\end{table}

\subsubsection{Comparison with Prior Point-Addition Results}
\label{sec:external-pa-comparison}

\Cref{tab:external-pa-comparison} compares representative operating points from Babbush et al., Schrottenloher, TrailMix, and {\challengename}, including their interfaces, correctness evidence, and windowing support \cite{google26,s26,trailmix}; \Cref{fig:compare-frontier-with-recent-work} places them alongside the challenge frontier. Among the window-supporting entries, the {\challengename} \textit{windowed} \texttt{8e9c9a2} circuit has the lowest tabulated retry-adjusted proxy, $Q\times T/\hat p\approx1.961\,\mathrm{B}$, while providing a coherent single-call window-indexed point-addition interface. External rows without common-corpus measurements are evaluated optimistically with $\hat p=1$.

The mixed-addition challenge incumbent is also numerically below the $1{,}175$-qubit Babbush et al.\ point and the TrailMix low-qubit route on both $Q$ and $T$ axes, while the low-width branch reaches $Q=825$ at approximately $489$M executed Toffoli gates. These comparisons provide numerical context rather than establish formal dominance because interfaces, correctness assumptions, and accounting conventions differ; in particular, the coherent challenge construction is a point-addition kernel rather than a complete implementation of Shor's algorithm. The challenge additionally provides a public source history connecting each operating point to its preceding branches, optimization mechanisms, and local correctness obligations.

\subsection{Best {\challengename} Point-Addition Circuits at the Data Cutoff}
\label{sec:final-circuit-designs}

This subsection describes the architectures and principal optimization mechanisms of the best {\challengename} point-addition circuits at the data cutoff. We present the best $Q \times T$-scoring circuit and the best low-qubit construction in \Cref{sec:best-QxT-scoring-circuit} and \Cref{sec:best-low-Q-circuit}, emphasizing their distinct resource objectives and architectural trade-offs. The key circuit optimization techniques underlying these circuits are provided in~\Cref{sec:key-circuit-optimization-techniques}.

\subsubsection{Best \texorpdfstring{$Q \times T$}{Q times T}-Scoring Circuit} 
\label{sec:best-QxT-scoring-circuit}

\input{figs/best-qxt-circuit-8e9c9a2-pipeline}

As of the data cutoff, the promoted submission with the best $Q \times T$ score was \texttt{8e9c9a2}, with $Q=1{,}151$, $T=1{,}299{,}453$, and $Q \times T=1{,}495{,}670{,}403$. Its promoted source commit is \texttt{60d61859fa69}. We place the cutoff date at 26 July 2026, 09:21:55 UTC, when that submission's API record was last updated. This subsection provides an overview of the circuit, where a more comprehensive analysis of its arithmetic components and principal optimizations is available in \href{https://github.com/jieyilong/ecdsafail-supplemental-materials/blob/ef7ecd2b27cf27881874468e003e53e16ba48075/circuit_details/QxT_track_circuit_8e9c9a2/QxT_track_circuit_8e9c9a2_details.pdf}{Supplemental Note D: Technical Anatomy of Submission 8e9c9a2}.

\Cref{fig:best-qxt-circuit-pipeline} presents the six-stage in-place computation of $R'=R+A$, where $R=(x_R,y_R)$ is the quantum accumulator and $A=(x_A,y_A)$ is the selected addend. Stage~1 forms $d_x=x_R-x_A$ and $d_y=y_R-y_A$, and Stage~2 computes the slope $\lambda=d_y d_x^{-1}\bmod p$. Stage~3 updates the recycled $X$ register to $d_x+3x_A=x_R+2x_A$, after which Stage~4 subtracts $\lambda^2$ to obtain $X=x_A-x_{R'}$. Stage~5 computes $Y=\lambda(x_A-x_{R'})$, and Stage~6 recovers $y_{R'}=Y-y_A$ and $x_{R'}=x_A-X$. The same $X$ and $Y$ registers are therefore reused throughout, avoiding separate output-coordinate registers.

Stages~2 and~5 use dialog-GCD, the record-and-replay Euclidean architecture introduced by Khattar et al.~\cite{ksgz25} and subsequently adapted to ECDLP point addition by Schrottenloher~\cite{s26}. Its forward Euclidean traversal evolves a shrinking operand pair $(u,v)$ and records the parity, comparison, swap, subtraction, and halving decisions needed for reversal. The Jump-2 GCD groups consecutive Euclidean operations into larger steps (\Cref{sec:jump-two-gcd}), while the base-$5$ codec exploits the five reachable control symbols to encode three steps in seven qubits rather than nine (\Cref{sec:base-five-codec}). Reverse replay decodes the transcript to update the B\'ezout registers, restore the Euclidean operand, and clear the transcript. The resulting reversible map implements $Y\mapsto YX^{-1}$ in Stage~2 and, when used in the opposite computational direction, $Y\mapsto YX$ in Stage~5. Storage released as $(u,v)$ shrink is reused for the growing transcript and temporary arithmetic state, reducing peak width.

The remaining stages use complementary optimizations. Stages~1 and~6 serialize the materialization of classical coordinates and employ truncated pseudo-Mersenne correction and measurement-assisted carry cleanup. Stage~3 materializes the classically computed value $3x_A\bmod p$ only for the required addition and immediately unloads it (\Cref{sec:consprop}). Stage~4 uses streamed, symmetry-aware Karatsuba squaring together with $2^{256}\equiv2^{32}+977\pmod p$ (\Cref{sec:specialized-squaring}). Constant propagation and post-generation dead-code and redundancy elimination apply across the complete circuit (\Cref{sec:consprop,sec:dead-code-elimination}), with additional optimizations documented in \href{https://github.com/jieyilong/ecdsafail-supplemental-materials/blob/ef7ecd2b27cf27881874468e003e53e16ba48075/appendices/anatomy_of_agent_identified_quantum_circuit_optimizations.pdf}{Supplemental Note B: List of agent-identified optimizations}. Finally, the QROM boundaries in \Cref{fig:best-qxt-circuit-pipeline} indicate the lookup/use/unlookup interface required for integration with windowed Shor, which we implement and evaluate for this circuit in~\Cref{sec:windowed-addition-compatible-circuit}.

The circuit with the best $Q \times T$ score at the data cutoff can be retrieved and evaluated using the artifact and pinned-repository procedure in \Cref{sec:reproducibility}.

\subsubsection{Best low-\texorpdfstring{$Q$}{Q} Circuit}
\label{sec:best-low-Q-circuit}

As of the data cutoff, the selected low-$Q$ endpoint was submission \texttt{b6f2b0a}, with $Q=825$, $T=489{,}161{,}900$, and $Q\times T=403{,}558{,}567{,}500$. This operating point explores the extreme low-width region of the reversible time--space trade-off: it uses approximately $28\%$ fewer qubits than the best $Q\times T$ circuit, but substantially more executed Toffoli gates. A subsequent public submission, \texttt{b4a51d2}, reported an $813$-qubit circuit. A comprehensive source-grounded analysis of the arithmetic stages, live-register allocation, and principal optimizations of the 825-qubit circuit is available in \href{https://github.com/jieyilong/ecdsafail-supplemental-materials/blob/ef7ecd2b27cf27881874468e003e53e16ba48075/circuit_details/low_Q_track_circuit_b6f2b0a/low_Q_track_circuit_b6f2b0a_details.pdf}{Supplemental Note E: Technical Anatomy of Submission b6f2b0a}.

The circuit computes the in-place mixed addition $R'=R+A$ using the six stages shown in \Cref{fig:best-low-q-circuit-pipeline}. We retain the paper-wide definitions $d_x=x_R-x_A$ and $d_y=y_R-y_A$, while the low-$Q$ implementation stores their negations, $\widetilde d_x=x_A-x_R=-d_x$ and $\widetilde d_y=y_A-y_R=-d_y$, in the live registers $X$ and $Y$. Reversing both signs leaves the slope unchanged: $\lambda=\widetilde d_y\widetilde d_x^{-1}=(-d_y)(-d_x)^{-1}=d_y d_x^{-1}\pmod p$.

Stage~1 forms $X=\widetilde d_x$ and $Y=\widetilde d_y$, and Stage~2 computes $\lambda=\widetilde d_y\widetilde d_x^{-1}=d_y d_x^{-1}\bmod p$ while reversibly restoring the source-level differences needed for cleanup. Stage~3 erases the stored numerator through $Y\gets Y-\lambda X=\widetilde d_y-\lambda\widetilde d_x=-d_y+\lambda d_x=0$ and computes $x_{R'}=\lambda^2-x_R-x_A$, allowing the released $Y$ register to serve as arithmetic workspace. Stage~4 constructs the output-side witnesses $d_x'=x_A-x_{R'}$ and $d_y'=\lambda d_x'=y_{R'}+y_A$. Hence $\lambda=d_y d_x^{-1}=d_y'(d_x')^{-1}\pmod p$. Stage~5 uses the second identity to erase the slope through a fresh reversible division, without retaining the inversion history from Stage~2. Finally, Stage~6 recovers $x_{R'}=x_A-d_x'$ and $y_{R'}=d_y'-y_A$ in the original coordinate registers. The two inversion stages determine the 825-qubit peak; the remaining coordinate-arithmetic stages remain below it.

\input{figs/best-low-q-circuit-b6f2b0a-pipeline}

Unlike the best $Q\times T$ circuit, this design computes $\lambda=d_y d_x^{-1}=\widetilde d_y\widetilde d_x^{-1}\bmod p$ without dialog-GCD or a compressed transcript retained for later replay. Instead, it descends from TrailMix's \texttt{shrunken-PZ} architecture~\cite{trailmix}, which implements a Proos--Zalka-style extended Euclidean algorithm as a fixed reversible state machine~\cite{proos-zalka03}. The shrinking Euclidean remainders, growing cofactors, quotient bits, and associated length metadata share two 259-qubit physical banks whose ownership boundaries move during the computation. During each inversion peak, these banks coexist with a 257-qubit passenger coordinate and 37 persistent metadata qubits:
\begin{equation}
    \underbrace{259+259+257+37}_{812\ \mathrm{persistent\ qubits}}
    +\underbrace{13}_{\mathrm{maximum\ local\ workspace}}
    =825.
\end{equation}
Step-dependent widths, dynamic cursors, and reversible borrowing allow inactive high-order lanes to host quotient, comparison, carry, and control information, provided that each lane is restored before its logical owner becomes active again. The underlying inversion algorithm, packing invariants, and register-sharing schedule are detailed in \Cref{sec:register-shared-eea}.

The reduction from TrailMix's reported $1{,}050$-qubit construction to the first $980$-qubit challenge circuit retained the same shrunken-PZ inversion architecture but tightened its live-information schedule. The principal savings came from support-calibrated, step-dependent bounds for the Euclidean state, narrower shift and quotient metadata, and removal of a separate loaded-coordinate register at competing arithmetic peaks. This $70$-qubit reduction was therefore primarily a scheduling and liveness improvement within the inherited architecture, rather than the introduction of the register-shared EEA described below.

A subsequent architectural transition reduced the circuit from $947$ to $851$ qubits by reimplementing the shrunken-PZ extended Euclidean computation with substantially more aggressive register sharing. Shrinking remainders, growing B\'ezout coefficients, and temporary quotient state were packed into two shared physical work registers, with dynamic cursor management and movable ownership boundaries used to address the active subranges. This construction, described in \Cref{sec:register-shared-eea}, established the architecture from which the later $851$-to-$825$ reductions descended. Those reductions added denser lane hosting, measurement-assisted release and recomputation, narrower metadata, and increasingly local scratch schedules. The final $826$-to-$825$ transition removed one persistent quotient-length bit and reconstructed its omitted high-bit effect only where it was consumed.

The algebraic point-addition identities and reversible cleanup steps are exact under the generic-addition preconditions. Some step-dependent width and metadata bounds are calibrated to the challenge's finite validation support and should therefore be interpreted as validated benchmark envelopes rather than all-input width theorems. As with the best $Q\times T$ circuit, the benchmark supplies the addend $A$ classically; integration into windowed Shor requires a separately implemented and validated QROM lookup/use/unlookup interface.

The circuit can be retrieved and evaluated using the artifact and pinned-repository procedure in \Cref{sec:reproducibility}.

\subsubsection{Windowed-Addition Compatibility of the Best-Scoring Circuit}
\label{sec:windowed-addition-compatible-circuit}

The benchmark supplies the addend $A$ classically, whereas windowed Shor arithmetic selects it coherently from a precomputed table using a quantum address (\Cref{sec:ecc-quantum,sec:task}). To measure the resulting interface overhead, we constructed a windowed-compatible variant of \texttt{8e9c9a2}. It exposes the table as $2\cdot2^w$ classical coordinate inputs alongside the quantum coordinate registers $X,Y$ and a $w$-qubit address register $J$. Every table-dependent operation is conditioned on these input bits, making the circuit topology independent of the table contents; shifted $G$- or $P$-tables can therefore be substituted without recompilation. Each of the three stages that uses $A$ loads its selected coordinates through QROM immediately before use and uncomputes them afterward, so the selected point is not live during the dialog-GCD or multiplication peaks. Derived values such as $3x_A$ are computed within the circuit rather than stored in additional table columns, while the identity row $A=\mathcal O$ is handled coherently by conditioning the arithmetic on a nonzero-address predicate.

Because each arithmetic stage preserves the loaded payload until it is no longer needed, reversing the QROM traversal returns its payload registers exactly to zero. Routing ancillas are then removed by measurement-based uncomputation with a conditional phase correction, leaving no address-dependent payload, ancilla, or selection phase. The resulting lookup/use/unlookup map is therefore coherent by linearity and exact reversible replay. Although the evaluator simulates only computational-basis addresses, we tested individual addresses and correlated four-call sequences for every $w\in\{0,\ldots,5\}$, confirming workspace cleanup after each call.

For $w=16$, the resulting circuit uses $Q=1{,}162$ logical qubits and $T=1{,}684{,}161$ average executed Toffoli gates on the same $100{,}000$-input corpus used for the baseline. Relative to the original \texttt{8e9c9a2}, this adds $11$ qubits ($0.96\%$) and approximately $29.6\%$ average executed Toffoli per addition. Its any-channel error rate was $0.191\%$, versus $0.192\%$ for the baseline on the same independently seeded $100{,}000$-input corpus. The paired comparison had 29 windowed-only and 30 baseline-only failures (two-sided exact sign test, $p=1.0$), providing no detectable difference on this corpus; no additional classical-output failures were attributable to coherent selection.

These counts follow the challenge's $Q\times T$ convention and exclude the fault-tolerant costs of table-payload CNOTs, classical table storage, measurements, and feed-forward. Those operations increase the static operation stream by a factor of $23.36$ and may be significant under other architectures. The construction provides a reusable window-indexed point-addition kernel that coherently handles the identity-table row $A=\mathcal O$ by bypassing the arithmetic. For $A\neq\mathcal O$, the construction inherits the source circuit's approximate arithmetic, lack of a coherent failure flag, and generic-affine domain from \Cref{sec:challenge-and-benchmark}; in particular, it requires $R\neq\mathcal O$ and $R\notin\{A,-A,-2A\}$. Instantiating every shifted table and integrating the complete window schedule, Fourier transforms, and classical postprocessing also remain future work. Nevertheless, the result shows that the benchmark's classically supplied addend is an interface choice rather than a fundamental capability gap, with the compatible variant remaining in the low-qubit, low-Toffoli region of \Cref{tab:external-pa-comparison}. \href{https://github.com/jieyilong/ecdsafail-supplemental-materials/blob/ef7ecd2b27cf27881874468e003e53e16ba48075/circuit_details/windowed_QxT_track_circuit_8e9c9a2/windowed_QxT_track_circuit_8e9c9a2_details.pdf}{Supplemental Note F} provides the complete construction, resource breakdown, and validation; the implementation is available in the \href{https://github.com/jieyilong/ecdsafail-challenge/tree/15a29ac9b1e97fc21eb81b9a6f0f6d9c4dc7a92e}{source repository}.



\subsection{Principal Circuit Optimization Techniques}
\label{sec:key-circuit-optimization-techniques}

This subsection examines the principal techniques underlying the circuits discussed in the previous subsection, including the Jump-2 Euclidean algorithm (\Cref{sec:jump-two-gcd}), base-$5$ transcript encoding (\Cref{sec:base-five-codec}), specialized squaring (\Cref{sec:specialized-squaring}), register-shared extended Euclidean inversion (\Cref{sec:register-shared-eea}), constant propagation (\Cref{sec:consprop}), and dead-code and redundancy elimination (\Cref{sec:dead-code-elimination}). We additionally document the post-cutoff comparison-free ping-pong dialog-GCD architecture (\Cref{sec:ping-pong-dialog-gcd}), which replaces rather than extends the earlier Jump-2/base-$5$ backend.

\subsubsection{The Jump-2 Euclidean Algorithm}
\label{sec:jump-two-gcd}

The baseline dialog-GCD record phase in \Cref{alg:dialog-gcd-record} records one symbol $\sigma_i=(b_i,s_i)$ per ordinary divstep, where $b_i$ indicates subtraction and $s_i$ indicates a preceding swap. Jump-2 preserves the resulting Euclidean transformation but retimes consecutive divsteps into larger macro-steps. The division that necessarily follows a subtraction of two odd operands is moved to the beginning of the next macro-step; if the intermediate value remains even, a second division is performed before the next possible subtraction.

At every normal macro-step boundary, $v$ is even. The first division is therefore unconditional, and the macro-step realizes
\begin{equation}
  v\longmapsto
  \begin{cases}
    v/2, & v/2\text{ is odd},\\
    v/4, & v/2\text{ is even}.
  \end{cases}
  \label{eq:jump-two-core}
\end{equation}
It then performs the next required comparison, swap, and subtraction. The ordinary symbol $(b_i,s_i)$ is extended to $\sigma_i=(b_i,s_i,s_{2,i})$, where $s_{2,i}$ records whether the second division in \eqref{eq:jump-two-core} occurred.

\begin{algorithm}[t]
  \caption{Jump-2 Euclidean macro-steps}
  \label{alg:jump-two-euclidean}
  \begin{algorithmic}[1]
    \Statex \textbf{Input:} $(u,v)=(p,x)$ with $x\in\F_p^{*}$ and odd $u$; fixed macro-step budget $L$
    \Statex \textbf{Output:} transcript $\tau=(\sigma_0,\ldots,\sigma_{L-1})$
    \For{$i=0,\ldots,L-1$}
      \If{$i=0$}
        \State $t_1\gets[v\text{ is even}]$;\quad
               $v\gets v/2$ if $t_1=1$
      \Else
        \State $v\gets v/2$
        \Comment{unconditional at a normal boundary}
      \EndIf
      \State $s_{2,i}\gets[v\text{ is even}]$;\quad
             $v\gets v/2$ if $s_{2,i}=1$
      \State $b_i\gets[v\text{ is odd}]$
        \Comment{subtraction flag}
      \State $s_i\gets b_i$ if $i=0$, otherwise
             $s_i\gets b_i\wedge[v<u]$
        \Comment{swap flag}
      \State $(u,v)\gets(v,u)$ if $s_i=1$
      \State $v\gets v-u$ if $b_i=1$
      \State append $\sigma_i=(b_i,s_i,s_{2,i})$ to $\tau$
        \Comment{step $0$ also records $t_1$}
    \EndFor
  \end{algorithmic}
\end{algorithm}

Every macro-step again leaves $v$ even. If $b_i=0$, the value remains even after the divisions; if $b_i=1$, both active operands are odd and their difference is even. The next normal macro-step may therefore begin with its unconditional division. The initial macro-step is exceptional because the parity of $x$ is unknown, so $t_1$ records whether its first division occurs. Jump-2 is an exact retiming of the ordinary dialog-GCD path: it reduces the number of macro-step boundaries, comparisons, swaps, and subtraction bodies without changing the induced arithmetic transformation.

\paragraph{Use in dialog-GCD.} For the transcript generated by a nonzero operand $x$, let $D_x$ denote the induced Euclidean transformation. Once the transcript is fixed, $D_x$ is a linear map over $\F_p$ because it consists only of swaps, additions or subtractions, and multiplication by powers of $2^{-1}$. Successful convergence gives $D_x(p,x)=(1,0)$, which reduces modulo $p$ to $D_x(0,x)=(1,0)$. Since $x\neq0$, its inverse $x^{-1}$ exists in $\F_p$ and any $y\in\F_p$ can be written as $y=(yx^{-1})x$. Linearity allows the scalar $yx^{-1}$ to be pulled through $D_x$, giving $D_x(0,y)=D_x\!\left(0,(yx^{-1})x\right)=yx^{-1}D_x(0,x)=(yx^{-1},0)$, which is the division map in \Cref{eq:jump-two-linear}. Likewise, scaling by $y$ gives $D_x(0,yx)=(y,0)$, and applying $D_x^{-1}$ gives the corresponding multiplication map:
\begin{equation}
  D_x(0,y)=(yx^{-1},0),
  \qquad
  D_x^{-1}(y,0)=(0,yx).
  \label{eq:jump-two-linear}
\end{equation}
Stage~2 uses the first map with $x=d_x$ and $y=d_y$ to obtain $\lambda=d_y d_x^{-1}$, while Stage~5 uses the inverse map with the then-current $X$ operand to perform in-place multiplication.

During reverse replay, each symbol controls the inverse macro-step
\begin{equation}
  \begin{aligned}
    &\text{if }b_i:\quad z\gets z+y\pmod p,\\
    &\text{if }s_i:\quad (y,z)\gets(z,y),\\
    &z\gets2^{k_i}z\pmod p,
  \end{aligned}
  \qquad
  k_i=
  \begin{cases}
    t_1+s_{2,0}, & i=0,\\
    1+s_{2,i}, & i>0.
  \end{cases}
  \label{eq:jump-two-reconstruct}
\end{equation}
Modular addition reverses subtraction, swapping reverses itself, and multiplication by $2^{k_i}$ reverses the corresponding divisions because $p$ is odd. Each decoded symbol is used to update the payload and reverse its Euclidean macro-step before being cleared. The doubling or quadrupling operation shares a pseudo-Mersenne correction using $2^{256}\equiv2^{32}+977\pmod p$, while the transcript encoding and streamed replay schedule are described in \Cref{sec:base-five-codec}.

A reversible circuit cannot terminate after an input-dependent number of macro-steps, so the implementation uses a fixed budget. Increasing it from $258$ to $261$ macro-steps reduced the observed nonconvergence frequency from $2.839\times10^{-4}$ to $2.86\times10^{-5}$ over ten million inputs. Conditional on convergence, Jump-2 exactly reproduces the baseline dialog-GCD transformation; approximation arises only from the finite schedule or separately documented width truncations.

\subsubsection{Base-5 Codec for the Jump-2 Transcript}
\label{sec:base-five-codec}

Because the Euclidean operands may be in superposition, the Jump-2 symbols can be entangled with the input and must be stored coherently in the transcript $\tau$. A raw symbol $\sigma_i=(b_i,s_i,s_{2,i})$ occupies three qubits, but the update rules permit only five of the eight Boolean triples. If $s_{2,i}=0$, the value after the first division is odd and therefore $b_i=1$; moreover, a swap requires a subtraction, so $s_i=1$ implies $b_i=1$. The reachable alphabet is
\begin{equation}
  \mathcal S=
  \left\{
    (0,0,1),
    (1,0,0),
    (1,1,0),
    (1,0,1),
    (1,1,1)
  \right\}.
  \label{eq:jump-two-alphabet}
\end{equation}
A fixed bijection $\phi:\mathcal S\rightarrow\{0,1,2,3,4\}$ therefore represents each symbol as a base-$5$ digit $d_i=\phi(\sigma_i)$.

Three consecutive digits can be packed into the integer
\begin{equation}
  z=d_0+5d_1+25d_2,
  \qquad
  0\leq z<5^3=125<2^7.
  \label{eq:base-five-triple}
\end{equation}
Thus, although three raw symbols occupy nine qubits, their $5^3=125$ reachable combinations fit in a seven-qubit register; similarly, two symbols have $5^2=25$ combinations and require five rather than six qubits. Writing $\ket{a}_m$ for the $m$-qubit computational-basis encoding of $a$, the reversible encoder acts on the reachable support as
\begin{equation}
  \ket{\sigma_0}_{3}\ket{\sigma_1}_{3}\ket{\sigma_2}_{3}
  \longmapsto
  \ket{z}_{7}\ket{0}^{\otimes2},
  \qquad
  z=d_0+5d_1+25d_2.
  \label{eq:base-five-unitary}
\end{equation}
The encoder therefore does not discard two qubits; it returns them coherently to $\ket{0}$ so they can be released or reused. Its inverse reconstructs all three symbols exactly. To define a valid reversible circuit on the complete nine-qubit Hilbert space, the implementation extends this action to a permutation of all computational-basis states, including those outside the reachable support $\mathcal S^3$.


The initial macro-step is encoded separately because it also contains $t_1$; its four reachable states fit in two qubits. Of the remaining $260$ ordinary symbols, $258$ are stored in $86$ seven-qubit triple blocks. The final two symbols form an incomplete group and are packed into one five-qubit pair block, since $5^2=25<2^5$. The complete $261$-step layout therefore uses
\begin{equation}
  2+86\cdot7+5=609
  \label{eq:base-five-dialog-size}
\end{equation}
qubits, compared with $261\cdot3=783$ qubits for raw storage, reducing the transcript by $174$ qubits.

Encoding and replay are streamed rather than performed only after the complete walk. As the Euclidean operands shrink, completed symbol groups are encoded into storage released by their inactive high-order bits. Reverse replay decodes one group at a time, applies its symbols to the payload and Euclidean state, and returns both the decoded controls and encoded block to zero. The codec therefore changes neither the Euclidean path nor the resulting multiplication or division; it exchanges reversible encoding work for a smaller persistent transcript and lower peak width.

\subsubsection{Comparison-Free Ping-Pong Dialog-GCD}
\label{sec:ping-pong-dialog-gcd}

After the paper's data cutoff, submission \href{https://github.com/Layr-Labs/ecdsafail-challenge/commit/897dda2b0cf267151ecd973252d2a5078cbf1b63}{\texttt{3616dbf}} replaced the Jump-2/base-$5$ backend with a different dialog-GCD architecture. An ordinary binary-GCD step compares two full-width quantum integers and may then swap them. A reversible comparison examines many bits, produces a control that must later be cleared, and controls a wide swap and subtraction. The ping-pong construction instead chooses addition or subtraction from only the two lowest bits and updates the operand registers in a fixed alternating order. It therefore removes full-width magnitude comparisons and data-dependent swaps from the Euclidean value walk. Here, ``comparison-free'' refers only to that walk; bounded carry and overflow tests remain inside the modular replay cells.

The construction combines two established ideas in a new circuit organization. Low-bit plus-minus GCD methods replace magnitude comparisons with tests modulo $4$~\cite{bigou-tisserand15}, while dialog-GCD records a Euclidean path for later coefficient replay~\cite{ksgz25,s26}. The challenge-specific specialization keeps both signed operands odd, alternates the updated operand deterministically, records one sign bit per round, and implements division and multiplication through fused replay cells.

\paragraph{Motivating example.} We illustrate how the ping-pong value walk and its transcript are used to obtain the inverse of $x=5$ modulo $p=13$. Starting from the odd pair $(13,5)$, the method alternately updates the second and first entries. In each round, the entry being updated is called the \emph{target}, and the other entry is called the \emph{source}. If the two entries have the same residue modulo $4$, the target is replaced by half their sum; otherwise, it is replaced by half the difference between the target and the source:
\begin{equation}
\begin{aligned}
(13,5)
&\xrightarrow{+}(13,9)
\xrightarrow{+}(11,9)
\xrightarrow{-}(11,-1)
\xrightarrow{+}(5,-1)\\
&\xrightarrow{-}(5,-3)
\xrightarrow{+}(1,-3)
\xrightarrow{+}(1,-1).
\end{aligned}
\label{eq:ping-pong-example}
\end{equation}
The target alternates at every arrow. For example, the first round updates the second entry as $(5+13)/2=9$, while the third updates it as $(9-11)/2=-1$. No round compares operand magnitudes or performs a data-dependent swap. The one-bit transcript is $\tau=(0,0,1,0,1,0,0)$, where $0$ denotes addition and $1$ denotes subtraction. Applying the transformations selected by $\tau$ modulo $13$ to a coefficient pair initialized as $(0,1)$ yields $(8,5)\equiv(8,-8)\pmod{13}$, where division by two is implemented as multiplication by $2^{-1}=7\pmod{13}$. Correcting the negative terminal sign gives $5^{-1}=8\pmod{13}$, consistent with $5\cdot8\equiv1\pmod{13}$.

\paragraph{Odd-preserving update and correctness.} For a general nonzero denominator $x\in\F_p$, the circuit begins with two odd operands by choosing
\begin{equation}
    x^\star=
    \begin{cases}
        x,   & x\text{ is odd},\\
        x-p, & x\text{ is even},
    \end{cases}
    \qquad
    x^\star\equiv x\pmod p.
    \label{eq:ping-pong-odd-lift}
\end{equation}
Since $p$ is odd, both $p$ and $x^\star$ are odd. In each round, let $s$ be the source and $t$ the predetermined target. For an odd signed integer $z$, let $z[1]$ distinguish $z\equiv1\pmod4$ from $z\equiv3\pmod4$. The circuit records
\begin{equation}
    e=t[1]\oplus s[1],
    \qquad
    t'\;=\frac{t+(-1)^e s}{2}.
    \label{eq:ping-pong-forward}
\end{equation}
If the low-bit patterns agree, the sum is $2\pmod4$; if they differ, the target-minus-source difference is $2\pmod4$. Thus, the numerator is divisible by two but not by four, so $t'$ is again odd. Moreover, $\gcd(s,t)=\gcd(s,t\pm s)=\gcd(s,(t\pm s)/2)$ because the common divisor of two odd operands is odd. Every round therefore preserves the GCD. The round is also reversible once $e$ is known, since $t=2t'-(-1)^e s$. Consequently, the transcript uniquely determines both the forward transformation and its inverse. Since $\gcd(p,x^\star)=1$, reaching a signed-unit pair certifies successful convergence.

\begin{algorithm}[t]
  \caption{Comparison-free ping-pong value walk}
  \label{alg:ping-pong-dialog-gcd}
  \begin{algorithmic}[1]
    \Statex \textbf{Input:} nonzero denominator $x\in\F_p$; fixed round budget $L$
    \Statex \textbf{Output:} terminal signed pair and transcript $\tau$
    \State $\rho_0\gets p$;\quad
           $\rho_1\gets x$ if $x$ is odd, otherwise
           $\rho_1\gets x-p$
    \State $\tau\gets()$
    \For{$i=0,\ldots,L-1$}
      \If{$i$ is even}
        \State $(s,t)\gets(\rho_0,\rho_1)$
          \Comment{update $\rho_1$}
      \Else
        \State $(s,t)\gets(\rho_1,\rho_0)$
          \Comment{update $\rho_0$}
      \EndIf
      \State $e_i\gets t[1]\oplus s[1]$
      \State $t\gets(t+(-1)^{e_i}s)/2$
      \State append $e_i$ to $\tau$
    \EndFor
    \Statex \textbf{Successful endpoint:}
      $(\rho_0,\rho_1)=(\eta_0,\eta_1)$ with
      $\eta_0,\eta_1\in\{-1,+1\}$
  \end{algorithmic}
\end{algorithm}

The implementation uses $L=704$ rounds and stores one transcript qubit per round. This is wider than the $609$-qubit base-$5$ Jump-2 transcript, but its one-bit symbols require cheaper replay. Signed-unit pairs are stable under subsequent rounds, so reversible padding preserves an endpoint reached before the fixed schedule ends.

\paragraph{Dialog replay.} For the transcript generated by $x$, let $D_x$ denote the resulting linear transformation. On a converged path,
\begin{equation}
    D_x(p,x^\star)=(\eta_0,\eta_1),
    \qquad
    \eta_0,\eta_1\in\{-1,+1\}.
    \label{eq:ping-pong-terminal}
\end{equation}
Modulo $p$, $(p,x^\star)$ is equivalent to $(0,x)$. Since every transcript-selected update is linear over $\F_p$, applying the same transformation to $(0,y)$ gives
\begin{equation}
    D_x(0,y)=yx^{-1}(\eta_0,\eta_1)\pmod p.
    \label{eq:ping-pong-division}
\end{equation}
After correcting the terminal signs, both coefficient registers contain $yx^{-1}$; one copy is retained and the other is cleared. Stage~2 applies this map with $x=d_x$ and $y=d_y$ to obtain $\lambda=d_y d_x^{-1}$.

Multiplication uses the inverse round derived from \Cref{eq:ping-pong-forward}. If $c_t'=(c_t+(-1)^{e_i}c_s)/2$, then
\begin{equation}
    c_t=2c_t'-(-1)^{e_i}c_s\pmod p.
    \label{eq:ping-pong-inverse-round}
\end{equation}
Reverse replay therefore maps the sign-corrected terminal pair to $(0,yx)$. Stage~5 uses this direction with $x=x_A-x_{R'}$ and $y=\lambda$. The circuit then reverses the value walk, restores $x$, undoes the odd lift, and clears the transcript and workspace.

\paragraph{Circuit realization and resource effect.} Each replay round is implemented as one fused modular cell. Division combines signed addition with modular halving, while multiplication combines modular doubling with the signed correction in \Cref{eq:ping-pong-inverse-round}. For $p=2^{256}-(2^{32}+977)$, overflow at bit $256$ becomes a sparse low-word correction. The implementation processes these corrections in short chunks and uses measurement-assisted cleanup for temporary products and carries.

Relative to its reproduced source parent, the ping-pong circuit increases peak width from $1{,}150$ to $1{,}321$ qubits but reduces average executed Toffoli count from $1{,}284{,}776$ to $952{,}707$, a reduction of $25.85\%$. Its $Q\times T$ score decreases from $1{,}477{,}492{,}400$ to $1{,}258{,}525{,}947$, or $14.82\%$. The construction therefore spends transcript width to obtain a simpler transition alphabet and cheaper replay.


\subsubsection{Karatsuba Squaring with Pseudo-Mersenne Modulo}
\label{sec:specialized-squaring}

The point-addition circuit requires the modular square-subtract transformation
\begin{equation}
    \ket{\lambda}\ket{X}\ket{0}_W
    \longmapsto
    \ket{\lambda}\ket{X-\lambda^2\bmod p}\ket{0}_W,
    \label{eq:square-subtract-map}
\end{equation}
where $W$ denotes temporary workspace and $p=2^{256}-F$ with $F=2^{32}+977$. Modular squaring is a standard component of quantum elliptic-curve point addition~\cite{rnsl17,hjn20,s26}, and the construction uses the established Karatsuba decomposition~\cite{parent2017karatsuba}. Its principal contribution is to specialize this decomposition to squaring, accumulate directly into the destination register, exploit the pseudo-Mersenne form of the \texttt{secp256k1} modulus, and schedule the computation reversibly without retaining a full-width product.

Let $\beta=2^{128}$ and split the slope as $\lambda=L+\beta H$, where $0\leq L,H<2^{128}$. Then $\lambda^2=L^2+2\beta LH+\beta^2H^2$. Define the three half-size squares $A=L^2$, $B=H^2$, and $C=(L+H)^2$. Since $C=A+2LH+B$, the cross term is $2LH=C-A-B$, yielding the one-level Karatsuba decomposition
\begin{equation}
    \lambda^2
    =A+\beta(C-A-B)+\beta^2B.
    \label{eq:karatsuba-square}
\end{equation}

The construction further specializes \Cref{eq:karatsuba-square} to the \texttt{secp256k1} prime. Since $\beta^2=2^{256}\equiv F\pmod p$,
\begin{equation}
    \lambda^2
    \equiv A+\beta(C-A-B)+FB
    \pmod p.
    \label{eq:secp-karatsuba-square}
\end{equation}
The sparse signed-binary representation $F=2^{32}+2^{10}-2^5-2^4+1$ allows multiplication by $F$ to use fixed shifts and signed additions rather than a generic 256-bit modular multiplication or reduction.

The circuit does not materialize the reduced value $\lambda^2$ in a separate register. Instead, it applies the three contributions in \Cref{eq:secp-karatsuba-square} directly to $X$:
\begin{align}
    C=(L+H)^2:&\qquad X\gets X-\beta C,
      \label{eq:apply-c}\\
    A=L^2:&\qquad X\gets X-A+\beta A,
      \label{eq:apply-a}\\
    B=H^2:&\qquad X\gets X+\beta B-FB.
      \label{eq:apply-b}
\end{align}
Their sum is $-\beta C-A+\beta A+\beta B-FB=-[A+\beta(C-A-B)+FB]$, which implements $X\gets X-\lambda^2\pmod p$.

The updates follow a build--apply--unbuild schedule. The circuit constructs $C$ in clean workspace, applies $-\beta C$ to $X$, and reverses the squaring network to clear the workspace; it then repeats this sequence for $A$ and $B$. The temporary register holding $L+H$ exists only while constructing $C$. Consequently, $A$, $B$, and $C$ are never simultaneously live, and no persistent 512-bit product register is required.

Each half-size product is implemented as a symmetric square rather than a generic multiplication. For $z=\sum_i z_i2^i$, symmetry gives $z^2=\sum_i z_i2^{2i}+2\sum_{i<j}z_iz_j2^{i+j}$. The diagonal terms satisfy $z_i^2=z_i$, while each off-diagonal product $z_iz_j$ is computed only once and contributes with a one-bit shift. A direct 256-bit symmetric square contains $256\cdot255/2=32{,}640$ off-diagonal products, whereas the two 128-bit squares and one 129-bit sum-square used here contain $2\binom{128}{2}+\binom{129}{2}=24{,}512$, approximately $24.9\%$ fewer. Temporary product bits and arithmetic carries can be uncomputed as soon as their final contribution to $X$ has been applied.

The construction therefore combines a square-specific Karatsuba decomposition, symmetry-aware partial-product generation, direct accumulation into the point-addition register, pseudo-Mersenne reduction, and phase-local reversible cleanup. The algebraic identity is exact; any approximation in a benchmark circuit arises only from separately documented finite-width modular-correction subroutines, not from the Karatsuba decomposition itself.

\subsubsection{Register-Shared EEA and Lane-Lifetime Refinements}
\label{sec:register-shared-eea}

The $947$-to-$851$-qubit transition replaced the preceding shrunken-PZ inversion backend with a source-level implementation of the register-shared extended Euclidean algorithm (EEA) of Luo et al.~\cite{luo26}. The underlying principle, i.e. sharing storage between shrinking Euclidean remainders and growing B\'ezout coefficients, was introduced by Proos and Zalka~\cite{proos-zalka03}; Luo et al.\ supplied an explicit reversible construction using two shared work registers, embedded quotient storage, moving field boundaries, a rotation cursor, and step-dependent active arithmetic. The challenge contribution is an implementation and refinement of this architecture within the benchmark point-addition circuit. Its additional reductions come from more aggressive lane lending, metadata narrowing, lifetime splitting, and route-specific control simplification.

For nonzero $x\in\F_p$, the circuit implements
\begin{equation}
    \ket{x}\ket{y}\ket{0}_{W}
    \longmapsto
    \ket{x}\ket{yx^{-1}\bmod p}\ket{0}_{W},
    \label{eq:register-shared-division-map}
\end{equation}
where $W$ is the Euclidean workspace. In Stage~2 of the low-$Q$ point-addition circuit, $x=\widetilde d_x=x_A-x_R=-d_x$ and $y=\widetilde d_y=y_A-y_R=-d_y$, giving $\lambda=\widetilde d_y\widetilde d_x^{-1}=d_y d_x^{-1}$. Stage~5 uses the same machinery with the output-side witnesses $x=d_x'=x_A-x_{R'}$ and $y=d_y'=y_{R'}+y_A$, whose ratio is again $\lambda$.

\paragraph{Cross-cofactor invariant.} The EEA begins with $\rho_0=p$ and $\rho_1=x$. At iteration $i$, it divides $\rho_{i-1}$ by $\rho_i$ and computes
\begin{equation}
    \rho_{i-1}=q_i\rho_i+\rho_{i+1},
    \qquad
    0\leq\rho_{i+1}<\rho_i,
\end{equation}
where $q_i$ is the quotient and $\rho_{i+1}$ is the remainder left by the division. The next iteration operates on $(\rho_i,\rho_{i+1})$. Thus, two consecutive remainders, meaning two adjacent values in the remainder sequence, must be stored simultaneously at each EEA boundary.

The extended algorithm also tracks how every remainder depends on the original inputs. Each remainder has the form $\rho_i=\beta_i p+\alpha_i x$ for integer coefficients $\alpha_i,\beta_i$, and therefore $\rho_i\equiv\alpha_i x\pmod p$. If the algorithm reaches $\rho_m=1$, then $\alpha_mx\equiv1\pmod p$, so $\alpha_m\equiv x^{-1}\pmod p$.

At a given boundary, denote the two simultaneously stored consecutive remainders by $\rho=\rho_i$ and $\rho'=\rho_{i+1}$, up to the ordering recorded by the role bit $\pi$, and denote their respective coefficients of $x$ by $\alpha=\alpha_i$ and $\alpha'=\alpha_{i+1}$. Following the two-coefficient representation of Luo et al.~\cite{luo26}, the packed implementation stores these coefficients as signed \emph{cross-cofactors}
\begin{equation}
    c=(-1)^\pi\alpha',
    \qquad
    c'=(-1)^{\pi+1}\alpha,
\end{equation}
where $\pi\in\{0,1\}$ records the current ordering of the two EEA states. They are crossed because $c$ is derived from the coefficient associated with $\rho'$, while $c'$ is derived from the coefficient associated with $\rho$. Consecutive EEA states satisfy the determinant identity $\rho\alpha'-\rho'\alpha=(-1)^\pi p$. Substituting the definitions of $c$ and $c'$ gives the maintained invariants
\begin{equation}
    \rho c+\rho'c'=p,
    \qquad
    \rho\equiv(-1)^{\pi+1}xc'\pmod p,
    \qquad
    \rho'\equiv(-1)^\pi xc\pmod p.
    \label{eq:register-shared-cross-invariants}
\end{equation}
The initial state $(\rho,\rho',c,c',\pi)=(p,x,1,0,0)$ satisfies these relations, and each paired remainder and cross-cofactor update preserves them throughout the EEA.

The representation makes each quotient update act on one remainder and the cross-cofactor associated with the opposite remainder. For a shifted quotient contribution $\kappa=2^{\ell_s}$, exactly one of the following paired updates is applied:
\begin{equation}
\begin{aligned}
    \rho  &\leftarrow \rho  \mp \kappa\rho',
    &\qquad
    c' &\leftarrow c' \pm \kappa c,
    &&\text{when $\rho$ is the target},\\
    \rho' &\leftarrow \rho' \mp \kappa\rho,
    &\qquad
    c  &\leftarrow c  \pm \kappa c',
    &&\text{when $\rho'$ is the target}.
\end{aligned}
\label{eq:register-shared-microstep}
\end{equation}
The two lines are symmetric and are not performed simultaneously. In the implementation, one orientation is used during a Euclidean division; when that division finishes, the remainder/cofactor roles are exchanged and $\pi$ is toggled.

The opposite signs preserve the central invariant. For example, the subtractive update in the first line gives $(\rho-\kappa\rho')c+\rho'(c'+\kappa c)=\rho c+\rho'c'=p$. The second line preserves the invariant by the same calculation.

\paragraph{Register sharing.} Luo et al.'s construction stores the active EEA state in two shared qubit banks:
\begin{equation}
    W_1=c\;\Vert\;q\;\Vert\;\rho,
    \qquad
    W_2=\operatorname{rot}_{\ell_s}\!\left(c'\;\Vert\;\rho'\right).
    \label{eq:register-shared-layout}
\end{equation}
Here $q$ is the partial quotient for the current division. The algorithm constructs $q_i=\lfloor\rho/\rho'\rfloor$ one binary digit at a time by testing shifted multiples $2^{\ell_s}\rho'$. Each accepted subtraction writes the corresponding quotient bit into $q$; after the completed quotient has controlled the associated cross-cofactor updates, its storage is cleared. The cursor $\ell_s$ records the circular offset of $(c'\Vert\rho')$ within $W_2$. Changing this offset aligns $\rho'$ as $2^{\ell_s}\rho'$ and aligns the corresponding cross-cofactor operation without allocating separate shifted copies.

Initially, the remainders occupy most of the banks and the cross-cofactors and partial quotient are small. As the remainders shrink, the other fields expand into their vacated lanes. Scheduled length bounds identify the active field boundaries and keep each bank within $n+3=259$ qubits for $n=256$. This two-bank layout, including quotient packing, cursor-based alignment, length metadata, and baseline active-window arithmetic, comes from Luo et al.~\cite{luo26}.

The challenge implementation integrates this layout into the complete mixed point-addition circuit and reduces its transient peak through more aggressive lifetime management. Temporarily inactive carry, comparison, coefficient, metadata, and coordinate lanes are loaned to local operations and restored before their owners become active again. It also narrows one length field, splits comparison and rotation lifetimes, fuses selected predicates and boundary operations, and specializes several guards to the validated route.

The implementation executes a fixed schedule of $1{,}479$ reversible microsteps, using reversible padding when the EEA terminates early. At the normalized endpoint, $\rho=1$, $\rho'=0$, and $q=0$. Equation~\eqref{eq:register-shared-cross-invariants} then gives $c'\equiv(-1)^{\pi+1}x^{-1}\pmod p$, so a sign correction recovers the inverse. After multiplying it into $y$, the circuit reverses the schedule to restore $x$ and clear the work banks and metadata.

The two banks and principal metadata use $567$ qubits. Together with the challenge-specific lifetime refinements, this architecture reduces the complete point-addition circuit from $947$ to $851$ logical qubits, while increasing average executed Toffoli count from $61{,}946{,}637$ to $464{,}495{,}956$. It is therefore a width-oriented trade-off. The EEA and cross-cofactor invariants are exact, whereas the submitted route's support-tuned active widths, truncated guards, and lane-hosting assumptions are validated only on the benchmark corpus and should not be interpreted as an all-input correctness result.

\subsubsection{Constant Propagation}
\label{sec:consprop}

The optimization referred to as \emph{constant propagation} (\emph{Constprop}) exploits control information that is classical for each evaluation instance rather than stored in quantum superposition. Let the classical multiplier be $y=\sum_{i=0}^{n-1} y_i2^i$, where $y_i\in\{0,1\}$. A quantum-by-classical schoolbook multiplication decomposes as
\begin{equation}
xy=\sum_{i=0}^{n-1} y_i(2^i x),
\end{equation}
so each multiplier bit selects a modular update of an accumulator $a$:
\begin{equation}
a\longmapsto
\begin{cases}
a+2^i x \pmod p, & y_i=1,\\
a, & y_i=0.
\end{cases}
\end{equation}
Because $y_i$ is classically available, the corresponding arithmetic block can be executed only when $y_i=1$. When $y_i=0$, the block implements the identity and may therefore be omitted without changing the circuit's input--output behavior. This is an exact specialization, not an approximation.

The same transformation would not apply if $y_i$ were held in quantum superposition, because the arithmetic operation would then require coherent quantum control. In the present setting, however, direct conditioning on the classical multiplier bits prevents inactive modular-addition and subtraction blocks from contributing to the executed-gate metric. This specialization reduced the multiplier's average executed Toffoli count from $55{,}277{,}050$ to $54{,}495{,}466$, a reduction of $781{,}584$, while leaving the peak logical width unchanged. More generally, the expected saving scales with the frequency with which the relevant classical control bits are zero over the evaluation distribution.

\subsubsection{Dead-Code Elimination}
\label{sec:dead-code-elimination}

Dead-code and redundancy elimination removes operations and temporary storage that cannot affect the required output on the declared input domain. For example, two adjacent applications of a self-inverse operation \(U\) cancel according to
\begin{equation*}
    U^2=I.
\end{equation*}
Such cancellations are valid provided that no intervening operation changes the relevant registers. Consistent operand ordering and sign conventions can similarly eliminate redundant negations, swaps, and reversible conversions. Modular-reduction folds, carry propagation, and conditional corrections may also be shortened or removed when an analytic invariant proves that their triggering conditions are unreachable. These transformations preserve the implemented map exactly and can reduce both the static and executed Toffoli counts.

Arithmetic operations can also be restricted to analytically established live bit ranges. If an intermediate nonnegative integer \(z\) satisfies
\begin{equation*}
    0\leq z<2^w
\end{equation*}
for every supported input, then
\begin{equation*}
    z_j=0
    \qquad\text{for all } j\geq w,
\end{equation*}
and operations acting exclusively on those high-order bits may be omitted. Similar range arguments can narrow comparisons and carry chains when the discarded bits cannot affect the branch decision or the retained output. Reducing a local operation's width lowers the peak logical width \(Q\) only when that operation belongs to a width-determining stage.

Not all truncations used by the submitted circuits have an all-input analytic justification. Some iteration-dependent active-width schedules and correction windows were calibrated and validated only on the benchmark's finite input test set. We therefore treat these changes as test-set-calibrated approximations rather than exact dead-code elimination. Their empirical status and associated limitations are discussed in \Cref{sec:evaluator} and \Cref{sec:limitations}.

\subsection{Evolution of the {\challengename} Point-Addition Circuit}
\label{sec:circuit-trajectory}

\begin{figure}[p]
  \centering
  \captionsetup[sub]{font=small,labelfont=bf,justification=justified,%
    singlelinecheck=false}

  \begin{subfigure}{\linewidth}
    \centering
    \resizebox{0.70\linewidth}{!}{\input{figs/qxt-error-dual-axis-plot}}
    \caption{Point-addition retry proxy. The left axis compares $Q\times T$ with
    the plug-in quantity $Q\times T/\hat p$; the latter has an expected-work
    interpretation only under an independently rerunnable classical-success
    model. The right axis reports sampled classical-output error
    $100(1-\hat p)$.}
    \label{fig:qxt-shor-comparison-pa}
  \end{subfigure}

  \vspace{0.15cm}

  \begin{subfigure}{\linewidth}
    \centering
    \resizebox{0.70\linewidth}{!}{\input{figs/shor-circuit-estimate-plot}}
    \caption{Windowed whole-circuit proxy. The raw estimate is $28Q'T'$; the
    adjusted curve divides by the plug-in union-bound proxy
    $\hat p'_{\mathrm{LB}}=1-28(1-\hat p)$. The right axis shows the
    corresponding composed-error proxy $28(1-\hat p)$.}
    \label{fig:qxt-shor-comparison-full}
  \end{subfigure}
  \caption{Resource-cost evolution in a later longitudinal sample of $42$ accepted commits (indices $1,11,\ldots,401$, plus $405$), extending through 20 July. Each commit is re-evaluated on the same fixed pseudorandom corpus of $50{,}000$ elliptic-curve point pairs; hence $\hat p$ and $T$ are sample estimates and can differ slightly from the official $9{,}024$-instance operating points elsewhere in this section. Following \eqref{eq:full-attack}, panel~(b) uses the analytic sensitivity proxy $Q'=Q+16$, $T'=T+3\cdot2^{16}$, and $28$ point additions at $n=256$ and $w=16$ (\Cref{sec:ecc-quantum}); it does not use the subsequently implemented windowed-addition-compatible circuit of \Cref{sec:windowed-addition-compatible-circuit}, which directly measures the coherent-lookup overhead for the best-scoring circuit. The error composition is illustrative: it assumes the sampled classical-output error transfers to the windowed attack setting. It incorporates neither sampling uncertainty nor the separately recorded ancilla- and phase-garbage outcomes.}
  \label{fig:qxt-shor-comparison}
\end{figure}

Over the course of the challenge, successive submissions reduced the point-addition score through a combination of major structural advances and sustained incremental refinement. In this section, we first trace the progression of the $Q \times T$ score over time (\Cref{sec:QxT-score-evolution}) and organize the major changes into three resource-model eras (\Cref{sec:product-trajectory}). We then examine the separate efforts to construct sub-$1{,}000$-qubit circuits and to explore the broader Pareto frontier (\Cref{sec:low-qubit-track}). Finally, we classify the optimization techniques applied across the successive submissions, distinguishing frequently used incremental refinements from less common structural transformations associated with major frontier advances (\Cref{sec:optimization-technique-breakdown}).

\subsubsection{Raw and Retry-Adjusted \texorpdfstring{$Q \times T$}{Q times T} Scores Across Submissions}\label{sec:QxT-score-evolution}

Because many submissions use approximate arithmetic, a reduction in raw $Q\times T$ may be accompanied by a lower probability of producing the correct output. To determine whether the observed resource improvements persist after accounting for this trade-off, we re-evaluate a longitudinal sample of $42$ accepted submissions on a common corpus of $50{,}000$ pseudorandom inputs and estimate each circuit's empirical success probability $\hat p$. Under an independently rerunnable classical-success model, obtaining one successful point addition requires $1/\hat p$ executions on average, so $Q\times T/\hat p$ estimates the expected work per successful result. Writing the empirical error rate as $\hat\epsilon=1-\hat p$, the repetition factor is $1/\hat p=1/(1-\hat\epsilon)\simeq1+\hat\epsilon$ for small $\hat\epsilon$, corresponding to approximately $100\hat\epsilon\%$ additional expected work. \Cref{fig:qxt-shor-comparison} therefore compares the raw score $Q\times T$, the retry-adjusted proxy $Q\times T/\hat p$, and the observed classical-output error rate $1-\hat p$ across the sampled submissions, ordered by their positions in the acceptance sequence.

As \Cref{fig:qxt-shor-comparison} shows, the point-addition score decreases from approximately $10.7$B to $1.5$B across the sampled trajectory, while the corresponding raw proxy for a windowed full Shor circuit decreases from roughly $320$B to $50$B. The trajectory is distinctly stepwise, with major reductions followed by extended plateaus and smaller incremental gains. At the point-addition level, the retry-adjusted score $Q \times T/\hat{p}$ generally tracks the raw $Q \times T$ score because the observed classical-output error rates remain small. At the full-circuit level, the adjusted proxy assumes 28 point additions and estimates their aggregate success probability using the union bound $\hat{p}'_{\mathrm{LB}}=1-28(1-\hat{p})$; under this model, the highlighted intervals show that some nominal resource improvements diminish or reverse after accounting for the accumulation of sampled error across the complete computation. These patterns motivate analyzing the trajectory as a succession of circuit-level resource models rather than as a smooth sequence of scalar-score reductions.

The separation between the curves in \Cref{fig:qxt-shor-comparison-full} is a sensitivity analysis, not a validated full-attack failure estimate. The plotted lower-bound proxy substitutes the empirical classical-output error for an applicable coherent per-addition error; it is neither a confidence bound nor evidence of coherent success. If it were applicable to an independently rerun computation, a $5\%$ composed error would imply a repetition factor of $1/(1-0.05)\simeq1.053$, or about $5.3\%$ additional expected work, not a fivefold overhead. The derivation and its scope conditions appear in \href{https://github.com/jieyilong/ecdsafail-supplemental-materials/blob/ef7ecd2b27cf27881874468e003e53e16ba48075/appendices/detailed_results_and_circuit_improvement_trajectory.pdf}{Supplemental Note A: Detailed results and circuit-improvement trajectory}.

Finally, note that what we analyzed above is just a sample of the submissions. The pinned 27 July API response contains $831$ rows, of which $826$ were created by the data cutoff: $412$ API-accepted, $253$ API-rejected, and $161$ API-failed entries.

The stepwise trajectory also reflects the collective organization of the search. Cross-referencing the frontier transitions with the public submission chronology shows that several plateau-ending advances came from participants other than those responsible for the preceding improvements, including both established and newly joined contributors. Once promoted, these advances became available for adoption and refinement by the wider community. The trajectory therefore records a sequence of public handoffs among partially independent search efforts, rather than the uninterrupted optimization path of a single participant or agent.

\subsubsection{Three Eras of Circuit Optimization for the \texorpdfstring{$Q \times T$}{Q times T} Score}
\label{sec:product-trajectory}

\begin{table}[tbp]
  \centering
  \small
  \setlength{\tabcolsep}{4pt}
  \begin{tabular}{@{}ll>{\raggedright\arraybackslash}p{3.2cm}>{\raggedright\arraybackslash}p{5.3cm}@{}}
    \toprule
    \textbf{Era} & \textbf{$Q$ trajectory} & \textbf{Inversion architecture} & \textbf{Principal architectural change} \\
    \midrule
    Era I & $2{,}715 \rightarrow 2{,}002$ & Two-pass Kaliski inversion & Packing, recomputation, and live-range reduction within the inherited architecture. \\
    Era II & $2{,}002 \rightarrow 1{,}168$ & Streamed dialog-GCD inversion & Replacement of the history stack by a compressed, replayed Euclidean transcript. \\
    Era III & $1{,}167 \rightarrow 1{,}151$ & Re-architected Jump-2 dialog-GCD & Off-peak addend materialization, base-$5$ coding, streamed replay, and coordinated local packing. \\
    \bottomrule
  \end{tabular}
  \vspace{5pt}
  \caption{Three architectural eras in the product-score trajectory.}
  \label{tab:trajectory-eras}
\end{table}

The promoted trajectory separates into three eras distinguished by their inversion architecture and dominant live-state constraint. Era~I retained the two-pass Kaliski inverter and reduced its peak through scratch packing, recomputation, and shorter live ranges, reaching approximately $2{,}002$ qubits without changing the underlying inversion method.

Era~II adopted the record-and-replay Euclidean architecture introduced by Khattar et al.~\cite{ksgz25} and adapted to ECDLP point addition by Schrottenloher~\cite{s26}. Replacing the persistent Kaliski history with a compressed, streamed transcript reduced the width to $1{,}168$ qubits, where the two coordinates of the classically fixed addend became a principal remaining peak owner.

Era~III began with a TrailMix-derived reimplementation~\cite{trailmix} built around a Jump-2 Euclidean recurrence. Its opening point of $1{,}167$ qubits improved the width record by a single qubit, but it removed the $1{,}168$-qubit wall: the fixed addend was now materialized only during off-peak coordinate operations. Combined with base-$5$ transcript encoding, streamed replay, specialized arithmetic, and tighter register lifetimes, this reorganization reached the $1{,}151$-qubit incumbent at the data cutoff. The trajectory therefore resembles a verified ratchet: occasional architectural changes opened new resource regimes, after which numerous local refinements consolidated the gains. The individual mechanisms are described in \Cref{sec:key-circuit-optimization-techniques}, while detailed transitions and neighboring controls appear in \href{https://github.com/jieyilong/ecdsafail-supplemental-materials/blob/ef7ecd2b27cf27881874468e003e53e16ba48075/appendices/detailed_results_and_circuit_improvement_trajectory.pdf}{Supplemental Note A}.

\subsubsection{Low-Qubit and Pareto Exploration}\label{sec:low-qubit-track} The low-qubit track targets aggressive width reduction below $1{,}000$ qubits by exchanging resident quantum state for recomputation, compact metadata, and extensive register reuse. An open low-width TrailMix route provided the principal implementation substrate \cite{trailmix}, which challenge participants further integrated and specialized to obtain admitted operating points at $Q=980$, $948$, $828$, $826$, and $825$. Along this branch, the work increase becomes abrupt, rising from $54.8$M Toffoli gates at $Q=948$ to $448.1$M at $Q=828$ and $489.2$M at $Q=825$. These endpoints demonstrate that the circuit can be compressed into a substantially lower-width regime and suggest a sharp width--work transition within the circuit families discovered so far. They do not, however, establish a fundamental hard-$Q$ boundary: the product-efficient and low-width branches use different inversion architectures, so the sparsely populated region between them may reflect this architectural discontinuity, a steep generic time--space trade-off, or the absence of a hybrid or otherwise undiscovered construction that bridges the two regimes. Within the existing extreme-width routes, source reconstruction identifies quotient metadata, hosted predicates, and denominator-witness lifetimes as immediate constraints, without implying that these constraints are unavoidable or assigning a unique mechanism to every movement of the frontier.

A separate Pareto-frontier exploration produced a sequence of circuits spanning $Q=1{,}150$ to $Q=1{,}133$ while retaining average executed Toffoli counts near $1.4$M. Representative points include $(Q,T)= (1,150,\;\,1,284,776)$, $(1,141,\;\,1,423,723)$, and $(1,133,\;\,1,460,511)$. Although these circuits do not attain the minimum $Q\times T$ score, they reduce logical width with only a modest increase in nonlinear work; for example, the $Q=1{,}133$ circuit uses approximately $12.4\%$ more Toffoli gates than the $Q=1{,}151$ incumbent while requiring $18$ fewer logical qubits. Subject to the circuit-interface and full-Shor compatibility considerations discussed elsewhere, these Pareto points may be particularly relevant to Q-day resource scenarios in which logical-qubit capacity is more restrictive than execution time. The selected low-qubit and Pareto sequences, together with the unsuccessful batching construction, are documented in \href{https://github.com/jieyilong/ecdsafail-supplemental-materials/blob/ef7ecd2b27cf27881874468e003e53e16ba48075/appendices/detailed_results_and_circuit_improvement_trajectory.pdf}{Supplemental Note A: Detailed results and circuit-improvement trajectory}.

\begin{figure}[tp]
  \centering
  \scalebox{0.90}{\begin{forest}
    for tree={
      grow=east, parent anchor=east, child anchor=west, anchor=west,
      font=\scriptsize, align=left, rounded corners=1.5pt,
      draw=arBorder, line width=0.4pt, inner sep=2.5pt,
      edge={arEdge, line width=0.4pt},
      edge path={\noexpand\path[\forestoption{edge}]
        (!u.parent anchor) -- +(5pt,0) |- (.child anchor)\forestoption{edge label};},
      l sep=9pt, s sep=2.2pt,
    }
    [{\footnotesize\textbf{400}\\ scored source\\ commits\\ \#883--1294},
       fill=arAccentBg, draw=arAccent, line width=0.7pt
      [Neutral / mechanical\\ \textbf{3} (0.8\%), fill=arWarmBg, draw=arWarm]
      [Verifier search \emph{(not a circuit opt.)}\\ nonce / island reroll --- \textbf{19} (4.8\%),
         fill=arWarmBg, draw=arWarm]
      [Algorithmic substitution\\ \textbf{2} (0.5\%) --- multiplier/squarer swap, fill=arAccentBg]
      [Codec / encoding\\ \textbf{7} (1.8\%) --- Trailmix/\texttt{K5} codec, fill=arAccentBg]
      [Windowing \textbf{8} (2.0\%), fill=arAccentBg
        [windowed carry/shift blocks --- 5]
        [K2 transcript pair-compressor --- 2]
        [two-jump inversion base --- 1]
      ]
      [Fusion \textbf{12} (3.0\%), fill=arAccentBg
        [gate / op fusion --- 7]
        [boundary \texttt{cswap}-merge --- 5]
      ]
      [Strength reduction \textbf{14} (3.5\%), fill=arAccentBg
        [fast-adder reroute --- 9]
        [Solinas / NAF fold --- 5]
      ]
      [Constant propagation\\ \textbf{19} (4.8\%) --- host-derived specialization, fill=arAccentBg]
      [Uncompute strategy\\ \textbf{22} (5.5\%) --- measurement uncompute, fill=arAccentBg]
      [Register / scratch alloc. \textbf{50} (12.5\%), fill=arAccentBg
        [ancilla-lifetime / venting --- 30]
        [scratch / ancilla reuse --- 17]
        [in-place mutation --- 3]
      ]
      [Dead-code \& redundancy\\ \textbf{244} (61.0\%), fill=arAccentBg
        [comparator / width trunc. --- 149]
        [fold / carry truncation --- 37]
        [iteration / threshold tuning --- 37]
        [dead-gate / self-inverse --- 21]
      ]
    ]
  \end{forest}}
  \caption{Anatomy of the 400 scored, accepted source commits from the public competition through 18 July (\#883--1294). Each commit is assigned to one primary optimization family (level~1) and, where applicable, one change type (level~2), based on an audit of its \texttt{git show} diff rather than its often-unreliable subject or lever name. Blue nodes represent nine circuit-optimization families: six conventional compiler categories and three reversibility-specific categories---uncompute strategy, register/scratch allocation, and \emph{codec/encoding}---with distinct resource and correctness obligations. Amber nodes represent verifier search for a clean Fiat--Shamir nonce, which is not a circuit optimization, and neutral or mechanical changes. Counts measure how often each mechanism was the primary change, not its contribution to score reduction; accompanying nonce changes are attributed to the associated circuit optimization. Pre-competition development commits are excluded, including the representation/coordinate-change family documented only in earlier research (see \href{https://github.com/jieyilong/ecdsafail-supplemental-materials/blob/ef7ecd2b27cf27881874468e003e53e16ba48075/appendices/detailed_results_and_circuit_improvement_trajectory.pdf}{Supplemental Note A: Detailed results and circuit-improvement trajectory}), because no corresponding competition submission was accepted. Four accepted but unscored \texttt{DUMMY\_TOFFOLIS} bootstrap commits are also excluded. The complete per-commit ledger, including those four commits, is provided in \href{https://github.com/jieyilong/ecdsafail-supplemental-materials/blob/ef7ecd2b27cf27881874468e003e53e16ba48075/appendices/optimization_census_of_accepted_competition_source_commits.pdf}{Supplemental Note C: List of accepted solutions}.}
  \label{fig:family-census-tree}
\end{figure}

\subsubsection{Breakdown of Optimization Techniques Across the Circuit Optimization Trajectory}
\label{sec:optimization-technique-breakdown}

Complementing the three-era chronology, \Cref{fig:family-census-tree} classifies the 400 scored accepted source commits of the public competition through 18 July by their primary optimization family and, where applicable, a more specific change type. Throughout this classification we distinguish the \emph{pre-competition development} history (commits \#1--882), in which most of the techniques were first implemented, from the \emph{public competition} that followed (accepted contestant commits \#883--1294), which the census counts; this split is by provenance in the source history and is independent of the three resource-model eras of \Cref{sec:product-trajectory}, which instead partition the promoted-submission trajectory by dominant width constraint. \href{https://github.com/jieyilong/ecdsafail-supplemental-materials/blob/ef7ecd2b27cf27881874468e003e53e16ba48075/appendices/anatomy_of_agent_identified_quantum_circuit_optimizations.pdf}{Supplemental Note B: List of agent-identified optimizations} provides detailed descriptions for the circuit optimization techniques listed in the figure. Each assignment is based on an audit of the implemented source diff rather than the commit title or configuration-lever name, which frequently provides an incomplete or misleading description of the actual circuit change. To produce a mutually exclusive census, each commit is assigned a single primary family even when it combines several mechanisms; secondary changes therefore do not contribute additional counts. The figure excludes the four accepted but unscored bootstrap commits and the earlier pre-competition development history, so it characterizes the techniques used to refine the circuit during the public competition rather than establishing where a technique originated or who first proposed it. Submission \texttt{3616dbf} and its ping-pong architecture lie outside both the 18 July census interval and the 26 July evidence cutoff. They therefore do not alter the two algorithmic-substitution commits or any other counts in \Cref{fig:family-census-tree}.

Dead-code and redundancy elimination is the most frequent family, accounting for 244 of the 400 commits ($61.0\%$). Most of these changes narrow comparators or arithmetic operations to the bits that can affect the result (149 commits), while iteration or threshold tuning and fold or carry truncation each account for 37 commits, and dead-gate or self-inverse cancellation accounts for 21. Register and scratch allocation is the second-largest family, with 50 commits ($12.5\%$), including 30 ancilla-lifetime or venting changes, 17 scratch-reuse changes, and three in-place mutations. The remaining families are less frequent but span a broader range of structural mechanisms: measurement-based uncomputation appears as the primary change in 22 commits; constant propagation in 19; strength reduction in 14; fusion in 12; windowing or iteration blocking in eight; codec and transcript encoding in seven; and algorithmic substitution in two. Within these smaller families, the census identifies concrete recurring patterns, including boundary controlled-swap merging, fast-adder rerouting, windowed carry and shift blocks, Jump-2 transcript compression, and TrailMix/K5 codec improvements.

Within the census interval, the distribution reflects how optimization proceeded after the then-current architectural components had stabilized. Most accepted commits performed localized refinement, particularly width truncation, threshold selection, and register-lifetime adjustment, whereas the larger structural transformations occurred much less frequently. Frequency, however, should not be interpreted as resource impact: the figure measures how often a technique was the primary change, not how much it reduced $Q$, $T$, or $Q\times T$. Several rare structural transformations, including Jump-2 inversion, fusion, and measurement-based uncomputation, are associated with disproportionately large movements of the frontier, while many high-frequency changes contributed smaller incremental gains. This distinction reinforces the three-era interpretation: a small number of architectural transitions created new resource regimes, after which numerous local optimizations exploited the resulting headroom.

Two residual categories are shown separately from the circuit-optimization taxonomy. Verifier search accounts for 19 commits ($4.8\%$) whose primary change is a nonce or validation-island reroll rather than a modification of the semantic circuit, while three neutral or mechanical commits ($0.8\%$) do not materially change the emitted circuit. When a commit combines a genuine circuit optimization with a nonce reroll, it is classified under the circuit transformation; verifier search is used only when no substantive circuit change is identified. The complete commit-level classifications, score joins, and attribution criteria are provided in \href{https://github.com/jieyilong/ecdsafail-supplemental-materials/blob/ef7ecd2b27cf27881874468e003e53e16ba48075/appendices/detailed_results_and_circuit_improvement_trajectory.pdf}{Supplemental Note A: Detailed results and circuit-improvement trajectory} and \href{https://github.com/jieyilong/ecdsafail-supplemental-materials/blob/ef7ecd2b27cf27881874468e003e53e16ba48075/appendices/optimization_census_of_accepted_competition_source_commits.pdf}{Supplemental Note C: List of accepted solutions}.

\subsection{Reproducibility of the Results}
\label{sec:reproducibility}

We support two complementary forms of reproducibility: replaying the reported circuit artifacts and reconstructing the frontier analysis from frozen public data. Circuit replay verifies the correctness and measured resources of a selected implementation, whereas analysis reproduction verifies the submission selection, Pareto construction, tables, and figures.

\paragraph{Circuit reproduction.} The companion manifest links each reported Pareto point to a submission identifier and an immutable source commit, so any point can be reproduced offline from the public repository, without the leaderboard service. After cloning the benchmark once,
\begin{verbatim}
git clone https://github.com/Layr-Labs/ecdsafail-challenge.git
cd ecdsafail-challenge
\end{verbatim}
each operating point is rebuilt by checking out its source commit and rerunning the pinned evaluator. For example, the best $Q\times T$ circuit \texttt{8e9c9a2} (the first promoted submission crossing the 50\% score-reduction threshold) and the post-cutoff ping-pong circuit \texttt{3616dbf} are reproduced by:
\begin{verbatim}
# best Q x T circuit (8e9c9a2, score 1,495,670,403)
git checkout 60d61859fa6965eba53634f91877b1141a6f9dce
./setup.sh && ./benchmark.sh

# post-cutoff ping-pong circuit (3616dbf, score 1,258,525,947)
git checkout 897dda2b0cf267151ecd973252d2a5078cbf1b63
./setup.sh && ./benchmark.sh
\end{verbatim}
Each checkout restores the complete historical repository state for that submission, including its circuit source, the evaluator, and the pinned toolchain; because the manifest pins every Pareto point to such a commit, each reported point has an immutable, self-contained reproduction. The evaluator attempts 9,024 Fiat--Shamir-derived draws, skips degenerate pairs, and reports the number actually tested as \texttt{n\_shots}. For each retained input, it executes the submitted operation stream once in the forward direction, checks classical outputs, phase cleanliness, and final ancilla cleanup, and reports $Q$, $T$, and $Q\times T$ in \texttt{score.json}.

\paragraph{Frozen data and frontier reconstruction.} The quantitative analyses use an official API response fetched on 27 July 2026 and a data cutoff of 26 July 2026 at 09:21:55 UTC, defined by the first promoted submission achieving a 50\% score reduction relative to the Google/Babbush low-gate reference. The frozen response contains 831 rows, including 826 created by the cutoff and 576 with structured $Q$ and $T$ metrics. It is joined with the 80-row contribution chronology used in \Cref{sec:circuit-trajectory}, which supplies the selected milestones and mechanism labels. The historical display contains 15 manifest-selected transitions and is an editorial reconstruction rather than a statistical sample.
The frozen public data, selection manifest, reconstruction code, tests, and generated outputs are available in the \href{https://github.com/jieyilong/ecdsafail-supplemental-materials/tree/d6cdcd4be0204b8df7b6d6c486ccd5702da0b7bd/frontier_reconstruction}{frontier reconstruction package}.

%% file: figs/pareto_frontier.tex
\begingroup%
\definecolor{minQ}{HTML}{D55E00}%
\definecolor{minT}{HTML}{0072B2}%
\definecolor{isoRed}{HTML}{CC79A7}%
\definecolor{ink}{HTML}{1A1A1A}%
\definecolor{wk0}{gray}{0.845}%
\definecolor{wk1}{gray}{0.751}%
\definecolor{wk2}{gray}{0.656}%
\definecolor{wk3}{gray}{0.562}%
\definecolor{wk4}{gray}{0.468}%
\definecolor{wk5}{gray}{0.374}%
\definecolor{wk6}{gray}{0.279}%
\definecolor{wk7}{gray}{0.185}%
\begin{tikzpicture}[font=\sffamily]

\begin{axis}[
  name=panelA, height=6.05cm,
  scale only axis, width=10.52cm,
  axis lines*=left,
  axis line style={black!58,line width=0.4pt},
  tick style={black!58,line width=0.35pt},
  tick align=outside, tick pos=left,
  tick label style={font=\sffamily\fontsize{6.3}{7}\selectfont,text=black!68},
  label style={font=\sffamily\fontsize{7.2}{8}\selectfont,text=black!80},
  xmajorgrids, ymajorgrids,
  grid style={black!9,line width=0.3pt},
  clip=false,
  xmode=log, ymode=log, log basis x=10, log basis y=10,
  xmin=770, xmax=3150,
  ymin=1000000, ymax=760000000,
  xtick={800,1000,1200,1500,2000,2800},
  xticklabels={800,1{,}000,1{,}200,1{,}500,2{,}000,2{,}800},
  ytick={1e6,3e6,1e7,3e7,1e8,3e8},
  yticklabels={1M,3M,10M,30M,100M,300M},
  xlabel={peak logical qubits\ \ $Q$\ \ (log)},
  ylabel={average executed Toffoli count\ \ $T$\ \ (log)},
  legend style={at={(0.995,0.995)}, anchor=north east, draw=none, fill=none,
    legend cell align=left, font=\sffamily\fontsize{6.9}{7.7}\selectfont,
    text=black!68, /tikz/every even column/.append style={column sep=2.5pt}},
]
    \addplot[forget plot,isoRed,line width=0.68pt,dash pattern=on 2.6pt off 2.2pt,mark=none] coordinates {(770,1943181.818) (788.6068639,1897333.22) (807.6633582,1852566.4) (827.180348,1808855.836) (847.1689612,1766176.605) (867.6405944,1724504.374) (888.6069196,1683815.382) (910.0798911,1644086.431) (932.0717517,1605294.868) (954.5950403,1567418.577) (977.6625987,1530435.962) (1001.287579,1494325.937) (1025.483451,1459067.914) (1050.264011,1424641.789) (1075.643387,1391027.935) (1101.63605,1358207.187) (1128.256819,1326160.831) (1155.520872,1294870.596) (1183.443754,1264318.642) (1212.041387,1234487.548) (1241.330074,1205360.308) (1271.326515,1176920.313) (1302.047813,1149151.348) (1333.511484,1122037.581) (1365.735467,1095563.553) (1398.738134,1069714.169) (1432.538303,1044474.69) (1467.155245,1019830.727) (1502.608697,995768.2285) (1516.232462,986820.9771)};

    \addplot[forget plot,minQ,line width=0.82pt,dash pattern=on 3.1pt off 1.9pt,mark=none] coordinates {(2310,3638710) (1285,1532135) (1170,1434070) (948,54781961) (825,489161900)};
    \addplot[forget plot,minT,line width=0.82pt,dash pattern=on 1.0pt off 1.3pt,mark=none] coordinates {(2708,3541837) (1302,1453867) (1285,1373922) (1159,1364380) (1152,1364230) (1152,1318724) (1152,1318717) (1158,1292651)};
    \draw[minQ,line width=0.82pt,solid,-{Stealth[length=3.3pt,width=2.5pt]}] (axis cs:1805.652868,2530313.146) -- (axis cs:1605.803838,2128355.596);
    \draw[minQ,line width=0.82pt,solid,-{Stealth[length=3.3pt,width=2.5pt]}] (axis cs:1235.383726,1490156.312) -- (axis cs:1212.434883,1470572.669);
    \draw[minQ,line width=0.82pt,solid,-{Stealth[length=3.3pt,width=2.5pt]}] (axis cs:1071.043973,6622753.406) -- (axis cs:1026.908613,13723093.72);
    \draw[minT,line width=0.82pt,solid,-{Stealth[length=3.3pt,width=2.5pt]}] (axis cs:1991.007226,2436761.005) -- (axis cs:1719.749932,2039257.92);
    \draw[minT,line width=0.82pt,solid,-{Stealth[length=3.3pt,width=2.5pt]}] (axis cs:1294.832777,1419738.588) -- (axis cs:1291.433694,1403769.707);
    \draw[minT,line width=0.82pt,solid,-{Stealth[length=3.3pt,width=2.5pt]}] (axis cs:1230.492178,1369906.259) -- (axis cs:1205.354848,1367998.128);
    \draw[minT,line width=0.82pt,solid,-{Stealth[length=3.3pt,width=2.5pt]}] (axis cs:1156.054834,1364316.998) -- (axis cs:1154.655006,1364286.998);
    \draw[minT,line width=0.82pt,solid,-{Stealth[length=3.3pt,width=2.5pt]}] (axis cs:1152,1344929.277) -- (axis cs:1152,1335834.669);
    \draw[minT,line width=0.82pt,solid,-{Stealth[length=3.3pt,width=2.5pt]}] (axis cs:1154.516204,1307705.864) -- (axis cs:1155.716328,1302494.828);
    \addplot[forget plot,only marks,mark=o,mark size=1.1pt,
      mark options={draw=minQ,fill=white,line width=0.6pt}] coordinates {(2310,3638710)};
    \addplot[forget plot,only marks,mark=o,mark size=1.1pt,
      mark options={draw=minT,fill=white,line width=0.6pt}] coordinates {(2708,3541837)};
    \addplot[forget plot,only marks,mark=o,mark size=3.4pt,
      mark options={draw=minQ,fill=none,line width=0.75pt}]
      coordinates {(825,489161900)};
    \addplot[forget plot,only marks,mark=o,mark size=3.4pt,
      mark options={draw=minT,fill=none,line width=0.75pt}]
      coordinates {(1158,1292651)};
    \node[anchor=south,text=minQ,font=\sffamily\fontsize{7.0}{7.8}\selectfont]
      at (axis cs:1075,3.65e8) {$Q=825$};
    \draw[minQ,line width=0.35pt]
      (axis cs:985,4.45e8) -- (axis cs:864,4.78e8);

    \addplot[forget plot,ink,line width=1.45pt,mark=*,mark size=2.3pt,mark options={fill=ink,draw=white,line width=0.28pt}] coordinates {(1158,1292651) (1151,1299453) (1133,1460511) (980,28847321) (948,54781961) (828,448102916) (826,479337428) (825,489161900)};

    \addplot[forget plot,only marks,mark=diamond*,mark size=2.2pt,
      mark options={fill=black!78,draw=white,line width=0.5pt}]
      coordinates {(1425,2100000)};

    \node[anchor=west,text=minT,font=\sffamily\bfseries\fontsize{7.8}{8.6}\selectfont]
      at (axis cs:1740,1.48e6) {min Toffolis};
    \node[anchor=west,text=minQ,font=\sffamily\bfseries\fontsize{7.8}{8.6}\selectfont]
      at (axis cs:2180,4.15e6) {min qubits};
    \node[anchor=south west,text=black!62,font=\sffamily\fontsize{6.9}{7.7}\selectfont]
      at (axis cs:1260,2.23e6) {Google/Babbush};
    \node[anchor=west,text=ink,font=\sffamily\bfseries\fontsize{8.0}{8.8}\selectfont]
      at (axis cs:1240,1.2e7) {frontier at the 26 Jul cutoff};
    \node[anchor=west,text=black!58,font=\sffamily\fontsize{6.8}{7.6}\selectfont]
      at (axis cs:1165,1.92e6) {(b)};
    \node[anchor=south west,text=ink,font=\sffamily\bfseries\fontsize{8.4}{9.2}\selectfont]
      at (rel axis cs:0.004,1.012) {(a)};

    \draw[black!48,line width=0.35pt]
      (axis cs:1116,1268000) rectangle (axis cs:1322,1560000);

    \addlegendimage{isoRed,line width=0.68pt,dash pattern=on 2.6pt off 2.2pt}
    \addlegendentry{half the Google/Babbush $Q \times T$}
\end{axis}

\begin{axis}[
  height=2.72cm,
  scale only axis, width=10.52cm,
  axis lines*=left,
  axis line style={black!58,line width=0.4pt},
  tick style={black!58,line width=0.35pt},
  tick align=outside, tick pos=left,
  tick label style={font=\sffamily\fontsize{6.3}{7}\selectfont,text=black!68},
  label style={font=\sffamily\fontsize{7.2}{8}\selectfont,text=black!80},
  xmajorgrids, ymajorgrids,
  grid style={black!9,line width=0.3pt},
  clip=false,
  at={(panelA.below south west)}, anchor=north west, yshift=-0.95cm,
  xmin=1116, xmax=1322,
  ymin=1.268, ymax=1.56,
  xtick={1150,1200,1250,1300},
  xticklabels={1{,}150,1{,}200,1{,}250,1{,}300},
  ytick={1.30,1.35,1.40,1.45,1.50,1.55},
  yticklabels={1.30M,1.35M,1.40M,1.45M,1.50M,1.55M},
  xlabel={peak logical qubits\ \ $Q$\ \ (linear)},
  ylabel={average executed Toffoli count\ \ $T$},
]
    \addplot[forget plot,isoRed,line width=0.68pt,dash pattern=on 2.6pt off 2.2pt,mark=none] coordinates {(1116,1.340725806) (1119.208751,1.33688197) (1122.426728,1.333049153) (1125.653958,1.329227325) (1128.890467,1.325416455) (1132.136281,1.32161651) (1135.391428,1.317827459) (1138.655934,1.314049271) (1141.929826,1.310281916) (1145.213131,1.306525361) (1148.505877,1.302779577) (1151.80809,1.299044531) (1155.119798,1.295320194) (1158.441027,1.291606534) (1161.771806,1.287903522) (1165.112161,1.284211125) (1168.462121,1.280529315) (1171.821713,1.276858061) (1175.190964,1.273197332) (1178.569903,1.269547098) (1180.553972,1.267416)};

    \addplot[forget plot,wk1,line width=0.68pt,mark=none] coordinates {(1302,1.453867) (1297,1.460157) (1285,1.532135)};
    \addplot[forget plot,wk1,mark=*,mark size=0.75pt,mark options={fill=wk1,draw=none},only marks] coordinates {(1302,1.453867) (1297,1.460157) (1285,1.532135)};
    \addplot[forget plot,wk2,line width=0.68pt,mark=none] coordinates {(1285,1.373922) (1220,1.401609) (1218,1.402192) (1215,1.403115) (1211,1.403512) (1203,1.410971) (1192,1.411983) (1185,1.418587) (1170,1.43407)};
    \addplot[forget plot,wk2,mark=*,mark size=0.75pt,mark options={fill=wk2,draw=none},only marks] coordinates {(1285,1.373922) (1220,1.401609) (1218,1.402192) (1215,1.403115) (1211,1.403512) (1203,1.410971) (1192,1.411983) (1185,1.418587) (1170,1.43407)};
    \addplot[forget plot,wk3,line width=0.68pt,mark=none] coordinates {(1159,1.36438) (1157,1.38089) (1156,1.381234) (1154.850743,1.560584)};
    \addplot[forget plot,wk3,mark=*,mark size=0.75pt,mark options={fill=wk3,draw=none},only marks] coordinates {(1159,1.36438) (1157,1.38089) (1156,1.381234)};
    \addplot[forget plot,wk4,line width=0.68pt,mark=none] coordinates {(1152,1.36423) (1133,1.460511) (1132.440929,1.560584)};
    \addplot[forget plot,wk4,mark=*,mark size=0.75pt,mark options={fill=wk4,draw=none},only marks] coordinates {(1152,1.36423) (1133,1.460511)};
    \addplot[forget plot,wk5,line width=0.68pt,mark=none] coordinates {(1152,1.36423) (1133,1.460511) (1132.440929,1.560584)};
    \addplot[forget plot,wk5,mark=*,mark size=0.75pt,mark options={fill=wk5,draw=none},only marks] coordinates {(1152,1.36423) (1133,1.460511)};
    \addplot[forget plot,wk6,line width=0.68pt,mark=none] coordinates {(1152,1.318724) (1133,1.460511) (1132.440929,1.560584)};
    \addplot[forget plot,wk6,mark=*,mark size=0.75pt,mark options={fill=wk6,draw=none},only marks] coordinates {(1152,1.318724) (1133,1.460511)};
    \addplot[forget plot,wk7,line width=0.68pt,mark=none] coordinates {(1152,1.318717) (1133,1.460511) (1132.440929,1.560584)};
    \addplot[forget plot,wk7,mark=*,mark size=0.75pt,mark options={fill=wk7,draw=none},only marks] coordinates {(1152,1.318717) (1133,1.460511)};

    \addplot[forget plot,minT,line width=0.82pt,dash pattern=on 1.0pt off 1.3pt,mark=o,mark size=1.05pt,mark options={fill=white,draw=minT,line width=0.6pt}] coordinates {(1302,1.453867) (1285,1.373922) (1159,1.36438) (1152,1.36423) (1152,1.318724) (1152,1.318717) (1158,1.292651)};
    \draw[minT,line width=0.82pt,solid,-{Stealth[length=3.3pt,width=2.5pt]}] (axis cs:1294.86,1.4202901) -- (axis cs:1291.46,1.4043011);
    \draw[minT,line width=0.82pt,solid,-{Stealth[length=3.3pt,width=2.5pt]}] (axis cs:1232.08,1.36991436) -- (axis cs:1206.88,1.36800596);
    \draw[minT,line width=0.82pt,solid,-{Stealth[length=3.3pt,width=2.5pt]}] (axis cs:1156.06,1.364317) -- (axis cs:1154.66,1.364287);
    \draw[minT,line width=0.82pt,solid,-{Stealth[length=3.3pt,width=2.5pt]}] (axis cs:1152,1.34511748) -- (axis cs:1152,1.33601628);
    \draw[minT,line width=0.82pt,solid,-{Stealth[length=3.3pt,width=2.5pt]}] (axis cs:1154.52,1.30776928) -- (axis cs:1155.72,1.30255608);

    \addplot[forget plot,ink,line width=1.45pt,mark=none] coordinates {(1158,1.292651) (1151,1.299453) (1133,1.460511) (1132.440929,1.560584)};
    \addplot[forget plot,ink,mark=*,mark size=2.3pt,mark options={fill=ink,draw=white,line width=0.28pt},only marks] coordinates {(1158,1.292651) (1151,1.299453) (1133,1.460511)};

    \node[anchor=south east,text=black!62,font=\sffamily\fontsize{6.9}{7.7}\selectfont]
      at (axis cs:1302,1.4539) {7 Jun};
    \node[anchor=south east,text=black!62,font=\sffamily\fontsize{6.9}{7.7}\selectfont]
      at (axis cs:1285,1.3739) {14 Jun};
    \node[anchor=south west,text=black!62,font=\sffamily\fontsize{6.9}{7.7}\selectfont]
      at (axis cs:1160,1.391) {21 Jun};
    \node[anchor=north east,text=ink,font=\sffamily\bfseries\fontsize{6.9}{7.7}\selectfont]
      at (axis cs:1158,1.2927) {26 Jul};
    \node[anchor=west,text=minT,font=\sffamily\bfseries\fontsize{6.9}{7.7}\selectfont]
      at (axis cs:1170,1.337) {min Toffolis};
    \node[anchor=south west,text=ink,font=\sffamily\bfseries\fontsize{8.4}{9.2}\selectfont]
      at (rel axis cs:0.004,1.012) {(b)};

    \node[anchor=east,text=black!68,font=\sffamily\fontsize{6.8}{7.6}\selectfont]
      at (rel axis cs:0.610,0.135) {frontier each Sunday};
    \node[anchor=east,text=black!55,font=\sffamily\fontsize{6.6}{7.4}\selectfont]
      at (rel axis cs:0.695,0.135) {7 Jun};
    \fill[wk1] (rel axis cs:0.7050,0.100) rectangle (rel axis cs:0.7200,0.170);
    \fill[wk2] (rel axis cs:0.7200,0.100) rectangle (rel axis cs:0.7350,0.170);
    \fill[wk3] (rel axis cs:0.7350,0.100) rectangle (rel axis cs:0.7500,0.170);
    \fill[wk4] (rel axis cs:0.7500,0.100) rectangle (rel axis cs:0.7650,0.170);
    \fill[wk5] (rel axis cs:0.7650,0.100) rectangle (rel axis cs:0.7800,0.170);
    \fill[wk6] (rel axis cs:0.7800,0.100) rectangle (rel axis cs:0.7950,0.170);
    \fill[wk7] (rel axis cs:0.7950,0.100) rectangle (rel axis cs:0.8100,0.170);
    \fill[ink] (rel axis cs:0.8100,0.100) rectangle (rel axis cs:0.8250,0.170);
    \node[anchor=west,text=black!75,font=\sffamily\fontsize{6.6}{7.4}\selectfont]
      at (rel axis cs:0.840,0.135) {26 Jul};
\end{axis}
\end{tikzpicture}
\endgroup

%% file: figs/best-qxt-circuit-8e9c9a2-pipeline.tex
\begin{figure}[!t]
  \centering
  \makebox[\linewidth][c]{%
    \resizebox{1.08\linewidth}{!}{%
      \begin{tikzpicture}[
        x=1cm,
        y=1cm,
        phasebox/.style={draw=black, line width=0.8pt, fill=white,
          rounded corners=1pt, minimum width=72mm, minimum height=31mm,
          text width=66mm, inner sep=4pt, align=center, font=\large},
        phaseviolet/.style={phasebox},
        phaseorange/.style={phasebox},
        qrombox/.style={draw=black, line width=0.8pt, fill=orange!12,
          rounded corners=1pt, minimum width=38mm, minimum height=13mm,
          inner sep=3pt, align=center, font=\large},
        wire/.style={line width=1pt},
        statebox/.style={draw=black!55, line width=0.7pt, fill=white,
          rounded corners=1pt, inner sep=5pt, align=center, font=\large},
      ]
        \path[use as bounding box] (-1.5,0) rectangle (27.4,16.0);

        \node[font=\Large, anchor=west] at (0.25,15.45)
          {Input: $R=(x_R,y_R)$, selected addend $A=(x_A,y_A)$};
        \node[qrombox] (qAin) at (4.2,13.65)
          {load/unload $A$\\$A=(x_A,y_A)$};
        \node[qrombox] (q3x) at (21.6,13.65)
          {load/unload $3x_A$};

        \node[phasebox] (s1) at (4.2,10.65)
          {\textbf{1. Coordinate differences}\\[-0.5mm]
           \textit{\Cref{alg:point-addition} line 1}\\[1mm]
           $X\gets x_R-x_A=d_x$\\
           $Y\gets y_R-y_A=d_y$};
        \node[phaseviolet] (s2) at (12.9,10.65)
          {\textbf{2. Dialog-GCD division}\\[-0.5mm]
           \textit{\Cref{alg:point-addition} line 2}\\[1mm]
           $Y\gets d_y d_x^{-1}=\lambda\pmod p$\\
           $X=d_x$; clean GCD workspace};
        \node[phasebox] (s3) at (21.6,10.65)
          {\textbf{3. Prepare the $X$ workspace}\\[-0.5mm]
           \textit{\Cref{alg:point-addition} line 3}\\[1mm]
           $X\gets d_x+3x_A=x_R+2x_A$\\
           $Y=\lambda$; then uncompute $3x_A$};
        \draw[->, dashed, line width=0.8pt] (qAin.south) -- (s1.north);
        \draw[->, dashed, line width=0.8pt] (q3x.south) -- (s3.north);

        \node[phaseorange] (s4) at (4.2,5.55)
          {\textbf{4. Specialized modular square}\\[-0.5mm]
           \textit{\Cref{alg:point-addition} line 4}\\[1mm]
           $X\gets X-\lambda^2=x_A-x_{R'}$\\
           $Y=\lambda$};
        \node[phaseviolet] (s5) at (12.9,5.55)
          {\textbf{5. Forward multiplication}\\[-0.5mm]
           \textit{\Cref{alg:point-addition} line 5}\\[1mm]
           $Y\gets\lambda X=\lambda(x_A-x_{R'})$\\
           $X=x_A-x_{R'}$; erase transcript};
        \node[phasebox] (s6) at (21.6,5.55)
          {\textbf{6. Recover the output}\\[-0.5mm]
           \textit{\Cref{alg:point-addition} lines 6--7}\\[1mm]
           $Y\gets Y-y_A=y_{R'}$\\
           $X\gets x_A-X=x_{R'}$\\[1mm]
           Output: $(X,Y)=(x_{R'},y_{R'})$};
        \begin{scope}[on background layer]
          \draw[wire]
            (-0.05,11.15) -- (26.00,11.15) -- (26.00,8.55)
            -- (-0.05,8.55) -- (-0.05,6.05) -- (26.15,6.05);
          \draw[wire]
            (-0.05,10.15) -- (26.40,10.15) -- (26.40,8.10)
            -- (-0.45,8.10) -- (-0.45,5.05) -- (26.15,5.05);
        \end{scope}

        \node[font=\Large, anchor=east] at (-0.15,11.15)
          {$X:\ket{x_R}$};
        \node[font=\Large, anchor=east] at (-0.15,10.15)
          {$Y:\ket{y_R}$};
        \node[font=\Large, anchor=west] at (26.25,6.05)
          {$X=\ket{x_{R'}}$};
        \node[font=\Large, anchor=west] at (26.25,5.05)
          {$Y=\ket{y_{R'}}$};

        \node[qrombox] (qAout) at (21.6,2.75)
          {load/unload $A$\\recover $R'$ and return $A$ to zero};
        \draw[->, dashed, line width=0.8pt] (qAout.north) -- (s6.south);

      \end{tikzpicture}%
    }%
  }
  \caption{High-level architecture of the best $Q\times T$-scoring
  \texttt{8e9c9a2} mixed point-addition circuit. The six phases implement the
  in-place map $R\mapsto R'=R+A$ while recycling the $X$ and $Y$ coordinate
  registers. Under the challenge evaluator, the circuit uses $Q=1{,}151$
  logical qubits and $T=1{,}299{,}453$ average executed Toffoli gates. The
  lookup/unlookup boxes indicate where a fully windowed Shor construction
  would temporarily materialize the selected addend; in the challenge
  benchmark, $(x_A,y_A)$ are provided as classical inputs.}
  \label{fig:best-qxt-circuit-pipeline}
\end{figure}

%% file: figs/best-low-q-circuit-b6f2b0a-pipeline.tex
\begin{figure}[!t]
  \centering
  \makebox[\linewidth][c]{%
    \resizebox{1.08\linewidth}{!}{%
      \begin{tikzpicture}[
        x=1cm,
        y=1cm,
        wire/.style={draw=black, line width=0.8pt},
        stage/.style={
          draw=black,
          line width=0.8pt,
          fill=white,
          rounded corners=1pt,
          minimum width=73mm,
          minimum height=31mm,
          text width=67mm,
          inner sep=4pt,
          align=center,
          font=\large
        },
        inversion/.style={stage, fill=black!3},
        access/.style={
          draw=black,
          line width=0.7pt,
          fill=orange!8,
          rounded corners=1pt,
          minimum width=35mm,
          minimum height=11mm,
          inner sep=3pt,
          align=center,
          font=\large
        },
        summary/.style={
          draw=black!55,
          line width=0.7pt,
          fill=white,
          rounded corners=1pt,
          text width=232mm,
          minimum height=14mm,
          inner sep=5pt,
          align=center,
          font=\large
        },
        accessarrow/.style={
          ->,
          dashed,
          draw=black!65,
          line width=0.7pt
        }
      ]
        \path[use as bounding box] (-1.4,0) rectangle (26.7,16.1);

        \node[font=\Large, anchor=west] at (0.1,15.55)
          {Input: $R=(x_R,y_R)$ and classically specified addend
           $A=(x_A,y_A)$};

        \draw[wire] (-0.15,11.05) -- (25.25,11.05);
        \draw[wire] (-0.15,10.15) -- (25.25,10.15);

        \node[font=\Large, anchor=east] at (-0.30,11.05)
          {$X:\ket{x_R}$};
        \node[font=\Large, anchor=east] at (-0.30,10.15)
          {$Y:\ket{y_R}$};

        \node[stage] (s1) at (4.05,10.60)
          {\textbf{1. Coordinate differences}\\[1mm]
           $X\gets x_A-x_R=\widetilde d_x=-d_x$\\
           $Y\gets y_A-y_R=\widetilde d_y=-d_y$\\[1mm]
           Form the slope numerator and denominator};

        \node[inversion] (s2) at (12.75,10.60)
          {\textbf{2. Register-shared division}\\[1mm]
           $\lambda\gets\widetilde d_y\widetilde d_x^{-1}$\\
           $=d_y d_x^{-1}\pmod p$\\
           Restore $\widetilde d_x$ and $\widetilde d_y$\\[1mm]
           \textbf{Peak: $825$ qubits}};

        \node[stage] (s3) at (21.45,10.60)
          {\textbf{3. Erase numerator and square}\\[1mm]
           $Y\gets Y-\lambda X=0$\\
           $X\gets\lambda^2-x_R-x_A=x_{R'}$\\[1mm]
           Reuse $Y$ as arithmetic workspace};

        \node[access] (a1) at (4.05,13.65)
          {transient access to $A$\\$(x_A,y_A)$};
        \node[access] (a3) at (21.45,13.65)
          {transient access to $x_A$};
        \draw[accessarrow] (a1.south) -- (s1.north);
        \draw[accessarrow] (a3.south) -- (s3.north);

        \draw[wire]
          (25.10,11.05) -- (25.65,11.05) --
          (25.65,7.35) -- (-0.25,7.35) --
          (-0.25,5.40) -- (0.40,5.40);

        \draw[wire]
          (25.10,10.15) -- (25.95,10.15) --
          (25.95,6.90) -- (-0.55,6.90) --
          (-0.55,4.50) -- (0.40,4.50);

        \draw[wire] (0.40,5.40) -- (25.25,5.40);
        \draw[wire] (0.40,4.50) -- (25.25,4.50);

        \node[stage] (s4) at (4.05,4.95)
          {\textbf{4. Construct alternate differences}\\[1mm]
           $X\gets d_x'=x_A-x_{R'}$\\
           $Y\gets d_y'=\lambda d_x'$\\[1mm]
           $d_y'=y_{R'}+y_A$};

        \node[inversion] (s5) at (12.75,4.95)
          {\textbf{5. Cancel the slope}\\[1mm]
           $\lambda=d_y'(d_x')^{-1}$\\
           Recompute the output-side witness and erase $\lambda$\\[1mm]
           \textbf{Peak: $825$ qubits}};

        \node[stage] (s6) at (21.45,4.95)
          {\textbf{6. Recover the output}\\[1mm]
           $X\gets x_A-d_x'=x_{R'}$\\
           $Y\gets d_y'-y_A=y_{R'}$\\[1mm]
           Clear temporary and padding lanes};

        \node[access] (a4) at (4.05,2.25)
          {transient access to $x_A$};
        \node[access] (a6) at (21.45,2.25)
          {transient access to $A$\\$(x_A,y_A)$};
        \draw[accessarrow] (a4.north) -- (s4.south);
        \draw[accessarrow] (a6.north) -- (s6.south);

        \node[font=\Large, anchor=west] at (25.35,5.40)
          {$X:\ket{x_{R'}}$};
        \node[font=\Large, anchor=west] at (25.35,4.50)
          {$Y:\ket{y_{R'}}$};

        \node[summary] at (12.75,0.65)
          {Stages~2 and~5 use the register-shared shrunken-PZ extended
           Euclidean architecture. The first traversal computes the slope;
           the second uses the independently constructed output-side witness
           $\lambda=d_y'(d_x')^{-1}$ to erase it. No dialog transcript is retained.};
      \end{tikzpicture}%
    }%
  }

  \caption{High-level architecture of the low-$Q$ \texttt{b6f2b0a}
  mixed point-addition circuit. The paper-wide differences are
  $d_x=x_R-x_A$ and $d_y=y_R-y_A$, whereas the implementation stores
  their negations $\widetilde d_x=-d_x$ and $\widetilde d_y=-d_y$;
  their ratio therefore yields the same slope. The six stages implement
  $R\mapsto R'=R+A$ while recycling the $X$ and $Y$ coordinate registers.
  Register-shared extended-Euclidean computations in Stages~2 and~5 determine
  the peak width. Under the challenge evaluator, the circuit uses $Q=825$
  logical qubits and $T=489{,}161{,}900$ average executed Toffoli gates.
  The dashed arrows indicate transient access to the classically specified
  addend; integration with windowed Shor requires a separately implemented and
  validated QROM lookup/use/unlookup interface.}
  \label{fig:best-low-q-circuit-pipeline}
\end{figure}

%% file: figs/qxt-error-dual-axis-plot.tex
\definecolor{scoreblue}{HTML}{0072B2}
\definecolor{adjustedorange}{HTML}{D55E00}
\definecolor{penaltyorange}{HTML}{E8A15B}
\definecolor{gridgray}{HTML}{D9D9D9}
\pgfplotstableread[col sep=comma]{figs/qxt-history.csv}\historydata
\begin{tikzpicture}
  \begin{axis}[
    name=scoreaxis,
    scale only axis,
    width=13.5cm,
    height=7.8cm,
    xmin=0,
    xmax=410,
    ymin=1.35,
    ymax=11.25,
    xtick={1,51,101,151,201,251,301,351,405},
    xlabel={Accepted submission index},
    ylabel={Score (B)},
    title={Point-addition cost and sampled classical error},
    title style={font=\large},
    tick label style={font=\small},
    label style={font=\small},
    axis y line*=left,
    axis x line*=bottom,
    grid=both,
    major grid style={draw=gridgray, line width=0.35pt},
    minor grid style={draw=gridgray!55, line width=0.2pt},
    minor x tick num=4,
    minor y tick num=1,
    axis line style={draw=black!70},
    legend style={
      at={(0.5,-0.18)},
      anchor=north,
      legend columns=3,
      draw=black!25,
      fill=white,
      fill opacity=0.94,
      text opacity=1,
      font=\small,
      cells={anchor=west},
      /tikz/every even column/.append style={column sep=0.65cm}
    },
  ]
    \addplot+[
      scoreblue,
      line width=1.25pt,
      mark=*,
      mark size=1.55pt,
      mark options={solid, fill=white, line width=0.75pt},
    ] table[
      x=accepted_submission_index,
      y expr={\thisrow{qxt}/1000000000},
    ] {\historydata};
    \addlegendentry{$Q\!\times\!T$}

    \addplot+[
      adjustedorange,
      dashed,
      dash pattern=on 4pt off 2.2pt,
      line width=1.15pt,
      mark=triangle*,
      mark size=1.65pt,
      mark options={solid, fill=adjustedorange, line width=0.6pt},
    ] table[
      x=accepted_submission_index,
      y expr={\thisrow{qxt_over_p}/1000000000},
    ] {\historydata};
    \addlegendentry{$Q\!\times\!T/\hat p$}

    \addlegendimage{
      penaltyorange,
      opacity=0.58,
      densely dotted,
      line width=1.15pt,
      mark=diamond*,
      mark size=1.55pt,
      mark options={solid, fill=penaltyorange}
    }
    \addlegendentry{Sampled error}
  \end{axis}

  \begin{axis}[
    at={(scoreaxis.south west)},
    anchor=south west,
    scale only axis,
    width=13.5cm,
    height=7.8cm,
    xmin=0,
    xmax=410,
    ymin=0,
    ymax=0.18,
    ytick={0,0.05,0.10,0.15},
    yticklabels={0,0.05,0.10,0.15},
    ylabel={Sampled classical error (\%)},
    tick label style={font=\small, text=penaltyorange!80!black},
    label style={font=\small, text=penaltyorange!80!black},
    axis y line*=right,
    axis x line=none,
    axis line style={draw=penaltyorange!75!black},
    xtick=\empty,
  ]
    \addplot+[
      penaltyorange,
      opacity=0.58,
      densely dotted,
      line width=1.15pt,
      mark=diamond*,
      mark size=1.55pt,
      mark options={solid, fill=penaltyorange, line width=0.5pt},
    ] table[
      x=accepted_submission_index,
      y expr={100*(1-\thisrow{correctness_probability})},
    ] {\historydata};
  \end{axis}
\end{tikzpicture}

%% file: figs/shor-circuit-estimate-plot.tex
\definecolor{uncorrectedblue}{HTML}{0072B2}
\definecolor{correctedorange}{HTML}{D55E00}
\definecolor{errororange}{HTML}{E8A15B}
\definecolor{gridgray}{HTML}{D9D9D9}
\pgfplotstableread[col sep=comma]{figs/qxt-history.csv}\historydata
\begin{tikzpicture}
  \begin{axis}[
    name=scoreaxis,
    scale only axis,
    width=13.5cm,
    height=7.8cm,
    xmin=0,
    xmax=410,
    ymin=40,
    ymax=330,
    xtick={1,51,101,151,201,251,301,351,405},
    ytick={50,100,150,200,250,300},
    xlabel={Accepted submission index},
    ylabel={Estimated circuit score (B)},
    title={Windowed whole-circuit cost and composed-error proxies},
    title style={font=\large},
    tick label style={font=\small},
    label style={font=\small},
    axis y line*=left,
    axis x line*=bottom,
    grid=both,
    major grid style={draw=gridgray, line width=0.35pt},
    minor grid style={draw=gridgray!55, line width=0.2pt},
    minor x tick num=4,
    minor y tick num=1,
    axis line style={draw=black!70},
    legend style={
      at={(0.5,-0.18)},
      anchor=north,
      legend columns=3,
      draw=black!25,
      fill=white,
      fill opacity=0.94,
      text opacity=1,
      font=\small,
      cells={anchor=west},
      /tikz/every even column/.append style={column sep=0.55cm}
    },
  ]
    \addplot+[
      uncorrectedblue,
      line width=1.25pt,
      mark=*,
      mark size=1.55pt,
      mark options={solid, fill=white, line width=0.75pt},
    ] table[
      x=accepted_submission_index,
      y expr={
        (\thisrow{peak_qubits}+16)
        *(\thisrow{avg_executed_toffoli}+3*2^16)
        *28/1000000000
      },
    ] {\historydata};
    \addlegendentry{$28\times Q'\times T'$}

    \addplot+[
      correctedorange,
      dashed,
      dash pattern=on 4pt off 2.2pt,
      line width=1.15pt,
      mark=triangle*,
      mark size=1.65pt,
      mark options={solid, fill=correctedorange, line width=0.6pt},
    ] table[
      x=accepted_submission_index,
      y expr={
        (\thisrow{peak_qubits}+16)
        *(\thisrow{avg_executed_toffoli}+3*2^16)
        *28/1000000000
        /(1-28*(1-\thisrow{correctness_probability}))
      },
    ] {\historydata};
    \addlegendentry{$28\times Q'\times T'/\hat p'_{\rm LB}$}

    \addlegendimage{
      errororange,
      opacity=0.58,
      densely dotted,
      line width=1.15pt,
      mark=diamond*,
      mark size=1.55pt,
      mark options={solid, fill=errororange}
    }
    \addlegendentry{Composed-error proxy}

    \coordinate (zoomsourceA) at (axis cs:226,65.4);
    \coordinate (zoomsourceB) at (axis cs:276,62.1);
    \coordinate (zoomsourceC) at (axis cs:396,52.2);
  \end{axis}

  \begin{axis}[
    at={(scoreaxis.south west)},
    anchor=south west,
    scale only axis,
    width=13.5cm,
    height=7.8cm,
    xmin=0,
    xmax=410,
    ymin=0,
    ymax=5,
    ytick={0,1,2,3,4,5},
    ylabel={Proxy $28(1-\hat p)$ (\%)},
    tick label style={font=\small, text=errororange!80!black},
    label style={font=\small, text=errororange!80!black},
    axis y line*=right,
    axis x line=none,
    axis line style={draw=errororange!75!black},
    xtick=\empty,
  ]
    \addplot+[
      errororange,
      opacity=0.58,
      densely dotted,
      line width=1.15pt,
      mark=diamond*,
      mark size=1.55pt,
      mark options={solid, fill=errororange, line width=0.5pt},
    ] table[
      x=accepted_submission_index,
      y expr={100*28*(1-\thisrow{correctness_probability})},
    ] {\historydata};
  \end{axis}

  \coordinate (zoomcenterA) at ($(scoreaxis.south west)+(4.00cm,6.35cm)$);

  \draw[
    black!35,
    line width=0.55pt,
  ] ($(zoomsourceA)+(-0.23cm,0.12cm)$)
    -- ($(zoomcenterA)+(0.65cm,-1.12cm)$);
  \draw[
    black!35,
    line width=0.55pt,
  ] ($(zoomsourceA)+(0.23cm,0.12cm)$)
    -- ($(zoomcenterA)+(1.04cm,-0.78cm)$);
  \draw[
    fill=white,
    fill opacity=0.40,
    draw=black!55,
    line width=0.75pt,
  ] (zoomsourceA) ellipse[x radius=0.26cm,y radius=0.17cm];

  \fill[black,opacity=0.13]
    ($(zoomcenterA)+(0.06cm,-0.06cm)$) circle[radius=1.30cm];
  \fill[white] (zoomcenterA) circle[radius=1.30cm];

  \begin{scope}
    \clip (zoomcenterA) circle[radius=1.27cm];
    \begin{axis}[
      at={($(zoomcenterA)+(0.06cm,0)$)},
      anchor=center,
      scale only axis,
      width=1.72cm,
      height=1.50cm,
      xmin=217,
      xmax=235,
      ymin=64.2,
      ymax=66.55,
      xtick={221,231},
      ytick={64.5,65.5,66.5},
      xticklabel style={font=\tiny\bfseries},
      yticklabel style={
        font=\tiny,
        xshift=8pt,
        fill=white,
        inner sep=0.5pt,
      },
      axis line style={draw=black!45,line width=0.4pt},
      grid=both,
      major grid style={draw=gridgray,line width=0.3pt},
      minor grid style={draw=gridgray!60,line width=0.2pt},
      minor x tick num=1,
      minor y tick num=1,
      axis background/.style={fill=white},
      clip mode=individual,
    ]
      \addplot+[
        uncorrectedblue,
        line width=1.15pt,
        mark=*,
        mark size=1.75pt,
        mark options={solid,fill=white,line width=0.85pt},
        restrict x to domain=217:235,
      ] table[
        x=accepted_submission_index,
        y expr={
          (\thisrow{peak_qubits}+16)
          *(\thisrow{avg_executed_toffoli}+3*2^16)
          *28/1000000000
        },
      ] {\historydata};

      \addplot+[
        correctedorange,
        dashed,
        dash pattern=on 4pt off 2.2pt,
        line width=1.15pt,
        mark=triangle*,
        mark size=1.90pt,
        mark options={solid,fill=correctedorange,line width=0.65pt},
        restrict x to domain=217:235,
      ] table[
        x=accepted_submission_index,
        y expr={
          (\thisrow{peak_qubits}+16)
          *(\thisrow{avg_executed_toffoli}+3*2^16)
          *28/1000000000
          /(1-28*(1-\thisrow{correctness_probability}))
        },
      ] {\historydata};

      \node[
        font=\tiny\bfseries,
        text=uncorrectedblue,
        fill=white,
        inner sep=1.2pt,
      ] at (axis cs:226,65.15) {$-0.507$ B};

      \node[
        font=\tiny\bfseries,
        text=correctedorange,
        fill=white,
        inner sep=1.2pt,
      ] at (axis cs:226,65.68) {$+0.427$ B};
    \end{axis}
  \end{scope}

  \draw[black!65,line width=0.9pt]
    (zoomcenterA) circle[radius=1.30cm];

  \coordinate (zoomcenterB) at ($(scoreaxis.south west)+(7.05cm,6.35cm)$);
  \draw[black!35,line width=0.55pt]
    ($(zoomsourceB)+(-0.23cm,0.12cm)$)
    -- ($(zoomcenterB)+(0.65cm,-1.12cm)$);
  \draw[black!35,line width=0.55pt]
    ($(zoomsourceB)+(0.23cm,0.12cm)$)
    -- ($(zoomcenterB)+(1.04cm,-0.78cm)$);
  \draw[
    fill=white,
    fill opacity=0.40,
    draw=black!55,
    line width=0.75pt,
  ] (zoomsourceB) ellipse[x radius=0.26cm,y radius=0.17cm];

  \fill[black,opacity=0.13]
    ($(zoomcenterB)+(0.06cm,-0.06cm)$) circle[radius=1.30cm];
  \fill[white] (zoomcenterB) circle[radius=1.30cm];

  \begin{scope}
    \clip (zoomcenterB) circle[radius=1.27cm];
    \begin{axis}[
      at={($(zoomcenterB)+(0.06cm,0)$)},
      anchor=center,
      scale only axis,
      width=1.72cm,
      height=1.50cm,
      xmin=267,
      xmax=285,
      ymin=60.55,
      ymax=63.65,
      xtick={271,281},
      ytick={61,62,63},
      xticklabel style={font=\tiny\bfseries},
      yticklabel style={
        font=\tiny,
        xshift=8pt,
        fill=white,
        inner sep=0.5pt,
      },
      axis line style={draw=black!45,line width=0.4pt},
      grid=both,
      major grid style={draw=gridgray,line width=0.3pt},
      minor grid style={draw=gridgray!60,line width=0.2pt},
      minor x tick num=1,
      minor y tick num=1,
      axis background/.style={fill=white},
      clip mode=individual,
    ]
      \addplot+[
        uncorrectedblue,
        line width=1.15pt,
        mark=*,
        mark size=1.75pt,
        mark options={solid,fill=white,line width=0.85pt},
        restrict x to domain=267:285,
      ] table[
        x=accepted_submission_index,
        y expr={
          (\thisrow{peak_qubits}+16)
          *(\thisrow{avg_executed_toffoli}+3*2^16)
          *28/1000000000
        },
      ] {\historydata};

      \addplot+[
        correctedorange,
        dashed,
        dash pattern=on 4pt off 2.2pt,
        line width=1.15pt,
        mark=triangle*,
        mark size=1.90pt,
        mark options={solid,fill=correctedorange,line width=0.65pt},
        restrict x to domain=267:285,
      ] table[
        x=accepted_submission_index,
        y expr={
          (\thisrow{peak_qubits}+16)
          *(\thisrow{avg_executed_toffoli}+3*2^16)
          *28/1000000000
          /(1-28*(1-\thisrow{correctness_probability}))
        },
      ] {\historydata};

      \node[
        font=\tiny\bfseries,
        text=uncorrectedblue,
        fill=white,
        inner sep=1.0pt,
      ] at (axis cs:276,61.65) {$-0.626$ B};

      \node[
        font=\tiny\bfseries,
        text=correctedorange,
        fill=white,
        inner sep=1.0pt,
      ] at (axis cs:276,62.58) {$+0.087$ B};
    \end{axis}
  \end{scope}

  \draw[black!65,line width=0.9pt]
    (zoomcenterB) circle[radius=1.30cm];

  \coordinate (zoomcenterC) at ($(scoreaxis.south west)+(9.65cm,2.00cm)$);
  \draw[black!35,line width=0.55pt]
    ($(zoomsourceC)+(-0.23cm,0.12cm)$)
    -- ($(zoomcenterC)+(0.60cm,-1.07cm)$);
  \draw[black!35,line width=0.55pt]
    ($(zoomsourceC)+(0.23cm,0.12cm)$)
    -- ($(zoomcenterC)+(1.00cm,-0.73cm)$);
  \draw[
    fill=white,
    fill opacity=0.40,
    draw=black!55,
    line width=0.75pt,
  ] (zoomsourceC) ellipse[x radius=0.26cm,y radius=0.17cm];

  \fill[black,opacity=0.13]
    ($(zoomcenterC)+(0.06cm,-0.06cm)$) circle[radius=1.25cm];
  \fill[white] (zoomcenterC) circle[radius=1.25cm];

  \begin{scope}
    \clip (zoomcenterC) circle[radius=1.22cm];
    \begin{axis}[
      at={($(zoomcenterC)+(0.06cm,0)$)},
      anchor=center,
      scale only axis,
      width=1.65cm,
      height=1.45cm,
      xmin=387,
      xmax=403,
      ymin=50.75,
      ymax=53.75,
      xtick={391,401},
      ytick={51,52,53},
      xticklabel style={font=\tiny\bfseries},
      yticklabel style={
        font=\tiny,
        xshift=8pt,
        fill=white,
        inner sep=0.5pt,
      },
      axis line style={draw=black!45,line width=0.4pt},
      grid=both,
      major grid style={draw=gridgray,line width=0.3pt},
      minor grid style={draw=gridgray!60,line width=0.2pt},
      minor x tick num=1,
      minor y tick num=1,
      axis background/.style={fill=white},
      clip mode=individual,
    ]
      \addplot+[
        uncorrectedblue,
        line width=1.15pt,
        mark=*,
        mark size=1.75pt,
        mark options={solid,fill=white,line width=0.85pt},
        restrict x to domain=387:403,
      ] table[
        x=accepted_submission_index,
        y expr={
          (\thisrow{peak_qubits}+16)
          *(\thisrow{avg_executed_toffoli}+3*2^16)
          *28/1000000000
        },
      ] {\historydata};

      \addplot+[
        correctedorange,
        dashed,
        dash pattern=on 4pt off 2.2pt,
        line width=1.15pt,
        mark=triangle*,
        mark size=1.90pt,
        mark options={solid,fill=correctedorange,line width=0.65pt},
        restrict x to domain=387:403,
      ] table[
        x=accepted_submission_index,
        y expr={
          (\thisrow{peak_qubits}+16)
          *(\thisrow{avg_executed_toffoli}+3*2^16)
          *28/1000000000
          /(1-28*(1-\thisrow{correctness_probability}))
        },
      ] {\historydata};

      \node[
        font=\tiny\bfseries,
        text=uncorrectedblue,
        fill=white,
        inner sep=1.0pt,
      ] at (axis cs:396,51.75) {$-0.351$ B};

      \node[
        font=\tiny\bfseries,
        text=correctedorange,
        fill=white,
        inner sep=1.0pt,
      ] at (axis cs:396,52.60) {$+0.195$ B};
    \end{axis}
  \end{scope}

  \draw[black!65,line width=0.9pt]
    (zoomcenterC) circle[radius=1.25cm];
\end{tikzpicture}

%% file: sections/06-limitations.tex
\section{Limitations, Safety, and Responsible Framing}
\label{sec:limitations}

The results of \Cref{sec:results} are operating points on a public competition leaderboard, not validated fault-tolerant resource estimates. Their interpretation therefore requires attention to measurement integrity, finite-support correctness, and attribution within human--agent workflows.

\paragraph{The score is sampled and can be optimized through support selection.} The benchmark measures average \emph{executed} Toffoli count over a deterministic, submission-dependent set of Fiat--Shamir inputs (\Cref{sec:evaluator}). Because the input set is derived by hashing the submitted operation stream, changing a submission's nonce can select a more favorable validation ``island'' on which the same semantic circuit executes fewer Toffoli gates. Such nonce search is not a circuit optimization, yet it can lower the reported score. It constitutes the \emph{verifier-search} bucket of \Cref{fig:family-census-tree} ($4.8\%$ of scored source commits), and approximately half of the audited commits include a reroll alongside a circuit modification. We distinguish rerolling from gate removal by comparing source changes and emitted operation streams, including static Toffoli counts after excluding nonce-tail and padding changes. An improvement is classified as verifier search only when these comparisons identify no semantic circuit-body change.

\paragraph{Approximate circuits can pass finite validation.} A transformation that fails on a small fraction of inputs may still pass the $9{,}024$-instance verifier if its nonce is selected to avoid the failing cases. Such a circuit is validated only over the selected support and does not establish all-input correctness. We therefore classify approximate reductions by their circuit mechanism while explicitly reporting their support-conditioned status. The archive also contains \texttt{DUMMY\_TOFFOLIS}, a self-canceling padding mechanism that changes the reported count without altering the circuit's semantics. These cases demonstrate that the leaderboard score is an empirical cost over finite, submission-selected support rather than a certified bound on correctness or cost over the complete input domain.

\paragraph{Human insight and agent automation are complementary.} AI agents proved effective at implementing, testing, and refining circuit optimizations, particularly across the many incremental changes that would otherwise require substantial manual effort. They also occasionally helped develop structural transformations. The post-cutoff comparison-free ping-pong dialog-GCD provides the clearest example: an agent-mediated workflow combined prior plus-minus GCD and dialog-replay ideas into a structurally distinct fixed-alternation backend that substantially reduced the Toffoli count (\Cref{sec:ping-pong-dialog-gcd}). At the same time, human research remained essential. The earlier dialog-GCD improvement followed the record-and-replay architecture introduced by Khattar et al.~\cite{ksgz25} and adapted to ECDLP point addition by Schrottenloher~\cite{s26}, while humans supplied research connections, objectives, interpretation, and technical review. Accordingly, \emph{agent-identified} denotes an optimization surfaced, adapted, combined, or developed within an agent-mediated workflow; it does not imply autonomous invention, exclusive model authorship, or independent novelty.

\paragraph{Windowed point-addition compatibility.} The benchmark optimizes mixed addition with a classically known addend, whereas windowed Shor (\Cref{sec:ecc-quantum}) selects the addend coherently through a QROM lookup~\eqref{eq:windowed-add}, so optimizations that specialize on classical coordinates must be replaced by coherent operations. We implemented a windowed-compatible variant of the best-scoring circuit (\Cref{sec:windowed-addition-compatible-circuit}) that retains the arithmetic core within a lookup/use/unlookup construction, at an overhead of $11$ logical qubits and about $29.6\%$ average executed Toffoli per addition and no detectable error-rate difference on the paired corpus ($29$ windowed-only versus $30$ baseline-only failures; two-sided exact sign test, $p=1.0$). This demonstrates the adaptation at the point-addition level but covers only a single call on the generic-affine domain of \Cref{sec:task}, which for nonidentity $A$ excludes $R\in\{A,-A,-2A\}$; coherent correctness is argued from linearity rather than established through direct simulation. We therefore reserve full-Shor claims until the complete coherent construction is assembled and evaluated end to end.

\paragraph{The public archive has an incomplete attempt denominator.}
The accepted-commit history records solutions that reached the public branch,
and the submissions API records candidates that were sent to the service. It
does not exhaustively record local experiments, discarded branches, failed
builds, private agent sessions, or hypotheses that participants chose not to
submit. Even the population of rejected submissions is therefore a selected subset
of all attempted work. The optimization census and chronology characterize
observable artifacts, not the number or distribution of research attempts.
They cannot by themselves estimate success rates per attempt, discovery time,
the frequency of negative results, or the productivity of a participant or
tool. Claims about failure accumulation are limited to failures that were
actually preserved in a public note, submission, commit, or deposited session.

\paragraph{The trajectory is observational rather than a causal comparison.}
Participants self-selected their tools, starting commits, objectives, compute
budgets, and the results they exposed. Tool adoption also changed over time as
the circuit, community, and available base solutions evolved. There was no
random assignment to human-only and agent-assisted workflows, no common attempt
budget, and no complete prompt or wall-clock log. The observed record supports
the descriptive claim that substantial improvements occurred in an
agent-mediated open workflow. It does not isolate the counterfactual improvement
that the same participants would have achieved without agents, nor establish
that one model, prompting strategy, or collaboration structure caused a given
frontier movement. We therefore keep implementation provenance, contributor
statements, and causal effectiveness claims distinct.

\paragraph{Responsible framing.} This work improves a reversible point-addition primitive relevant to quantum attacks on elliptic-curve cryptography, but it does not constitute an executable attack or a complete fault-tolerant resource estimate. The reported circuits omit physical error correction, architecture-dependent compilation, full non-Clifford accounting, and validated integration into windowed Shor arithmetic. The results therefore refine the understanding of one logical-resource bottleneck rather than establish an immediate capability to break \texttt{secp256k1}. We report them openly to support reproducible research, informed cryptographic risk assessment, and the timely transition to post-quantum cryptography.

%% file: sections/07-conclusion.tex
\section{Conclusion and Outlook}

This work presented {\challengename}, an open effort to optimize reversible secp256k1 point-addition circuits relevant to Shor's algorithm for the elliptic-curve discrete logarithm problem. Starting from a baseline of $2{,}715$ logical qubits and $3{,}960{,}753$ average executed Toffoli gates, the challenge reduced the best $Q\times T$ operating point to $1{,}151$ qubits and $1{,}299{,}453$ average executed Toffoli gates, corresponding to an $86.1\%$ reduction in the product metric. A separate low-width search reached $825$ logical qubits at the data cutoff, subsequently improved to $813$, exposing a markedly different region of the reversible time--space trade-off.

Beyond the resulting circuits, {\challengename} provides a case study of \emph{Open Autoresearch}: a verifier-gated research process in which human participants and AI agents iteratively generate, implement, test, and share candidate improvements against a common measurable objective. The public repository, evaluator, leaderboard, and submission history allowed individually produced improvements to become common starting points for subsequent work, while the resulting source history made it possible to reconstruct both incremental refinements and larger architectural transitions. The experience suggests that human judgment, agent-scale experimentation, and open sharing can be complementary when progress can be evaluated rapidly and reproducibly, while the observational nature of the challenge does not by itself establish the causal contribution of any particular agent, model, or workflow.

The reported operating points should nevertheless be interpreted within the benchmark that produced them. Correctness is established over finite, submission-dependent validation support rather than over the complete input domain, and the scored primitive assumes a classically specified addend. We showed this last restriction to be an interface convenience rather than a fundamental barrier, building a coherent windowed-compatible variant of the best-scoring circuit at modest overhead (\Cref{sec:windowed-addition-compatible-circuit}); this demonstration is limited to a single point-addition call, however, and the scored results do not by themselves constitute complete fault-tolerant attack estimates.

Building on the window-compatible point-addition kernel, the next step is a complete windowed implementation of Shor's algorithm, including the shifted tables, full window schedule, Fourier layers, and postprocessing, with end-to-end validation and comparable resource accounting. Beyond \texttt{secp256k1}, an important question is whether Open Autoresearch transfers to other hard optimization problems with machine-checkable objectives. {\challengename} provides one concrete and source-auditable case. Testing the generality of that model is a natural direction for future work.


%% file: sections/08-acknowledgments.tex
\section*{Acknowledgments}

We thank Craig Gidney and Tanuj Khattar of Google Quantum AI for carefully reviewing earlier drafts of this manuscript and for their constructive comments and suggestions, which substantially improved its technical accuracy, clarity, and presentation. We also thank Joe Doyle of Trail of Bits for his helpful suggestions and insightful discussions. Any remaining errors are our own.

%% file: sections/09-contributors.tex
\section{Contributor Statement}
\label{app:credit-contributors}

\begingroup
\small
\setlength{\emergencystretch}{2em}
\noindent The categories below are nonexclusive. Individual circuit submissions and their provenance remain recorded on the public leaderboard.

\paragraph{\textbf{Paper contributors}.} Jieyi Long led the drafting, synthesis, coordination, and manuscript-wide integration of the paper. Theodore Pender, Zhao Huang, Manuel B. Santos, and Samrendra Kumar Singh contributed major technical and empirical sections, including the quantum-circuit background, circuit-optimization analysis, results, improvement chronology, and Pareto-frontier analysis. Justin Drake, Pierre-Luc Dallaire-Demers, and Bartosz Naskr\k{e}cki contributed technical review and revision, and Francesco Giannicola contributed detailed review, reproducibility study, and editorial refinement.

\paragraph{\textbf{Challenge organizers}.} Gajesh Naik, Gautham Anant, Soubhik Deb, and Justin Drake contributed to the conception, harness setup, organization, infrastructure, administration, or supervision of the challenge.

\paragraph{\textbf{Leaderboard contributors}.} The challenge's progress reflects the cumulative contributions of participants who performed circuit research, implementation, testing, validation, and submission work. Within this broader collective effort, several key optimization techniques and plateau-breaking improvements were identified by Theodore Pender, BitWonka, Bartosz Naskr\k{e}cki, Joe Doyle, Matt Zweil, and Gajesh Naik, working in collaboration with their AI agents. These particularly consequential contributions complemented the many other advances, refinements, and validation efforts made by the wider leaderboard community. \textbf{Named leaderboard contributors}: Akash Balasubramani, BitWonka, John Boyer, Xavier Butler, Pierre-Luc Dallaire-Demers, Bereket Dereje, Michael Dong, Joe Doyle, Oli Freuler, Vasily Gnuchev, Alexander Hertlein, Zhao Huang, Anto Joseph, Robert Kodra, Edison Lee, Jackie Chia-Hsun Lee, Lucas Levy, Alan Li, Jieyi Long, Gajesh Naik, Bartosz Naskr\k{e}cki, Jordan Newman, Ruben Marcus Luz Paschoarelli, Shaan Patel, Theodore Pender, JT Rose, Manuel B. Santos, Samrendra Kumar Singh, Okechukwu Wisdom. \textbf{Other Leaderboard Contributors (Github ID)}: 0xfoobar,\allowbreak{} 0xnirlin,\allowbreak{} 0xpg,\allowbreak{} 10d9e,\allowbreak{} abipalli,\allowbreak{} aburan28,\allowbreak{} agilexmarketing,\allowbreak{} alecstecpe-oss,\allowbreak{} AnalyticETH,\allowbreak{} Angel-Arevalo,\allowbreak{} anshu4321,\allowbreak{} anupsv,\allowbreak{} austinamissah,\allowbreak{} ayushgupta0610,\allowbreak{} blocksek,\allowbreak{} borelien,\allowbreak{} bulengerk,\allowbreak{} bxue-l2,\allowbreak{} chainyoda,\allowbreak{} Chris-Moller,\allowbreak{} cryptogakusei,\allowbreak{} darius-oai,\allowbreak{} DavetheSlayer,\allowbreak{} davidxia3,\allowbreak{} dennisonbertram,\allowbreak{} eferbarn,\allowbreak{} eiso,\allowbreak{} factory-sagar,\allowbreak{} fortunexbt,\allowbreak{} Gaijin-01,\allowbreak{} gopikannappan,\allowbreak{} harrigan,\allowbreak{} hydrogenbond007,\allowbreak{} inishc,\allowbreak{} IntrepidEnki,\allowbreak{} jacksonhblau,\allowbreak{} jamesotter99,\allowbreak{} Jamgiter,\allowbreak{} Joshgriste,\allowbreak{} josusanmartin,\allowbreak{} jtaroreh,\allowbreak{} junpenglao,\allowbreak{} kamilsa,\allowbreak{} kunalvg,\allowbreak{} lucasabu1988,\allowbreak{} Luigy-Lemon,\allowbreak{} Lulu-Zhou-EigenLabs,\allowbreak{} M3kko,\allowbreak{} makimakiver,\allowbreak{} megabyte0x,\allowbreak{} mmurrs,\allowbreak{} mochimodev,\allowbreak{} mpjunior92,\allowbreak{} nandy-technologies,\allowbreak{} nikhiljha,\allowbreak{} nnn-gif,\allowbreak{} novoyd,\allowbreak{} pavvann,\allowbreak{} phileigenlabs,\allowbreak{} philuponcrypto,\allowbreak{} pq-cybarg,\allowbreak{} prasiddhnaik,\allowbreak{} pschork,\allowbreak{} ptrdsh,\allowbreak{} runeape-sats,\allowbreak{} rwnq8,\allowbreak{} Sam-Scolari,\allowbreak{} Sampriv,\allowbreak{} saucegodbased,\allowbreak{} simonik11,\allowbreak{} skb93,\allowbreak{} solimander,\allowbreak{} stevenhao,\allowbreak{} steventhornton,\allowbreak{} swadhin,\allowbreak{} Tasfia-17,\allowbreak{} tekkac,\allowbreak{} TeS44,\allowbreak{} tmid1,\allowbreak{} twangodev,\allowbreak{} twmmason,\allowbreak{} vatsalkshah,\allowbreak{} vineetguptadev,\allowbreak{} wh1sky02,\allowbreak{} xadenryan,\allowbreak{} ygboucherk,\allowbreak{} YQ-Wang,\allowbreak{} yudduy,\allowbreak{} yudongcao,\allowbreak{} zeeshan8281,\allowbreak{} zigtur,\allowbreak{} zpano,\allowbreak{} zuiris.

\endgroup